\documentclass[twocolumn,english,amsmath,amssymb,superscriptaddress,nofootinbib]{revtex4-2}
\usepackage{amsmath}
\usepackage{graphicx}
\usepackage{amssymb}
\usepackage{dsfont}
\usepackage{color}
\usepackage{lipsum,kantlipsum}
\usepackage[bookmarks=true,colorlinks,linkcolor=blue,urlcolor=blue,citecolor=blue]{hyperref}
\usepackage{mathtools}
\usepackage{amsthm}
\usepackage{bm}
\usepackage[shortlabels]{enumitem}

\usepackage{mathrsfs}

\usepackage{ifthen}
\usepackage{tikz}
    \usetikzlibrary{decorations.pathreplacing}
    \usetikzlibrary{shapes.geometric}
    \usetikzlibrary{plotmarks}
    \usetikzlibrary{calc}
    \usetikzlibrary{fadings}
    \usetikzlibrary[
    decorations,
    decorations.pathmorphing,
    snakes,
]
\usepackage{pgfplots}
\usepgfplotslibrary{colormaps}
\usepgfplotslibrary{patchplots}

\pgfplotsset{
	compat=newest
}

\theoremstyle{plain}

\newcommand{\beq}{\begin{equation}}
\newcommand{\eeq}{\end{equation}}
\newcommand{\bea}{\begin{eqnarray}}
\newcommand{\eea}{\end{eqnarray}}

\newcommand{\ep}{\text{EP}}
\newcommand{\disc}{\text{disc}}

\newcommand{\eq}{\begin{equation}}
\newcommand{\en}{\end{equation}}
\newcommand{\ear}{\begin{eqnarray}}
\newcommand{\rae}{\end{eqnarray}}

\newcommand{\tr}{{\rm tr}\,}

\newcommand{\id}{\mathds{1}}
\newcommand{\ii}{\text{i}}
\newcommand{\be}{\begin{eqnarray}}
\newcommand{\ee}{\end{eqnarray}}
\newcommand{\non}{\nonumber}

\usepackage[most]{tcolorbox}

\newtheorem*{algo*}{Algorithm}

\numberwithin{equation}{section}

\makeindex

\begin{document}

\title{Exceptional Points, Bulk-Boundary Correspondence, and Entanglement Properties\\ for a Dimerized Hatano-Nelson Model with Staggered Potentials}

\author{Yasamin Mardani}
\email{mardaniy@myumanitoba.ca}
\affiliation{Department of Physics and Astronomy, University of Manitoba, Winnipeg R3T 2N2, Canada}

\author{Rodrigo A. Pimenta}
\email{rodrigo.pimenta@ufrgs.br}
\affiliation{Department of Physics and Astronomy, University of Manitoba, Winnipeg R3T 2N2, Canada}
\affiliation{Instituto de F\'isica, Universidade Federal do Rio Grande do Sul, Porto Alegre 91501-970, Brazil}

\author{Jesko Sirker}
\email{sirker@physics.umanitoba.ca}
\affiliation{Department of Physics and Astronomy, University of Manitoba, Winnipeg R3T 2N2, Canada}

\date{\today{}}

\begin{abstract}
It is well-known that the standard bulk-boundary correspondence does not hold for non-Hermitian systems in which also new phenomena such as exceptional points do occur. Here we study, mostly 
by analytical means, a paradigmatic one-dimensional non-Hermitian model with dimerization, asymmetric hopping, and imaginary staggered potentials. We present analytical solutions for the singular-value and the eigensystem of this model with both open and closed boundary conditions. We explicitly demonstrate that the proper bulk-boundary correspondence is between topological winding numbers in the periodic case and singular values, {\it not eigenvalues}, in the open case. These protected singular values are connected to hidden edge modes which only become exact zero-energy eigenmodes in the semi-infinite chain limit. We also show that a non-trivial topology leads to protected eigenvalues in the entanglement spectrum. In the $\mathcal{PT}$-symmetric case, we find that the model has a so far overlooked phase where exceptional points become dense in the thermodynamic limit. This phase shows unusual hyper-ballistic transport properties with a dynamical critical exponent $z=1/2$.
\end{abstract}

\maketitle

\section{Introduction}

It is well known that fundamental principles in quantum mechanics are formulated using Hermitian operators,
which assure for example unitary time evolution of closed systems.
However, practically, no quantum system is perfectly isolated. The way a system interacts with its environment is often represented through a Master equation like the Lindblad equation \cite{L1976}.
If quantum jumps can be neglected, then the time evolution described by a Lindblad equation can be replaced by an effective non-Hermitian (NH) Hamiltonian \cite{RPBC2022}. The validity of this effective
approach has been verified in different
experimental arrangements \cite{MiriAlu,SuEstrecho,YangWang}. Strictly speaking, it requires a post-selection of those time evolutions which stay in the considered manifold \cite{NaghilooAbbasi}. Another motivation to study non-Hermitian Hamiltonians came from realizing that hermiticity is not required to guarantee a real spectrum. The weaker condition of $\mathcal{PT}$-symmetry is sufficient, see e.g.~Ref.~\cite{Bender_2005} for a review.

Generic non-Hermitian systems exhibit distinct properties associated with
their complex spectrum and the difference between
left and right eigenvectors \cite{AGU2021}.
In this context, as with Hermitian systems, Gaussian non-Hermitian models can offer valuable insights into possible phases and phenomena. In particular, the
topological properties of Gaussian non-Hermitian 
models have been studied and new phenomena have been revealed
\cite{BL2002,ESHK2011,L2016,L2018,MABVFT18,YZ2018,YJLLC2018,L2018,KSUS2019,BBK2021,DFM2022}. For example, non-hermiticity leads to a proliferation
of possible symmetry classes: the famous
Altland-Zirnbauer
ten-fold classification \cite{AZ1997} for Hermitian
Gaussian models
is replaced by a 38-fold classification \cite{BL2002,KSUS2019} in the non-Hermitian
domain. The richer classification scheme is a direct consequence of the distinction between
transposition and conjugate transposition.
Other distinctive properties include the extreme sensitivity of the eigenspectrum to the boundary conditions \cite{MABVFT18,YZ2018,BBK2021} leading to the skin effect and the breakdown of the
conventional bulk-boundary correspondence \cite{L2016,YZ2018,GAKTHU2018,Kunst}. As has very recently been proven, the proper bulk-boundary correspondence in non-Hermitian systems is based on the theory of Toeplitz operators and is between winding numbers of the symbols of such operators in the periodic case and the singular-value spectrum in the open case \cite{MonkmanSirkerNH}. Physically, it is important to note that the topologically protected singular values in general belong to boundary states which are metastable for finite systems and only become exact eigenstates in the limit of a semi-infinite system. The Hermitian case can then be understood as a special limit of this more general theory because in this case the singular values are just the absolute values of the eigenvalues. Another remarkable aspect of non-Hermitian Hamiltonians is
the existence of so-called exceptional points \cite{H2012} where both
eigenvalues and eigenvectors coalesce. Such exceptional points can be detected in experiments \cite{NaghilooAbbasi}. 

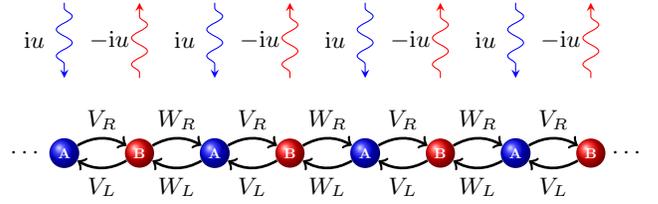
\begin{figure}[t]
\centering
\begin{tikzpicture}
\foreach \i in {1,3,5,7}
        \node (\i) [shade,shading=ball,circle,ball color=blue,minimum size=3pt,text=white,font=\bfseries,scale=0.6] at (\i,0)  {A};
\foreach \i in {2,4,6,8}        
        \node (\i) [shade,shading=ball,circle,ball color=red,minimum size=3pt,text=white,font=\bfseries,scale=0.6] at (\i,0)  {B};
\draw[line width=1pt,bend left,->]  (1) to node [auto] {$V_R$} (2);
\draw[line width=1pt,bend left,->]  (3) to node [auto] {$V_R$} (4);
\draw[line width=1pt,bend left,->]  (5) to node [auto] {$V_R$} (6);
\draw[line width=1pt,bend left,->]  (7) to node [auto] {$V_R$} (8);
\draw[line width=1pt,bend left,->]  (2) to node [auto] {$W_R$} (3);
\draw[line width=1pt,bend left,->]  (4) to node [auto] {$W_R$} (5);
\draw[line width=1pt,bend left,->]  (6) to node [auto] {$W_R$} (7);
\draw[line width=1pt,bend left,->]  (2) to node [auto] {$V_L$} (1);
\draw[line width=1pt,bend left,->]  (4) to node [auto] {$V_L$} (3);
\draw[line width=1pt,bend left,->]  (6) to node [auto] {$V_L$} (5);
\draw[line width=1pt,bend left,->]  (8) to node [auto] {$V_L$} (7);
\draw[line width=1pt,bend left,->]  (3) to node [auto] {$W_L$} (2);
\draw[line width=1pt,bend left,->]  (5) to node [auto] {$W_L$} (4);
\draw[line width=1pt,bend left,->]  (7) to node [auto] {$W_L$} (6);
\node at (0.5,0) {$\cdots$};
\node at (8.5,0) {$\cdots$};
\draw[-stealth,decorate,decoration={snake,amplitude=3pt,pre length=2pt,post length=3pt},color=blue]
(1,2) -- (1,1) ;
\node at (0.6,1.5) {$\ii u$};
\draw[-stealth,decorate,decoration={snake,amplitude=3pt,pre length=2pt,post length=3pt},color=blue]
(3,2) -- (3,1) ;
\node at (2.6,1.5) {$\ii u$};
\draw[-stealth,decorate,decoration={snake,amplitude=3pt,pre length=2pt,post length=3pt},color=blue]
(5,2) -- (5,1) ;
\node at (4.6,1.5) {$\ii u$};
\draw[-stealth,decorate,decoration={snake,amplitude=3pt,pre length=2pt,post length=3pt},color=blue]
(7,2) -- (7,1) ;
\node at (6.6,1.5) {$\ii u$};
\draw[-stealth,decorate,decoration={snake,amplitude=3pt,pre length=2pt,post length=3pt},color=red]
(2,1) -- (2,2);
\node at (1.6,1.5) {$-\ii u$};
\draw[-stealth,decorate,decoration={snake,amplitude=3pt,pre length=2pt,post length=3pt},color=red]
(4,1) -- (4,2);
\node at (3.6,1.5) {$-\ii u$};
\draw[-stealth,decorate,decoration={snake,amplitude=3pt,pre length=2pt,post length=3pt},color=red]
(6,1) -- (6,2);
\node at (5.6,1.5) {$-\ii u$};
\draw[-stealth,decorate,decoration={snake,amplitude=3pt,pre length=2pt,post length=3pt},color=red]
(8,1) -- (8,2);
\node at (7.6,1.5) {$-\ii u$};
\end{tikzpicture}
\caption{\label{fig:HNmodel} The considered minimal model has a two-site unit cell, asymmetric 
intra- ($V_L,V_R$) and inter-cell ($W_L,W_R$) hoppings as well as complex staggered potentials $\pm iu$.
}
\end{figure}

In order to highlight the different phenomena possible in Gaussian non-Hermitian models and to demonstrate in a fully analytical manner the recent novel results regarding their topology \cite{MonkmanSirkerNH}, we will study a minimal model which allows to turn various symmetries on and off. In the Hermitian case, one of the best-known one-dimensional systems with symmetry-protected topological order is the Su-Schrieffer-Heeger (SSH) model \cite{SSH1979} which has chiral symmetry and belongs to the BDI class. This model has alternating strong and weak hoppings and topological edge modes are present if the two sites at the end of an open chain are weakly coupled to their respective neighbors. Here, we make the model non-Hermitian by allowing for parity breaking hoppings, i.e., different hopping amplitudes to the left and to the right. In addition, we also introduce  staggered imaginary onsite potentials to allow for a non-Hermitian, $\mathcal{PT}$-symmetric limit. Physically, parity breaking can arise in an effective description of an SSH model with gain and loss. This phenomenon has been investigated in plasmonic chains \cite{Ling} and photonic lattices \cite{Longhi}. 
The model is depicted in Fig.~\ref{fig:HNmodel} 
and can be viewed either as a dimerized Hatano-Nelson model \cite{HN1996} or an SSH chain with asymmetric hopping.

The main motivation for our paper is twofold: (1) To demonstrate a mathematically rigorous bulk-boundary correspondence using the apparatus developed in Ref.~\cite{MonkmanSirkerNH} for a concrete, physically relevant model. (2) To re-examine the phase diagram of a model which contains many limiting cases which have been studied previously in the literature but to do so, as far as possible, analytically and in a rigorous manner instead of using exact diagonalizations. In the process, we discover a novel exceptional phase with unusual properties.

Our paper is organized as follows: In Sec.~\ref{Model}, we introduce the model and the difference equations the right and left eigenvectors have to fulfill. In Sec.~\ref{sec:EDCBC}, we diagonalize the model for periodic and aperiodic boundary conditions. Furthermore, we determine the parameters for which the excitation gap closes and for which exceptional points do occur. We also point out that exceptional points can occur in the complex plane of the model parameters. We introduce and determine the two winding numbers for this system. In the $\mathcal{PT}$-symmetric phase we discover a phase where exceptional points are dense leading to hyper-ballistic transport. In Sec.~\ref{sec:EDOBC}, we then analytically study the open boundary case. As expected, the spectrum and the eigenvectors change completely as compared to the closed boundary case. This is also true for the conditions for the closing of the excitation gap and for the location of exceptional points. Depending on the parity of the length of the chain, we find different families of exceptional points which are determined via certain transcendental equations, also reflected in a factorization of the discriminant of the associated characteristic polynomial. They are clearly different from the closed boundary case. In Sec.~\ref{BBC}, we present some of the central results of our paper. We analytically calculate the singular value spectrum of a finite open chain as well as the zero-energy eigenstates of a semi-infinite chain. Based on these results we explicitly establish a bulk-boundary correspondence for this specific model following results recently proven in general in Ref.~\cite{MonkmanSirkerNH}. We explicitly show in particular that the topologically protected boundary modes are metastable for finite system size and only become exact eigenstates in the limit of a semi-infinite system. In Sec.~\ref{Ent}, we then discuss in detail the entanglement entropy when cutting the system in half. The studied model is Gaussian and therefore its correlation matrix can be obtained
\cite{
HRB2019,CYWR2020,GYHYCLX2021,L2022,CPLL2022,
HFZ2023,FAC2023,GTS2023}. We find that in topological phases there are topologically protected eigenvalues in the entanglement spectrum. We note that the entanglement entropy is complex in general in the non-Hermitian case and find interesting properties of the entanglement entropy which can be directly tied to the physical properties of the system. In particular, we find that in the $\mathcal{PT}$-phase where exceptional points are dense the entanglement entropy is a nowhere continuous function. The last section is devoted to a short summary and our conclusions.

\section{The dimerized Hatano-Nelson model}
\label{Model}
We consider a model with a two-site unit cell with asymmetric hopping in both
intra ($V_L,V_R$) and inter ($W_L,W_R$) cells, see Fig.~\ref{fig:HNmodel}. Complex staggered fields
$\pm\ii u$ act on the sites of the chain. This model can be viewed either as a dimerized Hatano-Nelson model or as a Su-Schrieffer-Heeger (SSH) model with asymmetric hopping. It is described by the Hamiltonian
\bea
\label{ham}
\mathcal{H}&=&
\sum_{j=1}^{\lfloor\frac{N-1}{2}\rfloor}
W_L c_{2j}^\dagger c_{2j+1}+W_R c_{2j+1}^\dagger c_{2j}
\non\\&+&
\gamma (W_R  c_{1}^\dagger c_{L}+W_L  c_{L}^\dagger c_{1})
\non\\&+&
\sum_{j=1}^{\lfloor\frac{N}{2}\rfloor}
V_L c_{2j-1}^\dagger c_{2j}+V_R c_{2j}^\dagger c_{2j-1}
\non\\&-&
\ii u\sum_{j=1}^{\lfloor\frac{N}{2}\rfloor}
 c_{2j}^\dagger c_{2j}+
\ii u\sum_{j=1}^{\lfloor\frac{N+1}{2}\rfloor}
 c_{2j-1}^\dagger c_{2j-1},
\eea
where $c_j$ and $c_j^\dagger$ are fermionic operators satisfying the algebra
\beq\label{calgebra}
\{c_j,c_k\}=\{c_j^\dagger,c_k^\dagger\}=0,
\quad 
\{c_j,c_k^\dagger\}=\delta_{j,k}
\eeq
and $\lfloor x \rfloor$ denotes the floor function. By writing the Hamiltonian in the form
(\ref{ham}), we can consider
different boundary conditions (controlled by the parameter $\gamma$) in a uniform fashion. In most parts of the paper, we assume that the parameters $W_{L,R}$, $V_{L,R}$ and $u$
are real. However, when discussing the exceptional points of this model it makes sense to consider complex parameters as well.

The Hamiltonian (\ref{ham}) can be written
in the compact form
\beq
\mathcal{H}=\boldsymbol{c}^\dagger \mathcal{T}_N \boldsymbol{c}
\eeq
where
$\mathcal{T}_N$ is an asymmetric (almost) tridiagonal matrix
with
dimension $N\times N$ with matrix elements
\be\label{tri}
&&t_{2j-1,2j}=V_L,\quad
t_{2j,2j-1}=V_R,\quad t_{2j,2j}=-\ii u,\non\\
&&t_{2j,2j+1}=W_L,\quad
t_{2j+1,2j}=W_R,\quad t_{2j-1,2j-1}=\ii u,\non\\
&&t_{1,N}=\gamma W_R,\quad t_{N,1}=\gamma W_L,
\ee
and $\boldsymbol{c}^\dagger=(c_1^\dagger,\dots,c_N^\dagger)$, $\boldsymbol{c}=(c_1,\dots,c_N)^T$ with $T$ denoting transposition. Note that in the semi-infinite case, the Hamilton operator has Toeplitz block form with $2\times 2$ matrices
\begin{equation}
h_0=\begin{pmatrix} iu & V_L \\ V_R & -iu \end{pmatrix} \; , \; 
h_1=\begin{pmatrix} 0 & 0 \\ W_L & 0 \end{pmatrix} \; , \;
h_{-1}=\begin{pmatrix} 0 & W_R \\ 0 & 0 \end{pmatrix}
\end{equation}
which repeat along the diagonals. This will allow us later to define a corresponding symbol and winding numbers.

To make this paper self-contained, we will recall the diagonalization of
the asymmetric SSH chain for both closed ($\gamma=\pm 1)$
and open ($\gamma=0$) boundary conditions, referred to respectively as CBC or OBC. In both cases,
the diagonalization of $\mathcal{T}_N$ with entries (\ref{tri}) follows from difference equations of the spectral problem. Here the lack of hermiticity of the single-body Hamiltonian implies that left and right eigenvectors have to be considered. The spectral problems read

\beq
\mathcal{T}_N~|\vec{r}\rangle = \epsilon~|\vec{r}\rangle,
\quad
\langle\vec{\ell}|~\mathcal{T}_N=\epsilon~\langle\vec{\ell}|,
\eeq
where $|\vec{r}\rangle=(r_{1},\dots,r_{N})^T$
and $\langle\vec{\ell}|=(\ell_{1},\dots,\ell_{N})$ are right and left eigenvectors of $\mathcal{T}_N$, respectively, with eigenvalue $\epsilon$.

The components of the right eigenvector satisfy
the coupled difference equations
\bea
&&V_L r_{2j}+W_R r_{2j-2}=(\epsilon-\ii u) r_{2j-1},\label{difeq1}\\
&&W_L r_{2j+1}+V_R r_{2j-1}=(\epsilon+\ii u) r_{2j},\label{difeq2}
\eea
with the boundary conditions
\beq\label{BC}
r_0=\gamma r_N,\quad r_{N+1}=\gamma  r_1.
\eeq
Exchanging $R\leftrightarrow L$ one obtains
the analogous equations for $\ell_j$, namely
\bea
&&V_R \ell_{2j}+W_L \ell_{2j-2}=(\epsilon-\ii u) \ell_{2j-1},\label{difeq1L}\\
&&W_R \ell_{2j+1}+V_L \ell_{2j-1}=(\epsilon+\ii u) \ell_{2j},\label{difeq2L}
\eea
with the boundary conditions,
\beq\label{BCL}
\ell_0=\gamma \ell_N,\quad \ell_{N+1}=\gamma  \ell_1.
\eeq
For $\gamma=0$, we consider (\ref{BC}, \ref{BCL}) for both parities of $N$,
while for $\gamma\neq0$ we only consider the case where $N$ is even. In the following sections, each boundary condition is considered separately.

\section{Closed boundary conditions}\label{sec:EDCBC}

\subsection{Diagonalization}

Let us briefly recapitulate the diagonalization of 
$\mathcal{T}_N$ for
closed boundary conditions (CBC), that is, $\gamma=\pm 1$. In this case, $\mathcal{T}_N$ is a block circulant
matrix and can be easily diagonalized. We only consider $N$ even for CBC.
The difference equations (\ref{difeq1},\ref{difeq2},\ref{BC})
and (\ref{difeq1L},\ref{difeq2L},\ref{BCL}) are easily
solved by the ansatz
\be\label{rcomp}
&&r_{2j-1}=a_R(k) e^{-\ii k j},\quad
r_{2j}=b_R(k) e^{-\ii k j}
,\\\label{lcomp}
&&\ell_{2j}=a_L(k) e^{\ii k j},\quad
\ell_{2j-1}=b_L(k) e^{\ii k j}
\ee
where $k$ is a free parameter. The parameters
$a_{R,L}(k)$ and $b_{R,L}(k)$ are fixed by the spectral problem of the two band Bloch Hamiltonian
\beq\label{twoband}
H(k)=\left(
\begin{array}{cc}
  \ii u & H_1(k) \\
 H_2(k) &  -\ii u \\
\end{array}
\right),
\eeq
where
\be
H_1(k)&=&V_L+W_R e^{\ii k},\\
H_2(k)&=&V_R+W_L e^{-\ii k}.
\ee
That is
\be\label{ABvec}
&&H(k)
|r^\pm(k)\rangle=\epsilon^\pm(k)|r^\pm(k)\rangle,
\non\\
&&\langle \ell^\pm(k)|H(k)
=\epsilon^\pm(k) \langle \ell^\pm(k)|,
\ee
where
the quasienergies are given by
\beq\label{quasiCBC}
\epsilon^{\pm}(k)=\pm
\sqrt{
H_1(k)H_2(k)-u^2},
\eeq
and the left and right eigenvectors by
\be
|r^\pm(k)\rangle&=&\frac{1}{\sqrt{N^\pm}}\left(
\begin{array}{c}
 a_R^\pm(k) \\
 b_R^\pm(k) \\
\end{array}
\right)
\non\\&
=&\label{rvec}
\frac{1}{\sqrt{N^\pm}}\left(
\begin{array}{c}
 \ii u\pm\sqrt{H_1(k) H_2(k)-u^2} \\
 H_2(k) \\
\end{array}
\right)
,\\ \langle \ell^\pm(k)|&=&
\frac{1}{\sqrt{N^\pm}}
\left(
\begin{array}{cc}
 a_L^\pm(k), &
 b_L^\pm(k) 
\end{array}\right)
,\non\\
&=&
\frac{1}{\sqrt{N^\pm}}\left(
\begin{array}{cc}
\ii u\pm\sqrt{H_1(k) H_2(k)-u^2},&
 H_1(k) \, .
\end{array}\right)\label{lvec}\non\\
\ee
Normalizing left and right eigenvectors according to the biorthogonality
condition
\beq
\langle \ell^{\epsilon}(k)|r^{\epsilon'}(k)\rangle=\delta_{\epsilon,\epsilon'},
\eeq
fixes the normalization constant
\beq
N^\pm=2\left(H_1(k)H_2(k)-u^2\pm\ii u\sqrt{H_1(k) H_2(k)-u^2}\right).
\eeq

To conclude the solution, the parameter $k$ is quantized by (\ref{BC}) and given by
\beq\label{kPBC}
k_m^{\text{(PBC)}}=\frac{2\pi (2m)}{N},\quad m=1,\dots,\frac{N}{2},
\eeq
for periodic boundaries and
\beq\label{kABC}
k_m^{\text{(APBC)}}=\frac{2\pi (2m+1)}{N},\quad m=1,\dots,\frac{N}{2},
\eeq
for antiperiodic boundaries.

\subsection{Gap closing and exceptional points}\label{sec:excp}

The bands (\ref{quasiCBC}) for the asymmetric SSH chain are in general complex, and a band touching
requires the vanishing of both the real and the imaginary
parts of the complex gap
\beq\label{comgap}
\Delta_c=(\epsilon^{+}(k)-\epsilon^{-}(k))/2=
\sqrt{
H_1(k)H_2(k)-u^2}.
\eeq
We define the non-Hermitian gap by \cite{BWN2023}
\beq\label{NHgap}
\Delta=\text{min}_k |\Delta_c(k)|
\eeq
which closes if
\beq\label{gapclosingPBC}
H_1(k)H_2(k)-u^2=0.
\eeq
For each value of $k$ quantized by (\ref{kPBC},\ref{kABC}), the condition
(\ref{gapclosingPBC})
leads to constraints in the model parameters and can determine a rich
variety of phase boundaries.
Furthermore, it follows
immediately from Eqs.~(\ref{rvec},\ref{lvec}) that at points where the complex gap (\ref{gapclosingPBC}) vanishes the eigenvectors $\{|r^+(k)\rangle,|r^-(k)\rangle\}$ and also
$\{\langle \ell^+(k)|,\langle \ell^-(k)|\}$
coalesce. Gap closing points are therefore also always exceptional points. 
Exceptional points and their relation
with the discriminant of the characteristic polynomial of the hopping matrix $\mathcal{T}_N$ are discussed further in App.~\ref{app:EPpolynomial}.

\begin{figure*}[!htbp]
\centering
\includegraphics[width=1.95\columnwidth]{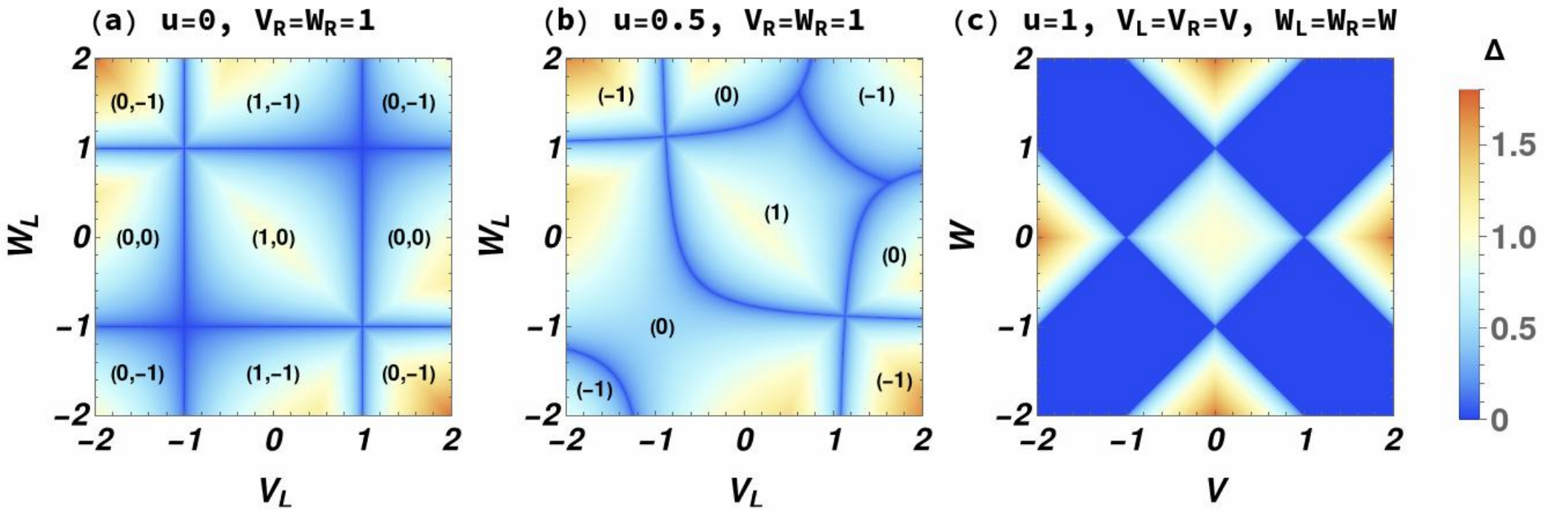}
\caption{\label{fig:gapclosingPBC} The non-Hermitian
gap $\Delta$, see Eq.~\eqref{NHgap}, as a function of the model parameters in the thermodynamic limit for periodic boundary conditions. In panel (a), the zero field  (sublattice symmetric) case, $u=0$, is shown for $V_R=W_R=1$ with two winding numbers $(\nu_1,\nu_2)$ characterizing each phase. In panel (b), we set
$u=0.5$, therefore breaking the sublattice symmetry and leaving only a single winding number $I_1$ labeling the emerging phases. In panel (c),
we consider the $\mathcal{PT}$-symmetric case $V_L=V_R=V$, $W_L=W_R=W$ and $u=1$. The regions in blue constitute a gapless phase where exceptional points are dense. The central ``diamond" is a complex (anti-$\mathcal{PT}$) phase. The remaining phases are $\mathcal{PT}$-unbroken phases, either trivial or topological.
}
\end{figure*}

To analyze the phase transitions and the manifold of exceptional points given by Eq.~(\ref{gapclosingPBC}) it is convenient
to first set $u=0$.
In this case,
the system has sublattice symmetry $\sigma^zH(k)\sigma^z=-H(k)
$.\footnote{In non-Hermitian systems, chiral and sublattice symmetries are not equivalent \cite{KSUS2019}. Indeed for $u=0$
the chiral symmetry $\sigma^zH(k)^{\dagger}\sigma^z=-H(k)$ is not satisfied since $\sigma^zH(k)^{\dagger}\sigma^z=-H(k)^\dagger$.}
Then, for every allowed quantized $k_m$ value and relaxing
the requirement of real couplings,
the solutions of $H_1=0$ or $H_2=0$ are
\beq\label{CBCratios}
\left(\frac{V_L}{W_R}\right)_\ep=-e^{\ii k_m}\quad\text{or}\quad
\left(\frac{W_L}{V_R}\right)_\ep=-e^{\ii k_m}
\eeq
where $k_m$ is given by (\ref{kPBC},\ref{kABC}). It follows that
for most values of $k_m$ the coupling parameters which lead to a gap closing/exceptional point
are complex. For finite $N$ and periodic boundaries,
one can see that $m=N/2$ ($k=2\pi$) and $m=N/4$ ($k=\pi$, if $N/2$ is even) 
are the only possible values of $k$ that lead to real ratios,
\beq\label{gapclosingrealCBC}
\left(\frac{V_L}{W_R}\right)_\ep=\pm 1\quad  \text{or} \quad\left(\frac{W_L}{V_R}\right)_\ep=\pm 1.
\eeq
Fixing $V_R=W_R=1$ (no loss in generality if $u=0$),
these exceptional lines can be visualized as a function of $V_L$
and $W_L$, see the left panel in Fig.~\ref{fig:gapclosingPBC}.
They separate gapped phases labeled by
two integer winding numbers $(\nu_1,\nu_2)$ whose definition is
discussed in subsection \ref{sec:wn}. With respect to the type of gap \cite{KSUS2019}, the phases
$(0,-1)$ and $(1,0)$ have a point gap while phases $(0,0)$ and $(1,-1)$ have a line gap (real or imaginary). For finite $N$ and antiperiodic
boundaries, the ratios (\ref{CBCratios})
are real for $m=1/2(N/2-1)$ ($k=\pi$, if $N/2-1$ is even).
In the thermodynamic
limit, the unit circles are filled and
real ratios occur at $k=\pi$ or $k=2\pi$ for both periodic and
antiperiodic boundary conditions. We remark that sublattice-symmetric
models have been widely
considered in the literature,
see \textit{e.g.}
Refs.~\cite{L2018,YJLLC2018,HBR2019,ALMS2020,HGB2023,ALB2023}.
Here we emphasize that
the gap closing points and exceptional points coincide and that they also occur in the complex parameter plane for every allowed quantized value $k_m$. 

Next, we consider a nonzero real field $u$.
The field couples $H_1(k)$ and $H_2(k)$
and Eq.~(\ref{gapclosingPBC}) no longer factorizes, giving
a constraint for all
model parameters for every possible
value of $k$. 
Eq. (\ref{gapclosingPBC}) can be solved for any of the model parameters, leading in general to a complex number.
As an example, in the central panel of Fig. \ref{fig:gapclosingPBC}, the non-Hermitian gap (\ref{NHgap}) is shown for
fixed parameters $V_R=W_R=1$ and $u=0.5$ as a function of $V_L$ and $W_L$. The emerging phases are characterized by an integer winding number $I_1$ defined in subsection \ref{sec:wn}.
We can observe, for example, that the field $u$ induces the merging of the zero field phases
$(\nu_1,\nu_2)=(0,0)$ and $(1,-1)$, with $V_L<-1$ and $W_L<-1$, into a single phase
characterized by $I_1=0$. This can be understood by looking at the point $V_L=W_{L}=-1$, where $H_1(k)H_2(k)-u^2=-(2\sin(k/2)+\ii u)(2\sin(k/2)-\ii u)$ cannot be canceled for any real $k$ and $u$; therefore, the gap is open.
On the other hand, the zero field phases $(0,0)$ and $(1,-1)$, with $V_L>1$ and $W_L>1$ do not merge at $u=0.5$, although the boundaries of the phases are bend. For instance, at $V_L=W_{L}=1$, one
has $H_1(k)H_2(k)-u^2=(2\cos(k/2)-u)(2\cos(k/2)+u)$ which can be canceled by properly choosing a real $k$ depending on the value of the field, provided $|u|\leq2$.

Another interesting case to consider is the $\mathcal{PT}$ symmetric case 
with $V_L=V_R=V$ and
$W_L=W_R=W$ and $u$ real but arbitrary, such that $\sigma^xH(k)\sigma^x=H(k)^*$.
In this case, the constraint
(\ref{gapclosingPBC}) reduces to,
\beq\label{gapclosingPTeq}
V^2+W^2+2 V W \cos (k_m)-u^2=0 \, .
\eeq
To visualize the gap closing,
we can set $u=1$ (no loss of generality)
and plot the gap (\ref{NHgap}) as a function of $V$ and $W$,
see the right panel of Fig.~\ref{fig:gapclosingPBC}. The blue regions $|V- W|<1<|V + W|$ or $|V+ W|<1<|V - W|$ are gapless in the thermodynamic limit. This is a so far overlooked phase where the exceptional points are dense.
In this phase, the $\mathcal{PT}$-symmetry is broken. The regions $|V|-|W|>1$ and $|W|-|V|>1$ are characterized by a purely real spectrum, and
are adiabatically connected to the standard SSH model via $u\rightarrow 0$. Then, the former is called the $\mathcal{PT}$-unbroken trivial phase and the latter the $\mathcal{PT}$-unbroken topological phase. In the phase
$|W|+|V|<1$ the spectrum is purely imaginary, and we name it the complex phase or anti-$\mathcal{PT}$ phase.

\begin{figure}[!htp]
    \centering
    \includegraphics[width=0.7\columnwidth]{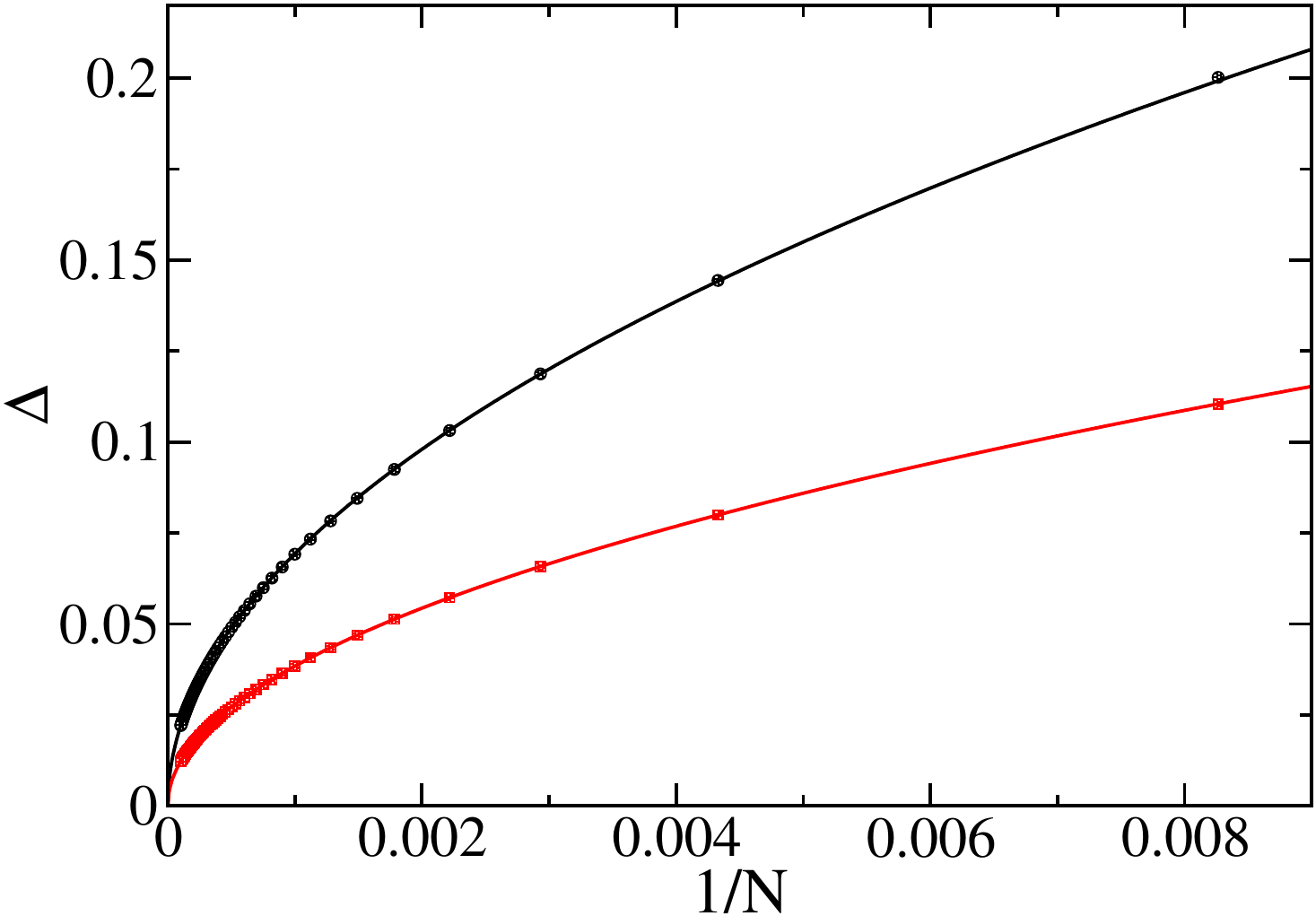}
    \caption{Scaling of the gap \eqref{NHgap} in the phase where exceptional points become dense for $V/W=(1-\sqrt{5})/2$ (black circles) and $V/W=(1+\sqrt{5})/2$ (red squares). The lines are fits $\Delta\sim N^{-1/2}$, i.e., the dynamical critical exponent is $z=1/2$.}
    \label{fig_scaling}
\end{figure}

The gapless phase where exceptional points are dense is peculiar because there is always a set of eigenvectors which is coalescent. I.e., this gapless phase is very different from a Hermitian phase where eigenvectors are orthogonal and one might wonder how this coalescence of eigenvectors affects its transport properties. To address this question, we investigate the scaling of the gap $\Delta\sim N^{-z}$ where $z$ is the dynamical critical exponent. A problem with such an analysis for a generic ratio of parameters $V/W$ is that the allowed momenta $4\pi m/N$ with $m=1,\cdots,N/2$ might or might not be commensurate or almost commensurate with the gap closing condition \eqref{gapclosingPTeq} depending on the system size $N$. In this case no clear scaling for generic $N$ emerges. To circumvent this issue, we choose a ratio $V/W$ which is badly approximable in the Diophantine sense by rational numbers thus making it possible to consider generic $N$. In Fig.~\ref{fig_scaling} we show, in particular, results for $V/W$ being the golden or the inverse golden ratio, respectively.
The results indicate that the dynamical critical exponent is $z=1/2$ which means that lengths scale as $\ell\sim t^2$. Transport in this phase is thus hyper-ballistic. We note that in a short-range Hermitian lattice model the Lieb-Robinson bounds make faster than ballistic transport impossible. However, such bounds do not apply in a non-Hermitian model where eigenvectors do not need to be orthogonal. In this phase, in particular, eigenvectors coalesce leading to an amplification.

\subsection{Winding numbers}\label{sec:wn}
To characterize the different phases we have found, we can use topological invariants. The eigenvalues of a non-Hermitian Hamiltonian are
in general complex and can trace, for periodic boundary conditions, a closed
path around an arbitrary reference complex energy $E_B$,
in which case a winding number can
be assigned \cite{KBRD2010,LBHCN2017,GAKTHU2018,SZF2018,HBR2019},
\beq\label{invI1}
I_{1}=\frac{1}{2\pi \ii}
\int_{0}^{2\pi}dk ~
\log\det(H(k)-E_B),
\eeq
provided that $H(k)$ has a point gap at $E_B$, that is, $\det(H(k)-E_B)\neq 0$. For concreteness we will chose $E_B=0$. Observe that the definition (\ref{invI1}) is universal since no symmetry requirements
are made.
Therefore, one
dimensional non-Hermitian systems---in contrast to Hermitian ones---can have non-trivial topology
in the absence of any spatial or non-spatial symmetries. 

If symmetries are present, then they lead to additional topological invariants \cite{KSUS2019,MonkmanSirkerNH}. So far, mostly the case of a two-band model with sub-lattice symmetry has been considered \cite{GW1988,DMT1990,M1999,LH2013,L2018,SZF2018}. The Hamiltonian of such a model takes the form
\beq
\label{Hk}
H(k)=\left(
\begin{array}{cc}
 0 & \hat H_1(k) \\
 \hat H_2(k) &  0 \\
\end{array}
\right) \, .
\eeq
Using the polar decomposition $\hat H_j(k)=d_j(k)q_j(k)$ with $q_j(k)$
unitary and $d_j(k)$ positive semi-definite, one can define 
two winding numbers 
\beq
\label{windings}
\nu_j = \frac{1}{2\pi \ii}
\int  dk ~ \tr\left( q_j^\dagger\partial_k q_j\right)=
\frac{1}{2\pi \ii}
\int  dk ~ \partial_k \log \det q_j \, .
\eeq
Instead of the winding numbers $\nu_{1,2}$, one can equivalently consider \cite{YJLLC2018,JYC2018}
\beq\label{invI1s}
I_1 = \nu_1+\nu_2,
\eeq
which is consistent with the definition of $I_1$ in Eq.~\eqref{invI1}, and
\be\label{invI2}
I_{2}=\nu_2-\nu_1
\ee
as independent winding numbers. Note that in the Hermitian case, $\hat H_2(k) = \hat H_1^\dagger(k)$, we have $\nu_2=-\nu_1$ implying that $I_1\equiv 0$ and only one non-trivial winding number, $I_2$, remains.
For our model with $u=0$, we can write $H_j(k)=|H_j(k)|e^{\ii \phi_j}$ and evaluating the integrals
(\ref{invI1s},\ref{invI2}) leads to
\beq
I_1(u=0)=\begin{cases}
 -1, & \text{if}\quad~ \left| \frac{W_L}{V_R}\right| > 1\land
 \left| \frac{V_L}{W_R}\right| > 1
  \\
 1, & \text{if}\quad
 \left| \frac{W_L}{V_R}\right| <1
 \land \left| \frac{V_L}{W_R}\right| <1\\
 0, & \text{otherwise} \, 
\end{cases}
\eeq
and
\beq
I_2=\begin{cases}
 -2 & \text{if}\quad~
 \left| \frac{V_L}{W_R}\right| <1 
 \land
 \left| \frac{W_L}{V_R}\right| > 1
 \\
 0 & \text{if}\quad
 \left| \frac{V_L}{W_R}\right| >1 
 \land
 \left| \frac{W_L}{V_R}\right| <1 \\
  -1 & \text{otherwise}.
\end{cases}
\eeq
Clearly, the winding numbers are not defined when $\left| \frac{V_L}{W_R}\right|=\left| \frac{W_L}{V_R}\right|=1$ which are precisely the gapless/exceptional lines discussed in the previous subsection.
The two winding numbers $\nu_{1,2}=(I_1\mp I_2)/2$ are shown in the left panel of Fig.~\ref{fig:gapclosingPBC} in each phase as a function of $V_L$ and $W_L$.


\section{Open boundary conditions}\label{sec:EDOBC}

\subsection{Diagonalization}

Let us also recall the diagonalization of 
$\mathcal{T}_N$ (\ref{tri}) for
open boundary conditions (OBC), that is, $\gamma=0$. In this case, $\mathcal{T}_N$ is an asymmetric tridiagonal
2-Toeplitz matrix whose eigenvalue problem was solved from different perspectives in Refs.~\cite{Gover94,shin_1997}.
In App.~\ref{app:OBC}, we provide some details on both approaches. The connection between these two approaches was pointed out in Ref.~\cite{Ikramov_1999} and the method \cite{shin_1997} was previously applied to the Hermitian SSH chain in Refs.~\cite{SMKS2014,CBC2023arXiv}.

The solution of the open boundary case
can also be obtained by a simple
similarity transformation to the standard (symmetric) SSH chain which leaves the spectrum invariant \cite{YZ2018},
namely,
\beq
S^{-1} \mathcal{T}_N S= \mathcal{T}_N^{SSH}
\eeq
where $S$ is a diagonal matrix with entries
\beq
s_{2j-1}=\frac{V_R^\frac{j-1}{2}W_R^\frac{j-1}{2}}{V_L^\frac{j-1}{2}W_L^{\frac{j-1}{2}}}
,
\quad 
s_{2j}=\frac{V_R^\frac{j}{2}W_R^\frac{j}{2}}{V_L^\frac{j}{2}W_L^{\frac{j}{2}}}
\frac{W_L^{\frac{1}{2}}}{W_R^{\frac{1}{2}}},
\eeq
and $\mathcal{T}_N^{SSH}$ is a symmetric hopping matrix of the standard SSH chain with effective couplings,
\beq
V_{\text{eff}}=\sqrt{V_L} \sqrt{V_R},
\quad W_{\text{eff}}=\sqrt{W_L} \sqrt{W_R}.
\eeq

The quasienergies are then given by
\beq\label{quasieOBC}
\epsilon^{\pm}=
\pm\sqrt{
\tilde{H}_1(\theta)\tilde{H}_2(\theta)-u^2},
\eeq
where
\beq
\tilde{H}_1(\theta)=V_L(e^{i\theta/2}+e^{-i\theta/2}\delta^{-1/2}),
\eeq
\beq
\tilde{H}_2(\theta)=V_R(e^{-i\theta/2}+e^{i\theta/2}\delta^{-1/2}),
\eeq
with
\beq\label{delta}
\sqrt{\delta}=\frac{\sqrt{V_L} \sqrt{V_R}}
{\sqrt{W_L} \sqrt{W_R}},
\eeq
and $\theta$ is a parameter that
is given by different expressions depending on the parity of $N$. We remark that, for odd $N$, there is an isolated root
$\epsilon=\ii u$ which is not contained in (\ref{quasieOBC}).

For odd $N$, the parameter $\theta$ is simply
\beq\label{tquant}
\theta_m^{(OBC)}=\frac{2\pi m}{N+1},\quad m=1,\dots,\frac{N-1}{2},
\eeq
which gives $N-1$ eigenvalues from (\ref{quasieOBC})
and (\ref{tquant}). Together with the isolated root $\epsilon=\ii u$, one has all the $N$ eigenvalues of the tridiagonal matrix
when $\gamma=0$.

For even $N$, the parameter $\theta$ is determined via
the transcendental equation
\beq\label{trans}
\sqrt{\delta}\sin\left(\left(\frac{N}{2}+1\right)\theta\right)+\sin\left(\frac{N}{2}\theta\right)=0 \, .
\eeq
We remark that the transcendental equation (\ref{trans}) has the same form
as the transcendental equation of the quantum Ising chain with
transverse field $\lambda=\sqrt{\delta}$ and $N/2$ sites \cite{P70}. Meanwhile,
the eigenvector
components can be written as
\bea
\label{vecOBC}
r_{2n}&=&\frac{V_R^{\frac{n-1}{2}}W_R^{\frac{n-1}{2}}}{V_L^{\frac{n-1}{2}}W_L^{\frac{n-1}{2}}}
(\epsilon-\ii u)\frac{\sin(n\theta)}{\sin(\theta)},\label{vLeven}\\
r_{2n-1}&=&\frac{V_R^{\frac{n-1}{2}}W_R^{\frac{n-1}{2}}}{V_L^{\frac{n-1}{2}}W_L^{\frac{n-1}{2}}}
\frac{\left(
\sin(n\theta)+\delta^{-\frac{1}{2}}\sin((n-1)\theta)
\right)}
{V_L^{-1}\sin(\theta)},\non\\\label{vLodd}
\eea
and
\bea
\ell_{2n}&=&\frac{V_L^{\frac{n-1}{2}}W_L^{\frac{n-1}{2}}}{V_R^{\frac{n-1}{2}}W_R^{\frac{n-1}{2}}}(\epsilon-\ii u)\frac{\sin(n\theta)}{\sin(\theta)},\label{vLevenL}\\
\ell_{2n-1}&=&\frac{V_L^{\frac{n-1}{2}}W_L^{\frac{n-1}{2}}}{V_R^{\frac{n-1}{2}}W_R^{\frac{n-1}{2}}}
\frac{\left(
\sin(n\theta)+\delta^{-\frac{1}{2}}\sin((n-1)\theta))
\right)}
{V_R^{-1}\sin(\theta)}.\non\\\label{vLoddL}
\eea
Note that all the eigenstates are, in general, localized at the boundaries for the open case in contrast to the closed case where they are extended Bloch waves. More specifically, the right eigenstates are localized at the left (right) boundary for $\Gamma<1$ ($\Gamma>1$) with $\Gamma=V_R W_R/V_L W_L$. Conversely, the left eigenstates are localized at the right (left) boundary for $\Gamma<1$ ($\Gamma>1$). There is no localization if $\Gamma=1$. The localization of all the eigenstates when switching from closed to open boundary conditions is called the non-Hermitian skin effect.

\subsection{Gap closing and exceptional points}

Similar to the case with closed boundary conditions,
one can define a complex gap
\begin{eqnarray}
\label{comgapOBC}       
&&\Delta_c^{(OBC)}=(\epsilon^{+}(\theta)-\epsilon^{-}(\theta))/2 =
\sqrt{
\tilde{H}_1(\theta)\tilde{H}_2(\theta)-u^2} \nonumber \\
&&=\!\!
\sqrt{\! V_L V_R+W_L W_R+2\sqrt{V_L} \sqrt{V_R} \sqrt{W_L} \sqrt{W_R} \cos\theta-u^2} \nonumber \\
\end{eqnarray}
as well as a minimal gap
\beq\label{NHgapOBC}
\Delta^{(OBC)}=\text{min}_\theta |\Delta_c^{(OBC)}(\theta)|,
\eeq
where $\theta$ is given by (\ref{tquant}) or by a solution of (\ref{trans}) depending on the parity
of $N$. A different analysis of the
complex gap is required
for each parity of $N$.

\subsubsection{Odd N}

Let us start with odd $N$
where the parameter $\theta\rightarrow\theta_m$
is explicitly given by (\ref{tquant}).
Also, for clarity, let us consider first $u=0$. In this case, 
the complex gap (\ref{comgapOBC}) will vanish if $\tilde{H}_1(\theta_m)\tilde{H}_2(\theta_m)=0$, which happens if any of the hopping parameters
vanishes
or if
\beq\label{OBCratios2}
\left(\sqrt{\delta}\right)_\ep=-e^{-\ii \theta_m}
~~\text{or}~~
\left(\sqrt{\delta}\right)_{\ep}=-e^{\ii \theta_m}.
\eeq
Therefore, for non-vanishing hopping
parameters, there are $N-1$ exceptional points
associated with the eigenvalue $\epsilon=0$.
For finite $N$,
these points always have an imaginary part.
In the thermodynamic limit,
however, the exceptional points will converge to the real axis
\beq
\lim_{N\rightarrow\infty}\left(\sqrt{\delta}\right)_\ep = \pm 1
\eeq
in the cases when $\theta_m\to 0$ or $\theta_m\to\pi$.

\begin{figure}[!htbp]
\centering
\includegraphics[width=0.99\columnwidth]{"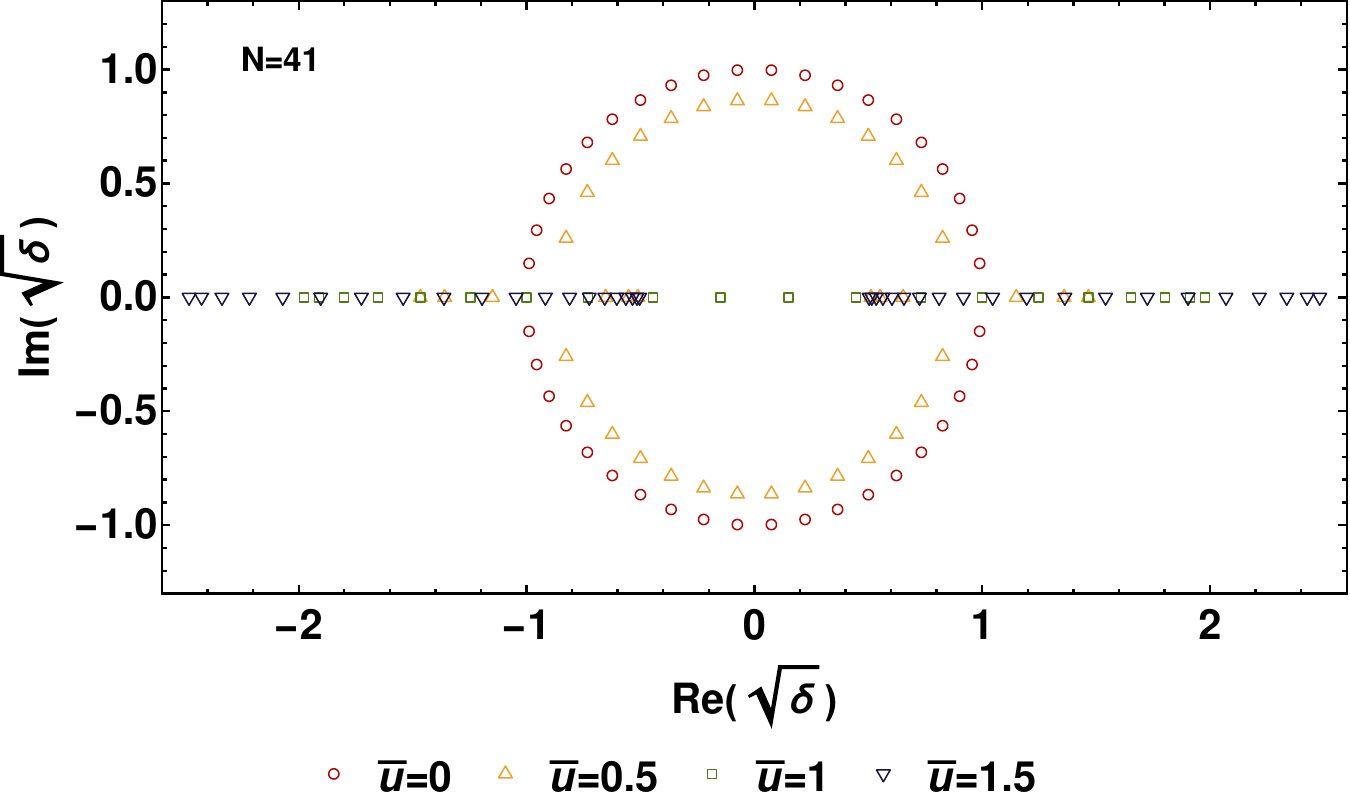"}
\caption{\label{fig:sqrtdubar} Loci of
exceptional points $\left(\sqrt{\delta}\right)_\ep$, see Eq.~\eqref{gapclosingLodd1}, for $N=41$ and different rescaled fields $\bar u$.}
\end{figure}

Similarly to the closed boundary case, turning
on the field $u$ couples $\tilde{H}_1(\theta_m)$
and $\tilde{H}_2(\theta_m)$. We observe that introducing the rescaled field,
\beq\label{rescu}
\bar u = u/(\sqrt{W_L}\sqrt{W_R})
\eeq
one can write
\bea
&&\tilde{H}_1(\theta_m)\tilde{H}_2(\theta_m)-u^2=
\sqrt{V_L}\sqrt{V_R}
\sqrt{W_L}\sqrt{W_R}\non\\&&\quad\times
\left(\sqrt{\delta}+\frac{1}{\sqrt{\delta}}(1-\bar u^2)+2\cos(\theta_m)\right) \, .
\eea
The exceptional points, which are still equivalent to the points where the gap closes,
are therefore now given by
\beq\label{gapclosingLodd1}
\left(\sqrt{\delta}\right)_\ep=
-\cos(\theta_m)\pm \sqrt{\bar u^2-\sin^2(\theta_m)}.
\eeq
Interestingly, the expression (\ref{gapclosingLodd1}) implies that
now real
exceptional points do exist even for finite $N$
provided that $\bar u\geq |\sin(\theta_m)|$ or
$\bar u\leq -|\sin(\theta_m)|$. In the thermodynamic limit, critical fields $\bar u\rightarrow\pm 1$ emerge when $\theta_m$ approaches
$0$ or $\pi$. In Fig.~\ref{fig:sqrtdubar}
some examples of
the loci of
exceptional points for different values of the rescaled field $\bar u$ are shown.

Recall
that when
$N$ is odd, there is, in addition, an isolated root $\epsilon=\ii u$. We can impose that this root coincides with
the quasienergies (\ref{quasieOBC}), that is, $\ii u=\epsilon^{\pm}$. For non-zero hopping parameters, this implies again the condition (\ref{OBCratios2}), but now
associated with the degenerate
quasienergy $\ii u$.

\subsubsection{Even N}
\begin{figure}[!htbp]
\centering
\includegraphics[width=0.99\columnwidth]{"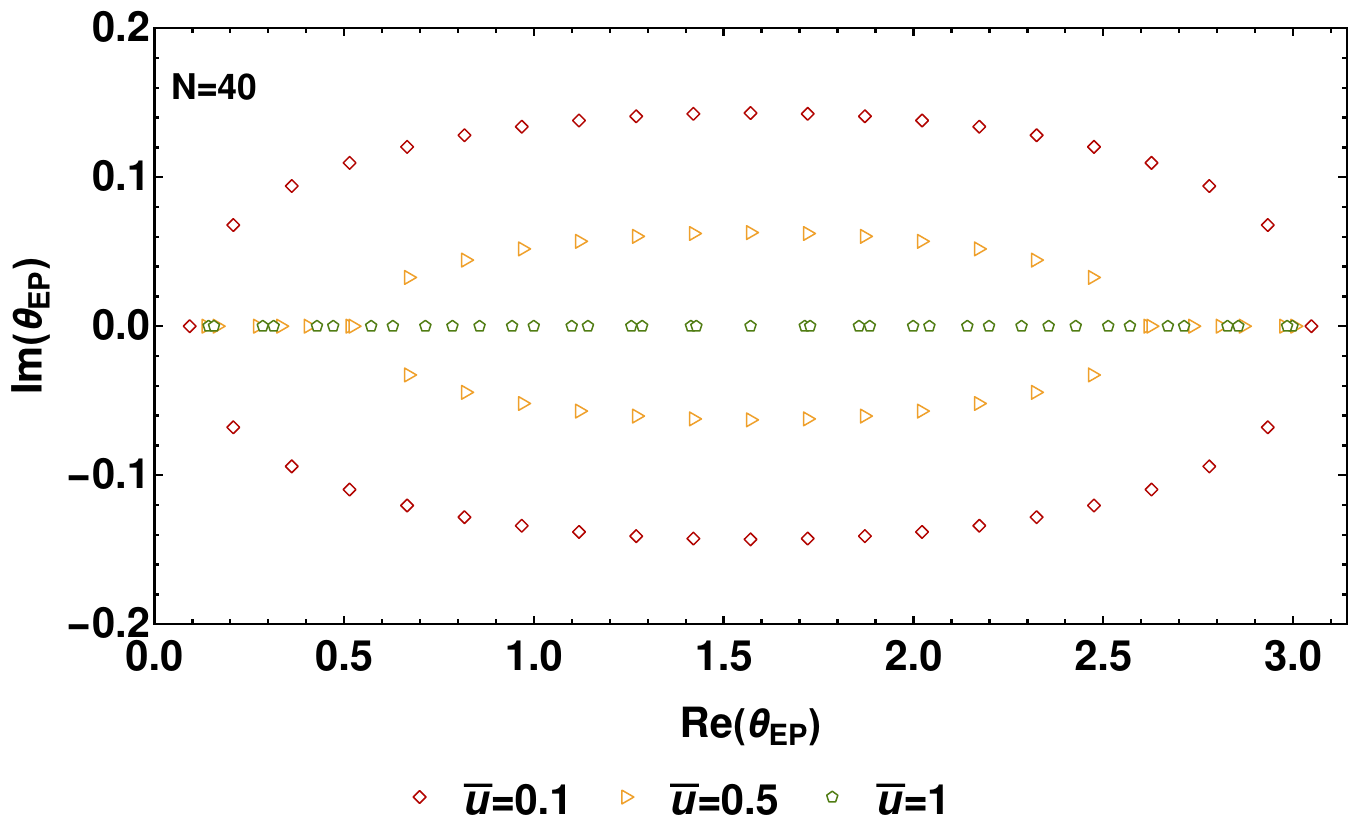"}
\includegraphics[width=0.99\columnwidth]{"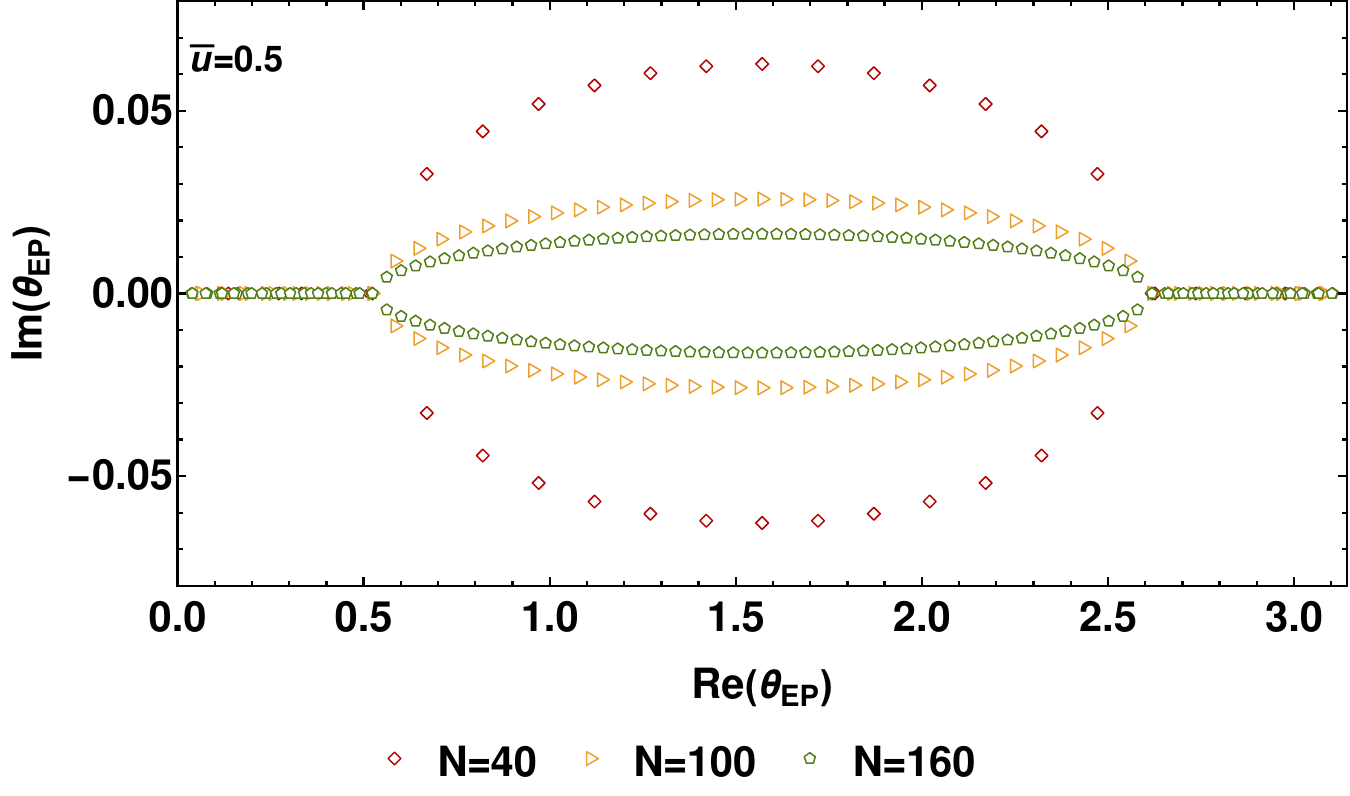"}
\includegraphics[width=0.99\columnwidth]{"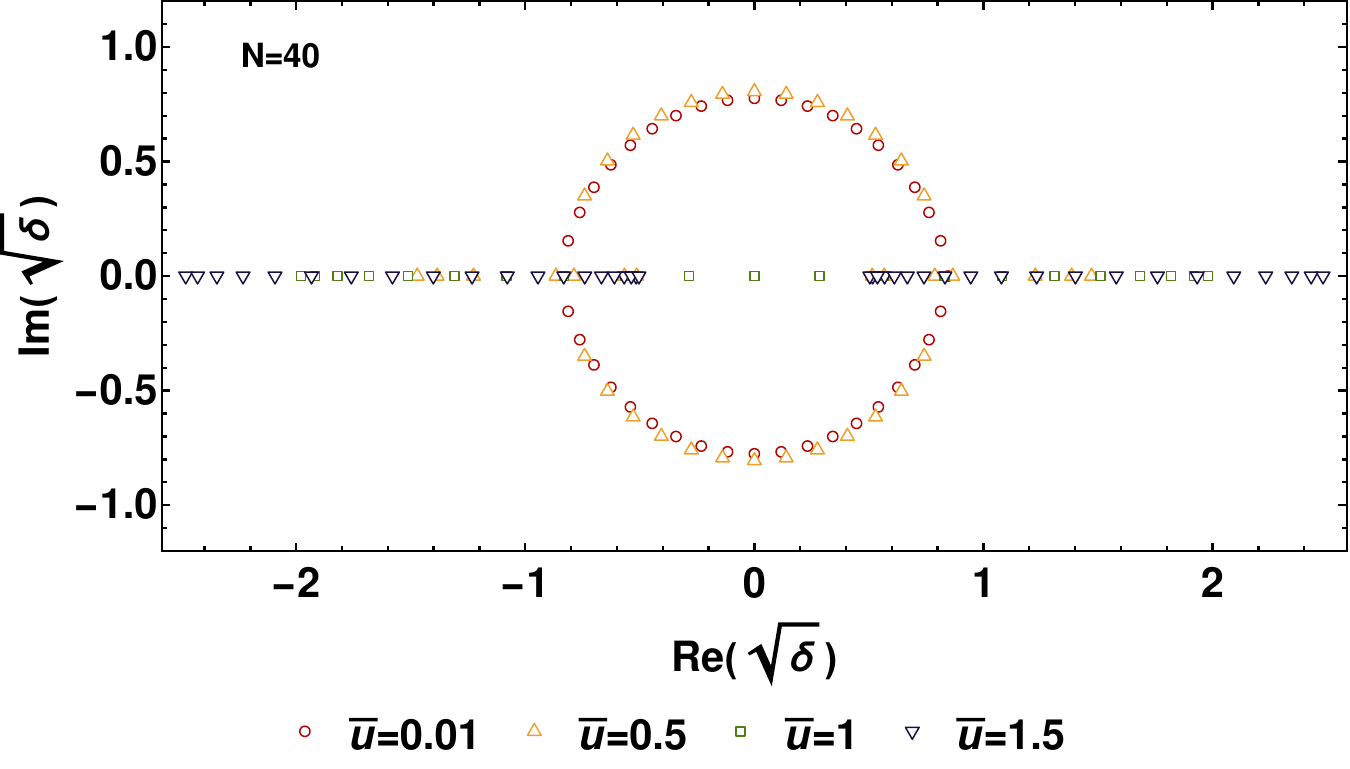"}
\caption{\label{fig:thetaEP1} In the upper panel, solutions of the
transcendental equation 
(\ref{trans2}) for 
different fields $\bar u$ and $N=40$ are shown. In the middle panel, we fix $\bar u=1/2$ and consider different lattice sizes. In the bottom panel, we show the exceptional points
associated with (\ref{trans2}).}
\end{figure}

Similar to the previously discussed case of odd $N$, 
the complex gap (\ref{comgapOBC}) will vanish if
\beq\label{gapclosingLeven}
\sqrt{\delta}+\frac{1}{\sqrt{\delta}}(1-\bar u^2)+2\cos(\theta)=0
\eeq
but with $\theta$ now given by
(\ref{trans}). Since Eq.~(\ref{trans}) also
involves $\sqrt{\delta}$, one has
to consider both equations simultaneously. Note
that both conditions are satisfied if
any of the hopping parameters
vanishes
similar to the case of $N$ odd. Otherwise,
after some manipulation, the intersection
of (\ref{gapclosingLeven}) and (\ref{trans})
leads to,
\beq\label{trans2}
\sin (\theta )\pm \bar{u} \sin\left(\left(\frac{N}{2}+1\right)\theta\right)=0,
\eeq
whose solutions $\theta=\theta_{EP}$
determine exceptional values of $\sqrt{\delta}$
from either (\ref{gapclosingLeven}) or (\ref{trans}).
That is, choosing (\ref{trans}),
the exceptional points are given by,
\beq\label{sqrtEPB}
\left(\sqrt{\delta}\right)_\ep=-\frac
{
\sin\left(\frac{N}{2}\theta_\ep\right)
}
{
\sin\left((\frac{N}{2}+1)\theta_\ep\right)
}.
\eeq
The transcendental equations
(\ref{trans2}) can be solved numerically
for different rescaled fields $\bar u$.
We observe
that each sign in (\ref{trans2})
has in general $N/2$ solutions with $\text{Re}(\theta_\ep)\geq 0$ which 
determine $N$ exceptional points (other solutions with $\text{Re}(\theta_\ep)<0$ do not
produce new exceptional points). An exception is the case $\bar u=1$ where the number of
solutions of Eq.~(\ref{trans2}) with a negative sign has one less solution. Nevertheless, all distinct exceptional points are obtained.
We show some examples in Fig.~\ref{fig:thetaEP1}.

\begin{figure}[!htbp]
\centering
\includegraphics[width=0.99\columnwidth]{"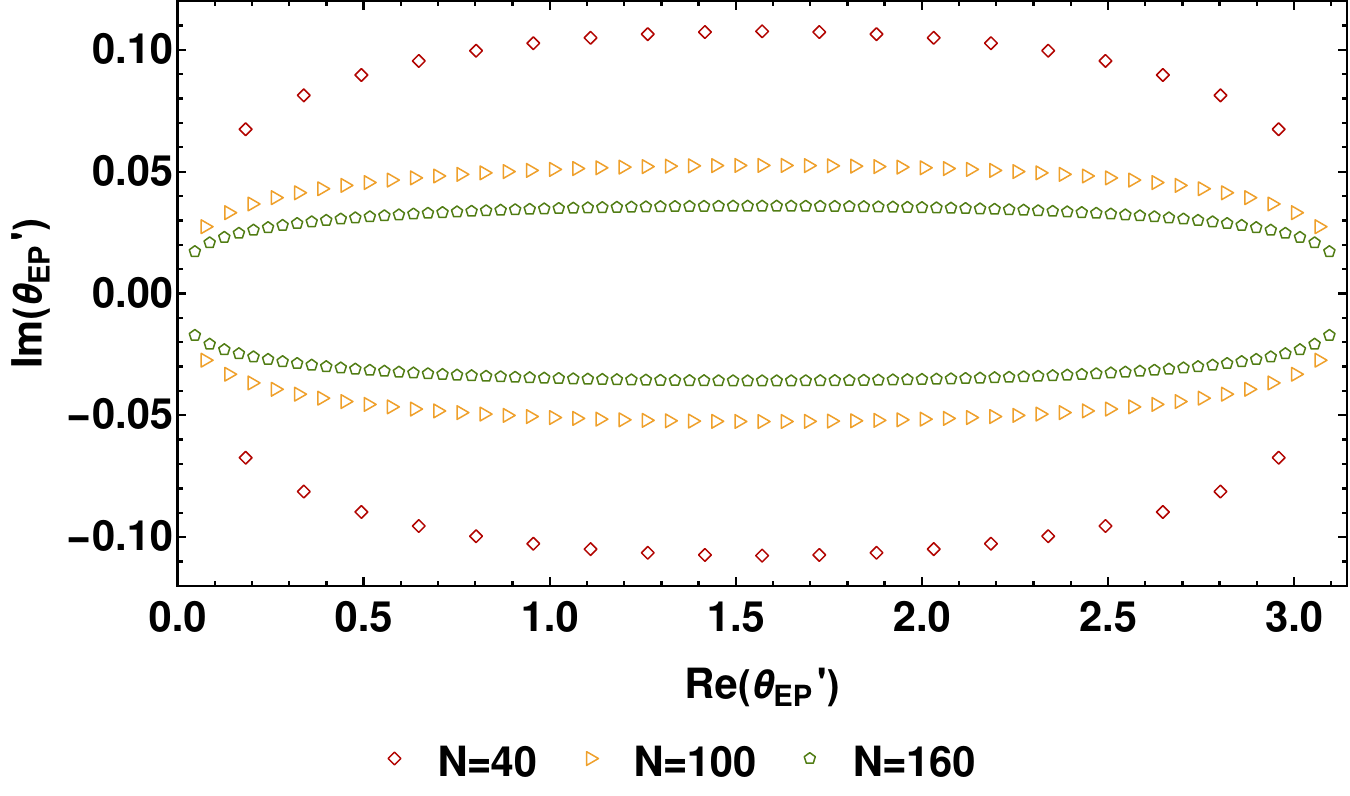"}
\includegraphics[width=0.99\columnwidth]{"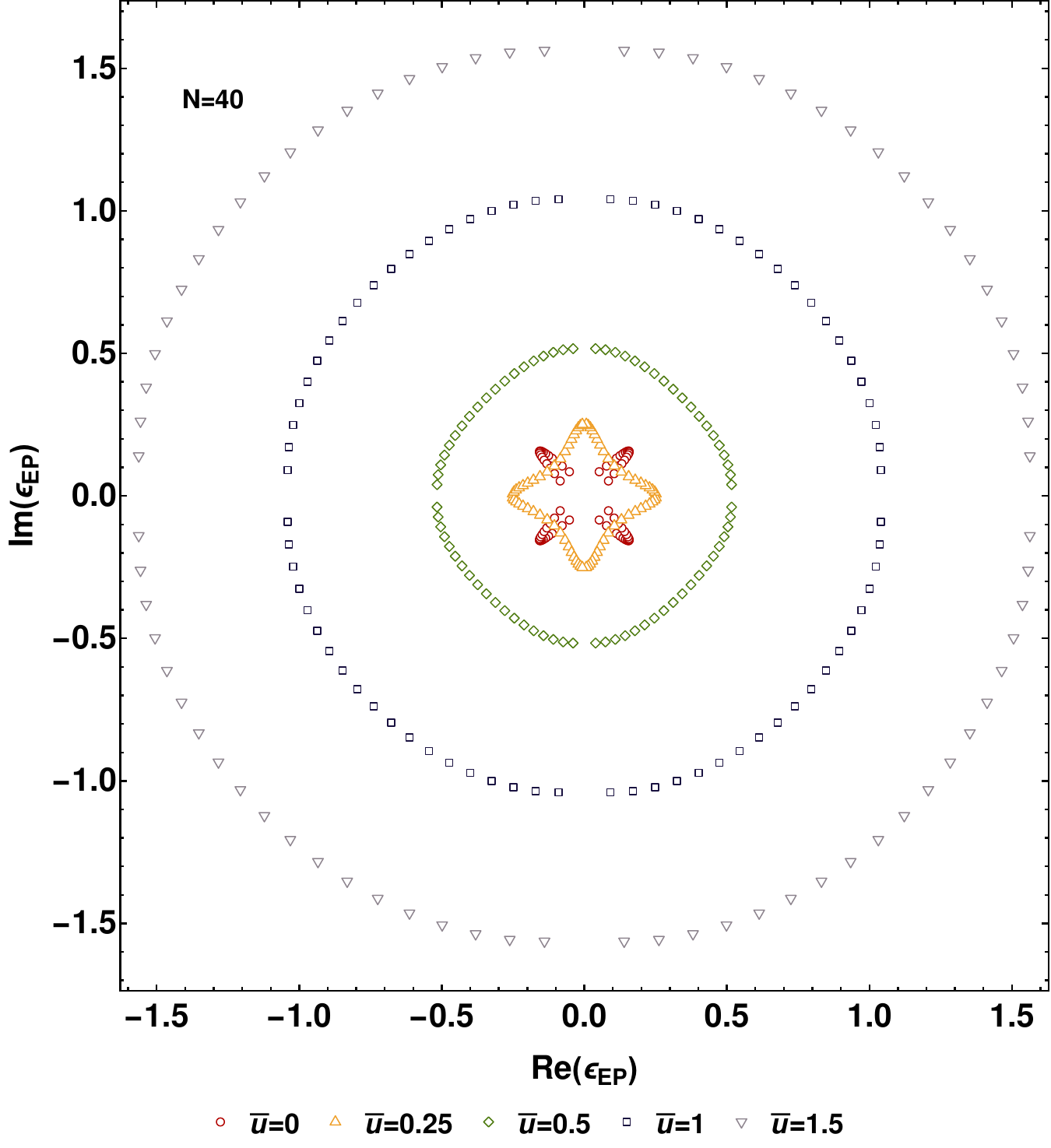"}
\caption{\label{fig:thetaEP2} (a) The solution of Eq.~\eqref{sqrtEPA} leads to additional, $\bar u$ independent exceptional points. (b) The corresponding energies $\varepsilon_{EP}$ lie along non-trivial curves in the complex plane.}
\end{figure}

There is an additional possibility to cancel the discriminant, first discovered in
the study of exceptional points of the Baxter free parafermionic model \cite{HB2023}. It consists
in imposing that
the derivative of (\ref{trans}) vanishes,
\beq\label{dtrans}
\sqrt{\delta}(N+2) \cos \left(\left(\frac{N}{2}+1\right)\theta\right)+
N\cos\left(\frac{N}{2}\theta\right)=0
\eeq
in addition to (\ref{trans}). The intersection
of (\ref{dtrans}) and (\ref{trans}) then gives \cite{HB2023},
\beq\label{HBcond}
\sin ((N+1)\theta)-(N+1) \sin (\theta)=0
\eeq
whose solutions $\theta=\theta_\ep'$ give extra exceptional points
\beq\label{sqrtEPA}
\left(\sqrt{\delta}\right)_\ep=-\frac
{
\sin\left(\frac{N}{2}\theta_\ep'\right)
}
{
\sin\left((\frac{N}{2}+1)\theta_\ep'\right)
}.
\eeq
The transcendental equation (\ref{HBcond}) is clearly independent
of the field $u$. It can be solved numerically; there are $N-2$ solutions \cite{HB2023} with $\text{Re}(\theta_\ep')\geq 0$ which fully describe
the field independent part of the discriminant. The numerical solution
for different lattice sizes is shown
in the upper panel of Fig.~\ref{fig:thetaEP2}.
Note that in this case the associated repeated quasienergy
is not zero. Instead, they trace
interesting curves in the complex plane,
see the examples shown in the bottom panel of Fig.~\ref{fig:thetaEP2}.

\section{Bulk-boundary correspondence}
\label{BBC}
For Hermitian systems, the bulk-boundary correspondence connects a topological invariant calculated for a system with periodic boundary conditions with the number of topologically protected edge modes in a system with open boundaries which have energies which converge exponentially to zero with system size. It is well-known that this type of bulk-boundary correspondence breaks down in the non-Hermitian case \cite{KSUS2019,YZ2018,Kunst}. Some authors have tried to devise new topological invariants based on, for example, 'non-Bloch theory' \cite{YZ2018} but these attempts lack a solid mathematical foundation. Furthermore, these theories are based on translational invariance, however,
"topological features must be properties of a translationally not invariant Hamiltonian" \cite{RyuSchnyder} and thus should be stable against perturbations which break translational invariance. From a mathematical point of view, none of this is at all surprising: it is well-known from Toeplitz theory that topological invariants are in general related to properties of the singular value, not the eigenvalue, spectrum. The importance of the singular value spectrum in the non-Hermitian case has been pointed out, for example, already in Refs.~\cite{HBR2019,BWN2023} but without establishing a rigorous bulk-boundary correspondence. An analysis of the singular value spectrum has also turned out to be useful in the classification of non-Hermitian random matrices \cite{KawabataXiao}. It is also worth noting that another way of understanding the importance of singular values is Hermitization \cite{Porras1} where the hermitian block matrix 
$$\bar H=\begin{pmatrix}0 & H\\ H^\dagger & 0 \end{pmatrix}$$
is constructed. A point gap in $H$ then becomes a spectral gap in $\bar H$ and the positive eigenvalues of $\bar H$ correspond to the singular values of $H$. However, this approach does introduce an unphysical additional chiral symmetry and is not needed here. The case where the Hamiltonian $H$ itself is Hermitian is special because the singular values are then just the absolute values of the eigenvalues so that index theorems for singular values also apply to the eigenvalues. Another important and so far overlooked issue is that the eigenvalues of a finite non-Hermitian matrix do, in general, not converge to the eigenvalues of the semi-infinite case. This is similar to a discontinuous point. 

All of these aspects can be fully understood if one starts from known index theorems in Toeplitz theory. In particular, the K-splitting theorem states \cite{BoettcherSilbermann} that a system with winding number $I$ has $K\geq |I|$ singular values which will converge to zero with increasing system size and which belong to states which are exponentially localized at the boundary. These topologically protected singular values are separated by a gap from the rest of the singular value spectrum. In the semi-infinite case, these states become exact eigenstates with exactly zero energy but for finite system size they only get mapped exponentially close to zero by the Hamiltonian but are not exact eigenstates. I.e., in a finite non-Hermitian system the topological zero-energy edge modes are 'hidden', in general, and instead constitute very long-lived metastable states. The latter aspect is extremely important for a proper understanding of the topological properties of non-Hermitian systems and for testing these properties experimentally. One possible scenario for such an experiment is to prepare a system in a hidden zero mode---which is localized at one of the boundaries---in a cold atomic gas, trapped ion system, or some other quantum simulation/computation platform and to monitor a generalized Loschmidt echo. This Loschmidt echo will stay close to $1$ for a time which is proportional to system size thus indicating the metastable character and its convergence to a true eigenstate in the thermodynamic limit. The corresponding protocol is discussed in more detail in the Suppl.~Mat.~of Ref.~\cite{MonkmanSirkerNH}. We note that the required postselection to trajectories without jumps has already been performed for
a single qubit \cite{NaghilooAbbasi}. Such
an experiment, generalized to multiple qubits, would be
able to detect the hidden zero modes.

We stress again that the formalism developed in Ref.~\cite{MonkmanSirkerNH} does encompass the standard bulk-boundary correspondence in the Hermitian case because for a Hermitian matrix the singular values are just the absolute values of the eigenvalues. Here we want to explicitly and fully analytically demonstrate this general bulk-boundary correspondence for the first time for the physically relevant minimal non-Hermitian model under consideration.

\subsection{Singular values}
To do so, we start by computing the eigenvalues of 
\beq\label{SIGN}
\Sigma_N=\mathcal{T}_N\mathcal{T}_N^\dagger
\eeq
when $\gamma=u=0$, that is, the open case with zero field. The singular values $\sigma$ of $\mathcal{T}_N$ are given by the square roots of the eigenvalues of $\Sigma_N$ which is
a symmetric band matrix with elements,
\bea\label{band}
&&(\Sigma_N)_{{2j-1,2j+1}}=
(\Sigma_N)_{{2j-1,2j-3}}=V_LW_R,\non\\
&&
(\Sigma_N)_{{2j,2j-2}}=
(\Sigma_N)_{{2j,2j+2}}=W_LV_R,\non\\
&&
(\Sigma_N)_{{2j-1,2j-1}}=V_L^2+W_R^2,
~~
(\Sigma_N)_{{2j,2j}}=W_L^2+V_R^2\non\\
&&
(\Sigma_N)_{{1,1}}=V_L^2,
\quad
(\Sigma_N)_{{L,L}}=\begin{cases}
    V_R^2,&\text{even~L}\\
    W_R^2,&\text{odd~L}
\end{cases}.
\eea
This means that $\Sigma_N$ represents a hopping model with next-nearest neighbour interaction.

The spectral problem $(\Sigma_N)\vec{s}=\sigma^2 \vec{s}$ reads,
\bea\label{difeqSVD}
&&s_{2j-3}+s_{2j+1}=\frac{\sigma^2-(V_L^2+W_R^2)}{V_LW_R}s_{2j-1}\non\\&&
s_{2j-2}+s_{2j+2}=\frac{\sigma^2-(W_L^2+V_R^2)}{W_LV_R}s_{2j}
\eea
with boundaries fixed by the last line in (\ref{band}).
Using the method reviewed in App.~\ref{app:OBC}, we find that there are two solutions to the difference equations
(\ref{difeqSVD}).

In the first (second) solution the even (odd) components of the 
vector $\vec{s}$ are zero $s_{2j}=0$
($s_{2j-1}=0$) and the left boundary
fixes the ratio $s_3/s_1$ ($s_4/s_2$). The first solution has singular values
\beq
\label{sigma_phi}
\sigma^2_\phi = V_L^2+W_R^2+2V_LW_R\cos(\phi)
\eeq
and vector components,
\beq
\frac{s_{2j-1}}{s_1}=\frac{W_R}{V_L}\frac{\sin((j-1)\phi)}{\sin(\phi)}+\frac{\sin(j\phi)}{\sin(\phi)},
\quad j\geq 3,
\eeq
\beq
\frac{s_3}{s_1}=
\frac{W_R}{V_L}+2\cos(\phi),
\eeq
where $\phi$ satisfies the transcendental
equation,
\beq\label{transphi}
\sin\left(\frac{N}{2}\phi\right)
\left(\frac{W_R}{V_L}+2\cos(\phi)\right)-
\sin\left(\left(\frac{N}{2}-1\right)\phi\right)=0.
\eeq
The second solution
has eigenvalues
\beq
\label{sigma_omega}
\sigma^2_\omega = W_L^2+V_R^2+2W_LV_R\cos(\omega)
\eeq
and vector components,
\beq
\frac{s_{2j}}{s_2}=\frac{\sin (j\omega)}{\sin(\omega)},
\quad j\geq 3,
\eeq
\beq
\frac{s_4}{s_2}=2\cos(\omega)
\eeq
with the parameter $\omega$ satisfying,
\beq\label{transom}
\sin\left(\left(\frac{N}{2}+1\right)\omega\right)
+\frac{W_L}{V_R}\sin\left(\frac{N}{2}\omega\right)=0.
\eeq

The spectrum of (\ref{SIGN})
is given by the union of $\sigma_\phi^2$
and $\sigma_\omega^2$. Remarkably,
the pairs of variables
$\{V_L,W_R\}$ and $\{V_R,W_L\}$
decouple from each other showing the value of an analytical solution. This means
that the characteristic polynomial
associated with (\ref{SIGN}) factorizes
into two independent factors depending solely on each of these pairs.

\begin{figure}[!htbp]
\centering
\includegraphics[width=0.99\columnwidth]{"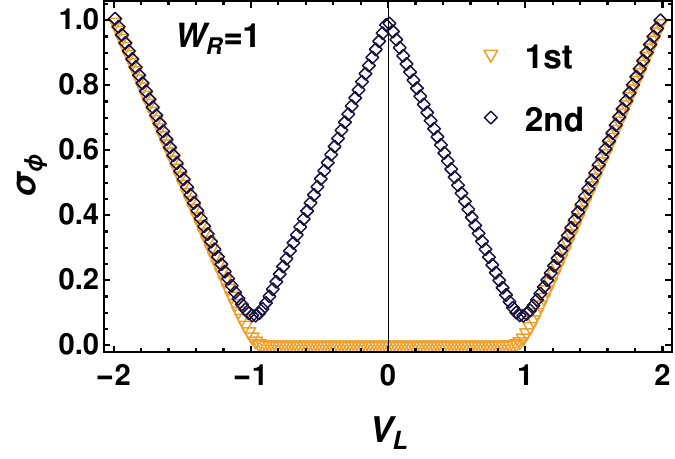"}
\includegraphics[width=0.99\columnwidth]{"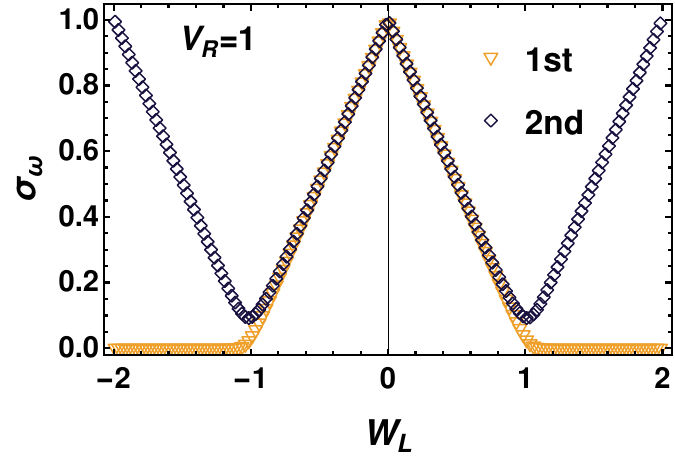"}
\caption{\label{fig:singularvalues} The two smallest singular values from Eq.~\eqref{sigma_phi} (upper panel) and from Eq.~\eqref{sigma_omega} (lower panel). In both cases the corresponding transcendental equation is solved numerically.}
\end{figure}

The transcendental equations
(\ref{transphi}) and (\ref{transom}) can
be solved numerically. We find that 
(\ref{transphi}) has $N/2$ real solutions in the interval $0<\phi<\pi$ when $|V_L/W_R|\geq 1$. When
$|V_L/W_R|< 1$ a complex conjugate pair emerges near $\pi$. This complex root
produces an exponentially small singular value $\sigma_\phi$. Similarly,
we find that (\ref{transom}) has $N/2$ real solutions in the interval $0<\omega<\pi$ when $|V_R/W_L|\geq 1$ and
$N/2-1$ real solutions plus a complex conjugate pair solution when $|V_R/W_L|< 1$. In Fig.~\ref{fig:singularvalues}, we plot the
smallest two singular values coming from each transcendental equation
considering a lattice with $N=100$
sites.

The Bloch Hamiltonian for the corresponding periodic problem with $u=0$ has block structure, see Eq.~\eqref{Hk}. There are therefore two winding numbers $\nu_{1,2}$ defined in Eq.~\eqref{windings}. The K-splitting theorem then predicts that there are $K\geq |\nu_1|+|\nu_2|$ singular values $\sigma$ with $\lim_{N\to\infty}\sigma = 0$ \cite{BoettcherSilbermann,MonkmanSirkerNH}. This is fully consistent with our results; compare Fig.~\ref{fig:singularvalues} with the phase diagram in the left panel of Fig.~\ref{fig:gapclosingPBC}.

\begin{figure*}[!htbp]
\centering
\includegraphics[width=0.99\textwidth]{"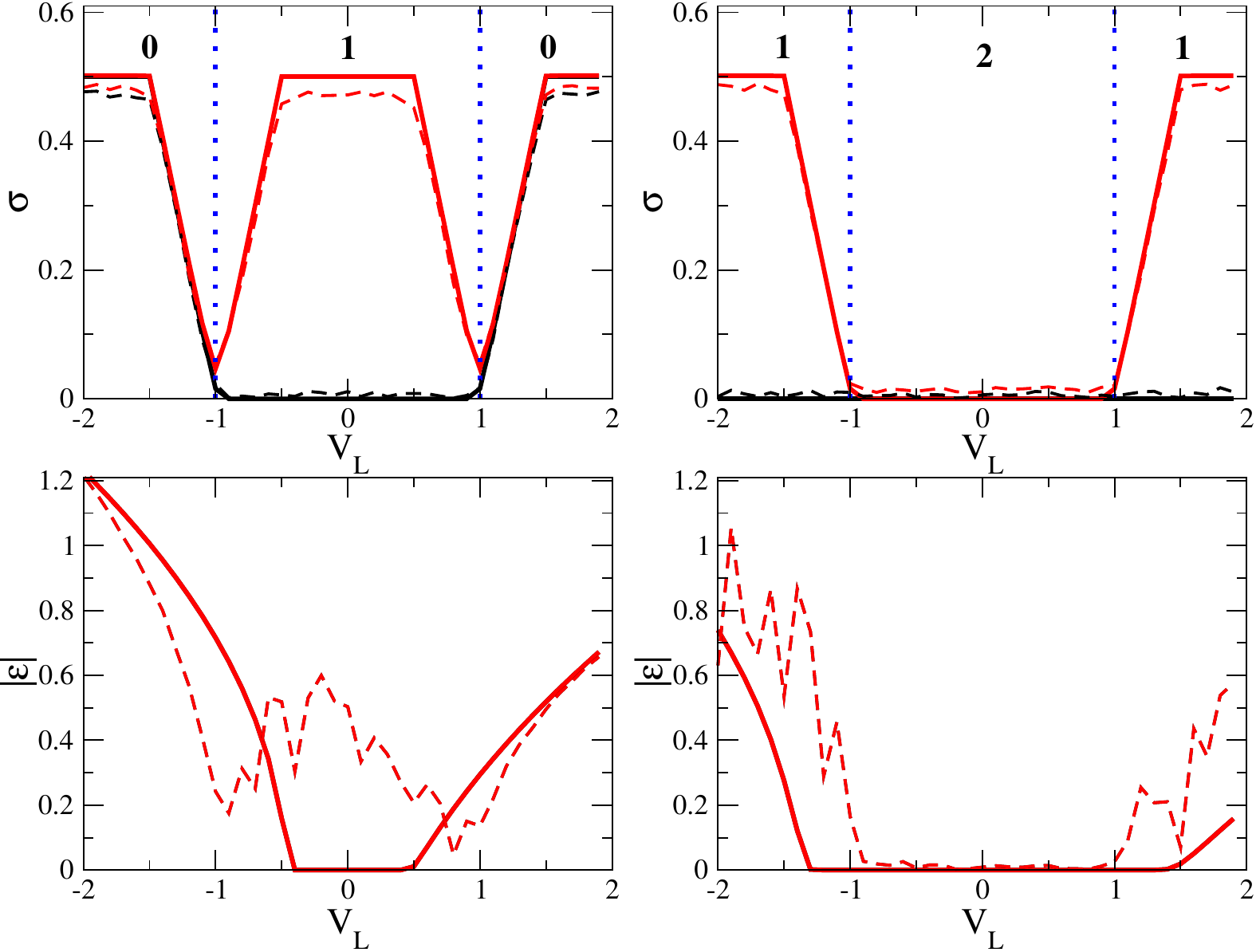"}
\caption{\label{fig:bbc} The smallest two singular values (top row) and the smallest two eigenvalues  (bottom row) for $W_L=0.5$ (left column) and $W_L=1.5$ (right column) with $V_R=W_R=1$ for a system with $N=200$. The two eigenvalues are always degenerate. The solid lines are for the unperturbed system, the dashed lines for a system where a random matrix with matrix elements $|a_{ij}|<0.02$ has been added. The numbers in the top row indicate the sum of the winding numbers $|\nu_1|+|\nu_2|$.}
\end{figure*}

\subsection{Topological protection and hidden zero modes}
In the limit of a semi-infinite chain, the system will have $K\geq |\nu_1|+|\nu_2|$ topologically protected edge modes with exactly zero energy. However, these states will only be eigenstates in this limit but not for a finite system. For a finite system, the singular values have to be considered as we have done above. Here we want to show explicitly how the exact edge states in the semi-infinite limit arise and how they become 'hidden' if the system is finite.

For a semi-infinite chain, we can find zero modes by looking for solutions of $Hv=0$ and $\tilde Hw=0$. Here $\tilde H=H(h_j \to h_{-j})$ is the reflected Hamiltonian with Fourier components $h_j$. The reason that we have to consider also the reflected Hamiltonian is that we are thinking about a system which extends up to infinity either from an edge to the left or to the right.

The condition $Hv=0$ leads to $v_2=0$ and $W_Rv_{2j}+V_Lv_{2j+2}=0$ implying that all even vector components are zero, $v_{2j}=0$. For the odd vector components we find the recurrence relation $V_R v_{2j-1}+W_L v_{2j+1}=0$ with the solution $v_{2j+1}=\left(-\frac{V_R}{W_L}\right)^j v_1$. In addition, we have to demand that the solution is normalizable, $||v||^2 =|v_1|^2\sum_{j=0}^\infty \left(-\frac{V_R}{W_L}\right)^{2j}=1$. This is possible if $|W_L|>|V_R|$. In this case, the semi-infinite chain has a zero energy edge mode $v$ exponentially localized at the left edge with vector components
\begin{equation}
    \label{zeromode1}
    v_{2j+1}=\sqrt{1-\left(\frac{V_R}{W_L}\right)^{2}}\left(-\frac{V_R}{W_L}\right)^{j},\quad v_{2j}=0 \, .
\end{equation}
Note that the only solution of $Hv=0$ for a finite system is the trivial vector $v=0$.

Similarly, we can consider zero modes of the reflected Hamiltonian, $\tilde Hw=0$. In this case we find that an exponentially localized zero mode exists if $|W_R|>|V_L|$ and has vector components
\begin{equation}
    \label{zeromode2}
    w_{2j+1}=0,\quad w_{2j+2}=\sqrt{1-\left(\frac{V_L}{W_R}\right)^{2}}\left(-\frac{V_L}{W_R}\right)^{j} \, .
\end{equation}
Again, for a finite system the only solution to $\tilde H w=0$ is $w=0$. The non-trivial edge mode with energy zero only exists in the thermodynamic limit. We note that we find the exact same separation of parameters as for the singular values: the first zero mode depends only on $\{V_R,W_L\}$ and the second zero mode only on $\{V_L,W_R\}$. For a finite system, we can truncate the vectors $v,w$. They are then no longer eigenvectors, $Hv\neq 0$, but they do get mapped exponentially close to zero, i.e. $||Hv||\to 0$ for $N\to\infty$ and similarly for $w$. This is what we mean by hidden zero modes.

For the example considered in Fig.~\ref{fig:bbc} with $V_R=W_R=1$ this means that the first zero mode is stable for $-1<V_L<1$ and the second zero mode for $|W_L|>1$. This is fully consistent with the singular values as well as with the winding numbers as is expected based on the K-splitting theorem. We thus do have explicitly confirmed the bulk-boundary correspondence in this case. 

To check that the singular values with $\lim_{N\to\infty}\sigma=0$ are topologically protected while zero eigenvalues are, in general, not protected for a finite non-Hermitian system, we also show in Fig.~\ref{fig:bbc} both quantities when calculated for a finite Hamiltonian $H$ ($N=200$) with a random complex matrix $A$ with elements $|a_{ij}|\leq 0.02$ added. The results show that the singular values are stable and thus indeed topologically protected while the eigenvalues are, in general, not protected. More specifically, we find that for the case $W_L=0.5$ shown in the lower left panel there are two degenerate zero eigenvalues in the unperturbed system for $-0.4\lesssim V_L\lesssim0.5$. However, the small perturbation immediately moves them away from zero energy. They are not protected. 

The situation is slightly more complicated in the case $W_L=1.5$ shown in the lower right panel of Fig.~\ref{fig:bbc}. In the unperturbed case, there are two degenerate zero energy eigenvalues for $-1.3\lesssim V_L\lesssim1.5$. Interestingly, adding a perturbation does not completely remove them from zero energy. Instead, they are stable for $-1<V_L<1$. This can be understood as follows: in this regime, the non-Hermitian Hamiltonian with $V_R=W_R=1$ and $W_L=1.5$ can be adiabatically connected to a chiral Hermitian Hamiltonian without closing the gap. This can be achieved, for example, by $W_R\to 1.5$ and then $V_R\to V_L$. For the obtained Hermitian Hamiltonian one can calculate the winding number of the upper block which turns out to be $I=1$. The standard bulk-boundary correspondence then predicts the existence of two protected zero-energy edge modes. Since the gap never closes and the sub-lattice symmetry remains intact, these zero modes survive in the non-Hermitian case. However, the full topology of the non-Hermitian model also in this case is only captured by the singular values, not by the eigenvalues.

\section{Entanglement entropy}
\label{Ent}
For a system in a topological phase, we expect that a non-zero lower bound for the entanglement entropy exists. This entanglement is protected and cannot be removed \cite{MonkmanSirker4}. To study this aspect of topology, we investigate in this section the
entanglement entropy for the
non-Hermitian model (\ref{tri})
with periodic boundary conditions. The other novel aspect which motivates our study of the entanglement entropy is the $\mathcal{PT}$-symmetric phase in which exceptional points are dense. As we will show, this leads to an entanglement entropy which is nowhere continuous.

We use the reduced density matrix approach for free fermions
\cite{P2003,VLRK2003}
which has been extended to the
non-Hermitian realm \cite{HRB2019,CYWR2020,GYHYCLX2021,L2022,CPLL2022,
HFZ2023,FAC2023,GTS2023,ShiDong} using the
generalized
density matrix $\rho = |r\rangle \langle \ell |$ where $\langle \ell|$ and $|r\rangle$
are bi-orthogonal left and right eigenstates
of (\ref{tri}). The key quantity
is the two-point correlation matrix with
elements
\beq\label{corrmatrix}
C_{ij}=\langle \ell|c_i^\dagger c_j|r\rangle
=\tr \left(\rho c_i^\dagger c_j\right) \, .
\eeq
Restricting the indices
to a given subsystem $\mathcal{A}$,
the von Neumann entanglement entropy
is given by,
\beq
S_{\mathcal{A}}=-\sum_{j}
\nu_j \log \nu_j+(1-\nu_j)
\log (1-\nu_j)
\eeq
where $\nu_j$ are the eigenvalues
of the correlation matrix (\ref{corrmatrix}) restricted 
to $\mathcal{A}$. In this paper
we only consider $\mathcal{A}=\{1,\dots,N/2\}$, that is, we cut
the system in half. For the model
considered in this paper, in contrast to Hermitian systems, the eigenvalues $\nu_j$ and therefore also $S_{\mathcal{A}}$ can be complex. Nevertheless, we will indiscriminately call $S_{\mathcal{A}}$ the entanglement entropy. In addition,
we choose the branch cut of the logarithm to be on the negative real axis. Before we proceed further, we might ask why $S_{\mathcal{A}}$ is complex and what the meaning of a complex entanglement entropy is. For the first part of the question it is important to remember that a non-Hermitian Hamiltonian is an effective approach obtained from a Master equation by dropping terms which induce quantum jumps. The density matrix in the proper Master equation is always positive semi-definite. $S_{\mathcal{A}}$ being complex is an artefact of the approximation. Nevertheless, the entanglement entropy can be a useful indicator for the correlations in the system and we expect, in particular, that it is sensitive to phase transitions and exceptional points. Physically, one can typically understand the real part of $S_{\mathcal{A}}$ as being still associated with the spreading of Schmidt values while the imaginary part is related to dissipation.

Also, in non-Hermitian
systems,
the notion of a ground state is not well
defined since the spectrum
is complex in general. We have to distinguish two cases. If there is a line gap, then we fill the states on one side of the line gap so that there is an excitation gap. If there is no line gap then constructing a ground state is somewhat arbitrary but we will describe in each case what the chosen ground state is.

To evaluate (\ref{corrmatrix}),
we first organize the eigenvalues in the desired ordering,
\beq
D=\text{diag}(\epsilon(k_{\sigma(1)}),\dots,
\epsilon(k_{\sigma(N)})
)
\eeq
where $\epsilon(k_{\sigma(j)})$
denotes one of the $N$ possible quasienergies.
Accordingly,
define the 
column vectors,
\beq
\vec{r}(k_{\sigma(j)}) =\sqrt{\frac{2}{N}} \begin{pmatrix}
r_1(k_{\sigma(j)}) \\
r_2(k_{\sigma(j)}) \\
\vdots \\
r_N(k_{\sigma(j)}) \\
\end{pmatrix},
\eeq
\beq
\vec{\ell}(k_{\sigma(j)}) =\sqrt{\frac{2}{N}} \begin{pmatrix}
\ell_1(k_{\sigma(j)}) \\
\ell_2(k_{\sigma(j)}) \\
\vdots \\
\ell_N(k_{\sigma(j)}) \\
\end{pmatrix},
\eeq
with components given by (\ref{rcomp},\ref{lcomp}),
which are respectively right and left eigenvectors of $\mathcal{T}_N$ with eigenvalue $\epsilon(k_{\sigma(j)})$.
Introduce the matrices
\beq
R=\left(\vec{r}(k_{\sigma(1)}),\dots,
\vec{r}(k_{\sigma(N)})\right),
\eeq
\beq
L=\left(\vec{\ell}(k_{\sigma(1)}),\dots,
\vec{\ell}(k_{\sigma(N)})\right).
\eeq
It follows that,
\beq
\mathcal{T}_N R=RD,
\quad
L^T \mathcal{T}_N=DL^T,
\quad
RL^T=L^TR=\id.
\eeq
The
diagonalized hopping matrix is
then given
by
\beq
\mathcal{T}_N=\boldsymbol{\eta}^\dagger D
\boldsymbol{\eta},
\quad
\boldsymbol{\eta}=L^T \boldsymbol{c},
\quad 
\boldsymbol{\eta}^\dagger=
\boldsymbol{c}^\dagger
R
\eeq
implying the algebra (\ref{calgebra}) also
for the $\boldsymbol{\eta}$. We have,
\beq
c_j = \sum_{m=1}^{N} R_{jm}\eta_m,
\quad
c_i^\dagger =
\sum_{n=1}^{N} L_{in}\eta_n^\dagger.
\eeq
Finally, define the states,
\beq
\langle \ell|=
\langle 0|\prod_{a\in occ.}\eta_a , 
\quad
|r\rangle=\prod_{a\in occ.}\eta_a^\dagger |0\rangle, 
\eeq
and compute,
\beq\label{corrmatrix1}
C_{ij}=\langle \ell|c_i^\dagger c_j|r\rangle=
\sum_{a\in occ.}
L_{ia}R_{ja}.
\eeq

We now investigate the spectrum of
the correlation matrix (\ref{corrmatrix1}) and the associated
entanglement entropy for indices in (\ref{corrmatrix1})
restricted to $i,j\in\{1,\dots,N/2\}$.
We perform numerical calculations
for large lattices,
in which case we only
keep
eigenvalues of the correlation
matrix that satisfy $|\nu_j|>\Lambda$ and 
$|1-\nu_j|>\Lambda$. Additionally, we set to zero real or imaginary parts of $\nu_j$ and $1-\nu_j$ that are individually smaller in magnitude than the chosen cutoff $\Lambda=10^{-50}$. This is particularly important in cases where the eigenvalues
of the correlation matrix are negative with a tiny imaginary part, whose sign can numerically oscillate around the branch cut. By calculating the discarded weight $\sum_{|\nu_j|<\Lambda}|\nu_j| + \sum_{|1-\nu_j|<\Lambda}|1-\nu_j|$ we make sure that the cutoff does not affect the results on the scale shown in the figures provided that we stay on one branch of the logarithm. 
Before considering numerical results for large systems we will also study, as a warm-up, the 
2-cell model ($N=4$) analytically.
Both zero field $u=0$ and $\mathcal{PT}$-symmetric cases are considered.

\subsection{Sub-lattice symmetric case}

For the zero field case, setting $V_R=W_R=1$, let us consider transitions
between the phases (i) $(\nu_1,\nu_2)=(1,-1)$ (with $W_L>1$) and 
$(0,-1)$, (ii) between $(1,0)$
and $(0,0)$, and (iii) between 
$(1,-1)$ (with $W_L<-1$) and $(0,-1)$,  recall the phase diagram
shown in the left panel of Fig.~\ref{fig:gapclosingPBC}. For concreteness, let us consider only $V_L\geq0$. Recall that the spectrum of $\mathcal{T}_N$ in the phases $(1,-1)$ and $(0,0)$ is in general composed of two lobes in the complex plane either separated by the imaginary axis or real axis, therefore being characterized
by $I_1=0$. On the other hand, the spectrum in the phases $(1,0)$
and $(0,-1)$ traces a closed path around
$E_B=0$, therefore being characterized
by $I_1\neq 0$.

\begin{figure*}[!htbp]
\centering
\includegraphics[width=1.99\columnwidth]{"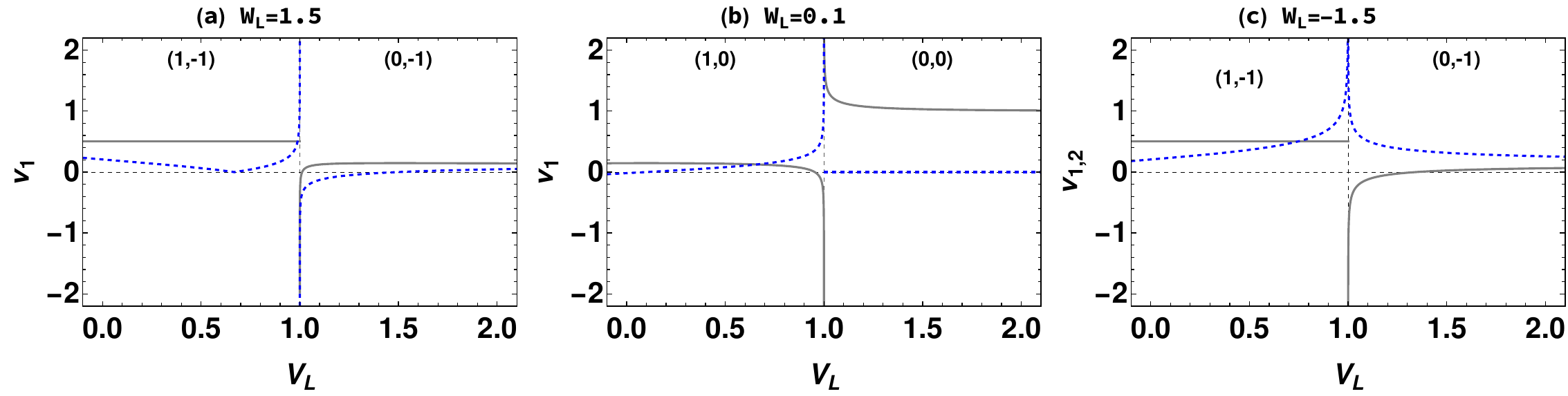"}
\includegraphics[width=1.99\columnwidth]{"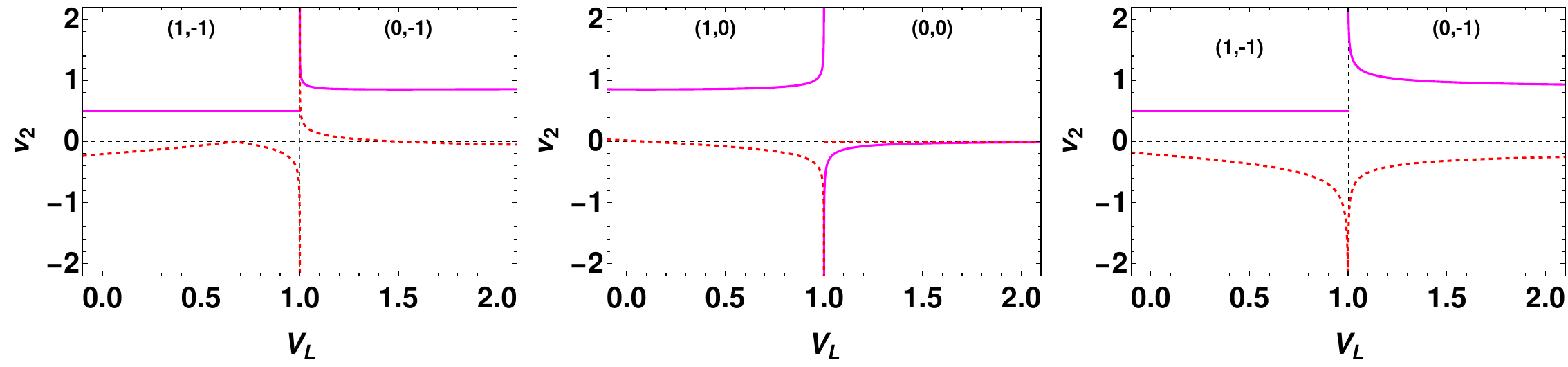"}
\includegraphics[width=1.99\columnwidth]{"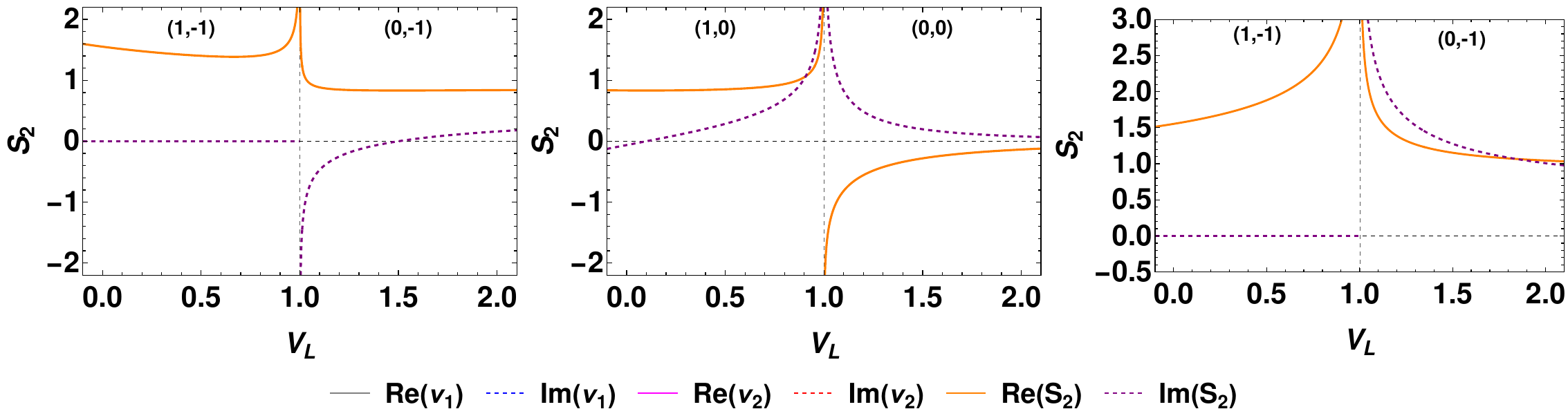"}
\caption{\label{fig:eigcorrsmall}Eigenvalues of the correlation matrix (top and middle panels)
and entanglement entropy (bottom panels)
for the two-cell model ($N=4$) with zero field $u=0$, $W_R=V_R=1$ and some fixed values of $W_L$ indicated in the labels (a)-(c) as a function of $V_L$. $S_2$ is discontinuous at the phase transitions.}
\end{figure*}

We start with
the 2-cell
model ($N=4$). For this lattice size, there is no lobe in the spectrum, and the eigenvalues are either purely real or purely imaginary. We fill the states with energy $-\sqrt{(V_L\mp1)(1\mp W_L)}$ and leave empty the
states with energy $\sqrt{(V_L\mp1)(1\mp W_L)}$. The associated correlation matrix is given by,
\begin{widetext}
\beq\label{C2}
C=\left(
\begin{array}{cc}
 \frac{1}{2} & -\frac{\sqrt{\left(V_L-1\right)
 \left(1-W_L\right)}}{4 \left(V_L-1\right)}-\frac{\sqrt{\left(V_L+1\right) \left(W_L+1\right)}}{4 \left(V_L+1\right)} \\
 -\frac{\sqrt{\left(V_L-1\right) \left(1-W_L\right)}}{4 \left(1-W_L\right)}-\frac{\sqrt{\left(V_L+1\right) \left(W_L+1\right)}}{4 \left(W_L+1\right)} & \frac{1}{2} \\
\end{array}
\right),
\eeq
\end{widetext}
which is non-Hermitian
and can have complex eigenvalues.
Indeed, we plot both real and imaginary
parts of the two eigenvalues of (\ref{C2}) in Fig.~\ref{fig:eigcorrsmall} for some
fixed values of $W_L$ as function of $V_L$.

We observe that within the phases characterized by $(1,-1)$ the eigenvalues
form conjugate pairs of the form 
$1/2\pm \ii\alpha$, where $\alpha$ is a given function of $V_L$ and $W_L$, see panels (a) and (c) in Fig.~\ref{fig:eigcorrsmall} for the fixed values $W_L=\pm 1.5$. The same form of eigenvalues is observed for other values of $|W_L|>1$. Thus, in this phase, the entanglement entropy is real and given by,
\beq\label{S2formula}
S_2=-\log\left(\frac{1}{4}+\alpha^2\right)+4\alpha \arctan(2\alpha)
\eeq
where $\alpha$ depends on $V_L$ and $W_L$. We also note
that $\alpha\rightarrow \infty$ as the transition point $V_L=1$ is approached and,
as a consequence,
$S_2$
 diverges at $V_L=1$. On the other hand, deep inside the phase, taking $V_L=0$ and $W_L\rightarrow\infty$, we have $\alpha\rightarrow0$ and therefore $S_2\rightarrow2\log 2$.

Increasing $V_L$ and crossing the transition point $V_L=1$, we observe that the eigenvalues of the correlation matrix in the phase $(0,-1)$ are complex, with the form $\beta_{1,2}\pm \ii \alpha$. That is, the eigenvalues have imaginary parts with opposite signs but with different real parts $\beta_{2}=1-\beta_1$. This leads to complex values for the entanglement entropy. However, for large $|W_L|$ 
the imaginary part tends to disappear when $V_L$ increases. In the limit $V_L\rightarrow \infty$ and $|W_L|\rightarrow \infty$, the exact eigenvalues
of the correlation matrix are $1/2\pm \sqrt{2}/4$, producing
a real entropy $S_2\approx 0.832991$.

In the phase
$(1,0)$, the eigenvalues once more have the form $\beta_{1,2}\pm \ii \alpha$ with $\beta_2=1-\beta_1$, and the entanglement entropy is complex (except if $V_L=W_L$),
see panel (b) in Fig.~\ref{fig:eigcorrsmall} for the fixed value $W_L=0.1$. Moving to the phase $(0,0)$, we observe that the eigenvalues of $C$ are real, however, they are greater than unity or smaller than zero, and therefore also produce a complex entanglement entropy. Only in the limit $V_L\rightarrow\infty$ and $W_L\rightarrow0$,
do the eigenvalues of the correlation matrix tend to $1$ and $0$, thus producing a zero entanglement
entropy.

\begin{figure*}[!htbp]
\centering
\includegraphics[width=1.99\columnwidth]{"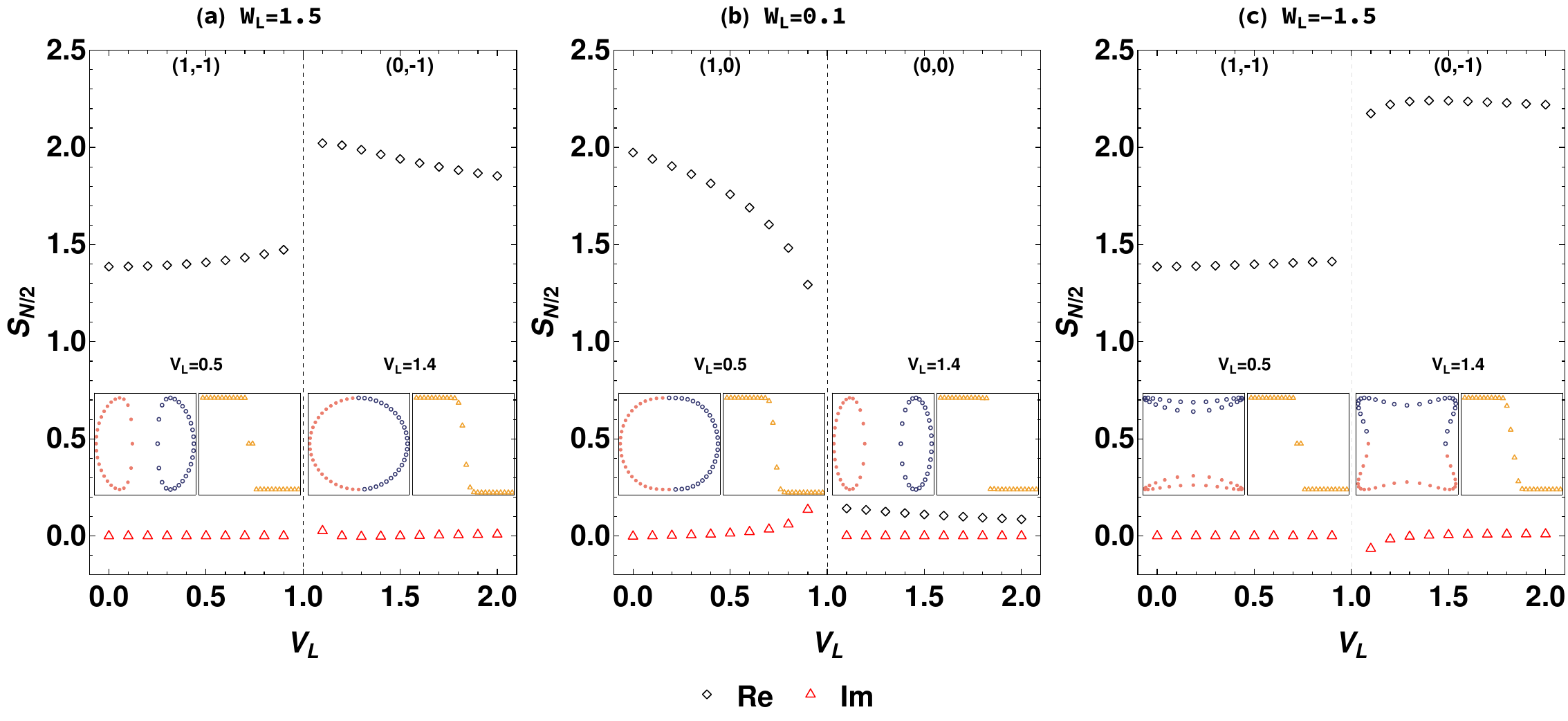"}
\caption{\label{fig:entPBCcutszerofieldim}Entanglement entropy for the model with sublattice symmetry (zero field $u=0$) for $V_R=W_R=1$ and some fixed values of $W_L$ indicated in the labels (a)--(c), as a function of $V_L$. The lattice size is $N=120$. $S_{N/2}$ is discontinuous at the phase transitions. In the insets, the typical form of the complex spectrum is shown; the occupied states are indicated by filled circles while the unoccupied states by open circles. The typical eigenvalues of the correlation matrix
are represented by empty up triangles (only the real part is shown since the complex part when present is small). For better visualization,
in the insets we consider $N=52$.}
\end{figure*}

Now, we consider larger lattice sizes using numerical diagonalization. We observe a variety of possible types of
eigenvalues in the correlation spectrum, similar to the 2-cell model. In some phases, we may find eigenvalues with $\text{Re}(\nu_j)>1$ as well as $\text{Im}(\nu_j)\neq 0$. Also, we note that
near the transition points
the behavior of the correlation matrix
eigenvalues is more intricate. Let us now analyze each phase
in detail. Results
are collected in Fig.~\ref{fig:entPBCcutszerofieldim} for a lattice composed of $N=120$ sites. In the
insets of Fig.~\ref{fig:entPBCcutszerofieldim}, we show the typical form of the complex spectrum indicating the filling adopted,
as well as the real part of the eigenvalues of the correlation matrix,
considering a smaller lattice with $N=52$
for better visualization of the results.

Starting with the phases $(-1,1)$, see panels (a) and (c) in Fig.~\ref{fig:entPBCcutszerofieldim} for the fixed values $W_L=\pm 1.5$, we observe that most of the eigenvalues
of the correlation matrix
are either close to $1$ or $0$. Some of the eigenvalues have the property $\nu_j>1$ or $\nu_j<0$, and there are two eigenvalues sitting
at $1/2\pm \ii \alpha$. As the lattice size increases, we observe that those eigenvalues with $\nu_j>1$ or $\nu_j<0$ tend to unity or zero, while $\alpha$ decreases. For instance, for $N=52$, the lattice size considered in the insets, the highest eigenvalue is $\nu_{\text{max}}\approx 1.000000000290214$, the lowest $\nu_{\text{min}}=1-\nu_{\text{max}}$ and $\alpha \approx 4.275663\times 10^{-4}$, while
for $N=120$ the highest eigenvalue for $W_L=1.5$ and $V_L=0.5$ is $\nu_{\text{max}}\approx
1.00000000000000000265940$, the lowest $\nu_{\text{min}}=1-\nu_{\text{max}}$ and $\alpha \approx 2.955440\times 10^{-7}$.
Therefore, the entanglement entropy comes essentially from the
conjugate
eigenvalues $1/2\pm \ii \alpha$, and its value is also given by (\ref{S2formula}).
For a fixed length of the chain, but going
deeper into the phase by increasing $|W_L|$,
we also note that $\alpha$ decreases.
Then, in the limit
$\alpha\rightarrow 0$, the entanglement entropy has the finite value $S_{N/2}\rightarrow 2\log 2$ deep inside the phase. Similar considerations can be made for the phase $(1,-1)$ with $W_L<-1$; note however that a different
filling must be used in this case.

\begin{figure}[!h]
\centering
\includegraphics[width=0.99\columnwidth]{"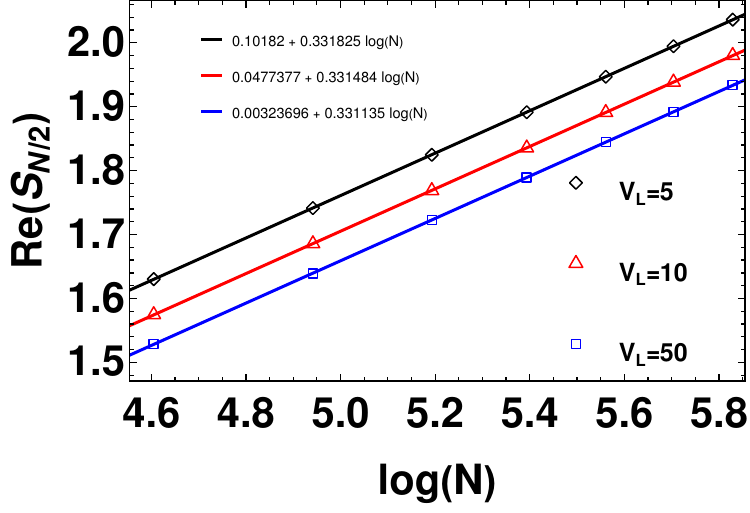"}
\includegraphics[width=0.99\columnwidth]{"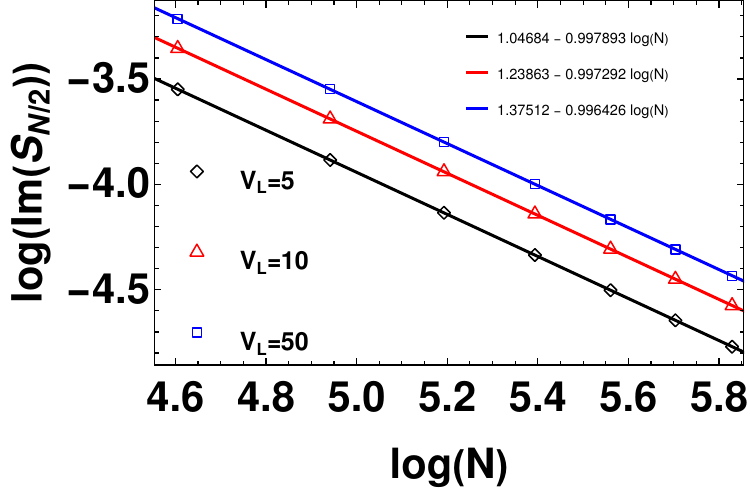"}
\caption{\label{fig:gSN} Scaling of the entanglement entropy in the phase $(0,-1)$ for $W_L=1.5$ and values of $V_L$ deep inside the phase.}
\end{figure}

Moving to the phases $(0,-1)$, see once more
panels (a) and (c) in Fig.~\ref{fig:entPBCcutszerofieldim},
we note that all eigenvalues of the correlation matrix satisfy $0<\text{Re}(\nu_j)<1$, with most eigenvalues being either close to unity or zero. Some intermediate eigenvalues appear, and these carry a small imaginary part. They produce a finite entanglement entropy, also complex,
but with a small imaginary part. However,
the number of intermediate states actually
grows with the length of the chain. This implies
a scaling of the entanglement entropy,
see Fig.~\ref{fig:gSN}, where we fix $W_L=1.5$
and some values of $V_L$ deep inside the phase.
We observe a scaling of the form $\text{Re}(S_N)\sim (c/3) \log(N)$ with $c=1$,
as well as a decay $\text{Im}(S_N)\sim \exp(-\log(N))$
of the imaginary part.
It follows that, remarkably, this phase
may be described by a CFT with central charge $c=1$, even though it has a point gapped spectrum.
Similar results can be obtained for $W_L<-1$ in the same phase $(0,-1)$.

\begin{figure} 
\centering
\includegraphics[width=0.99\columnwidth]{"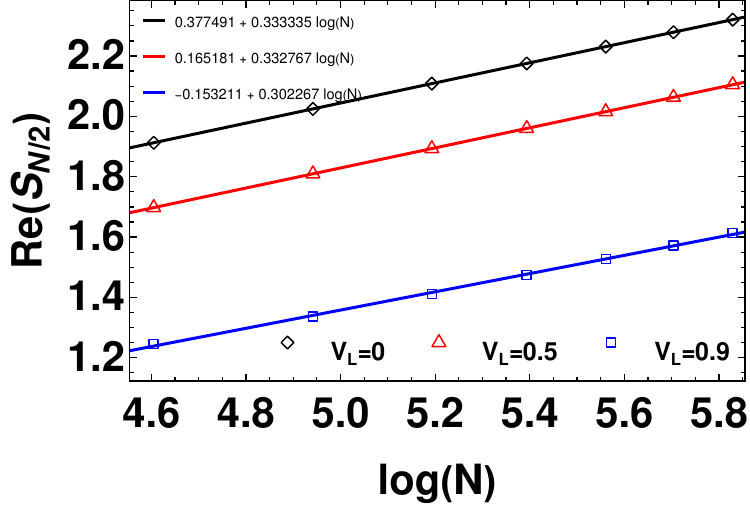"}
\includegraphics[width=0.99\columnwidth]{"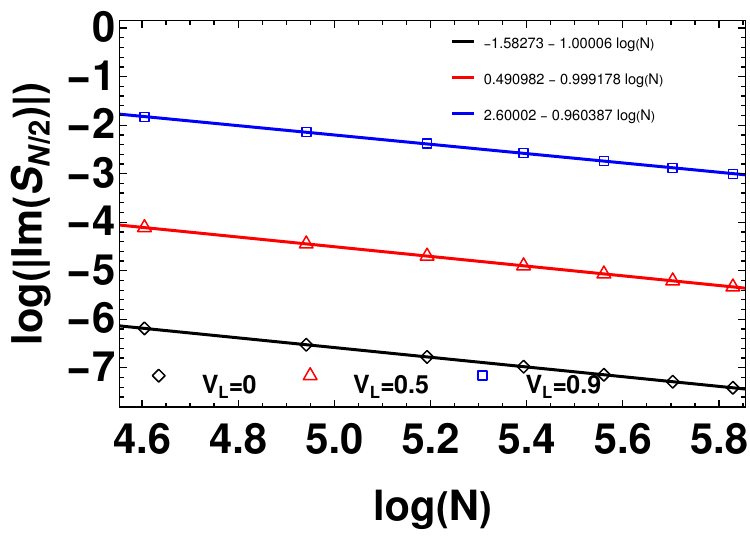"}
\caption{\label{fig:gSN10} Scaling of the entanglement entropy in the phase $(1,0)$ for $W_L=0.1$ and some values of $V_L$ inside the phase. Note that for $V_L=0$ the imaginary part is negative, so it is its absolute value that decreases with the length of the chain.}
\end{figure}

The phase $(1,0)$, see panel (b) in Fig.~\ref{fig:entPBCcutszerofieldim}, is similar to the phases $(0,-1)$. The scaling of the entanglement entropy for this case is shown in Fig.~\ref{fig:gSN10}. Also in this phase the scaling of the entanglement entropy is consistent
with a central charge $c=1$.

The last phase to be discussed in this subsection is the phase $(0,0)$. Here, we observe eigenvalues of the correlation matrix with $\text{Re}(\nu_j)>1$ and also some eigenvalues with
a small imaginary part. We note that,
for fixed parameters $W_L$ and $V_L$
in this phase, the imaginary part of the entanglement entropy decreases with the length of the chain, while the real part
slightly increases and saturates. However, as we go deeper into this phase
by increasing $V_L$, the entanglement entropy tends to disappear. This phase is therefore trivial.

In summary, for the zero field case and in the thermodynamic limit, the
entanglement entropy is real deep inside the quantum phases, and discontinuous
near the transition
points. The gapped non-trivial phases have two protected eigenvalues at $1/2$, contributing $S_N=2\log 2$ to the entanglement entropy.

\subsection{$\mathcal{PT}$-symmetric case}

Let us now consider the $\mathcal{PT}$
symmetric case with PBC, setting $u=1$.
For simplicity, 
we will restrict ourselves to the quadrant with $V\geq0$ and $W\geq0$ in the phase diagram shown in the right panel of
Fig.~\ref{fig:gapclosingPBC}. Then,
four regions must be analyzed.

\begin{figure*}[!htbp]
\centering
\includegraphics[width=1.99\columnwidth]{"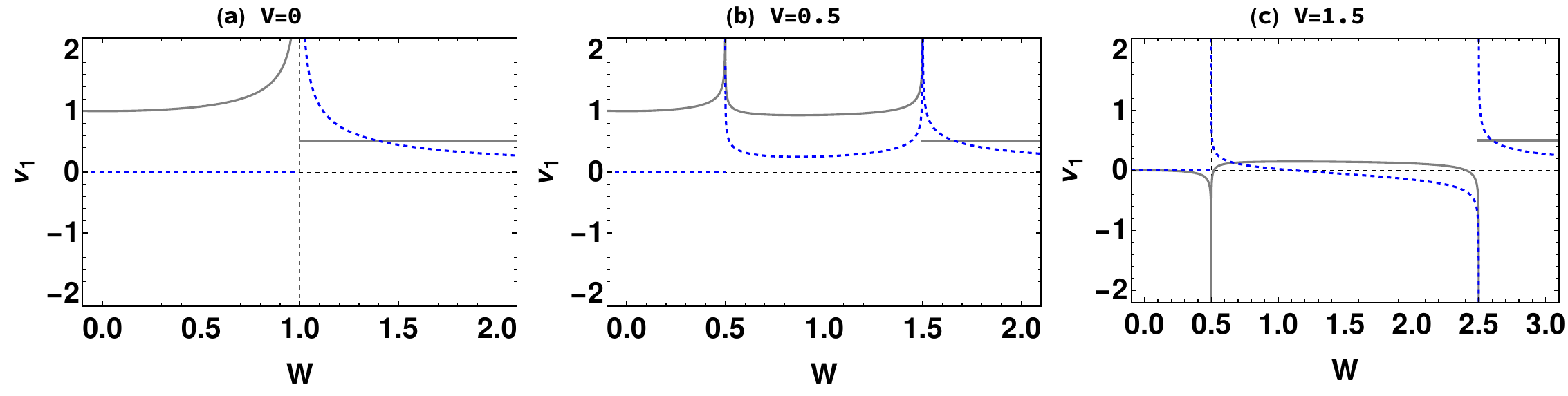"}
\includegraphics[width=1.99\columnwidth]{"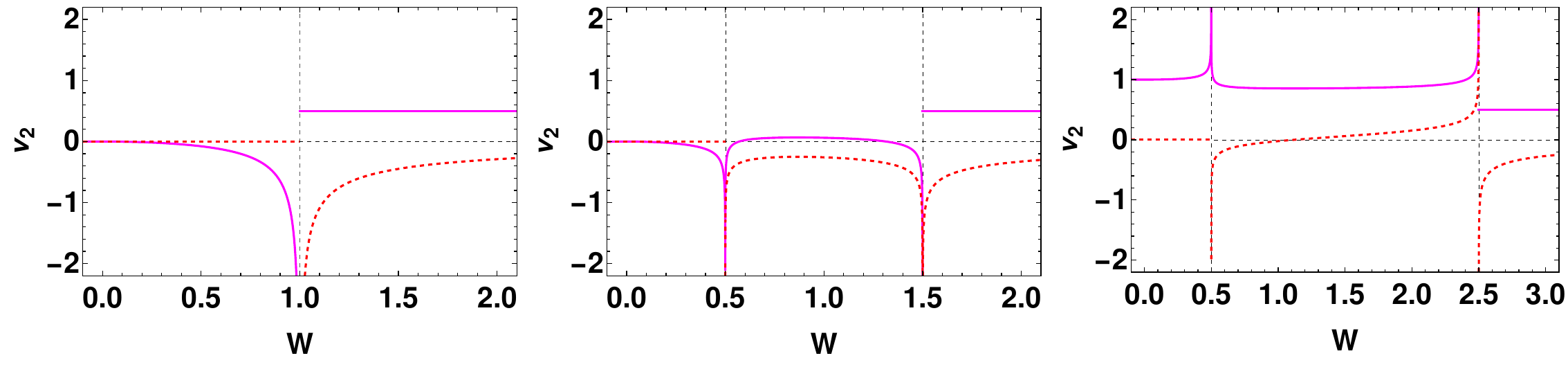"}
\includegraphics[width=1.99\columnwidth]{"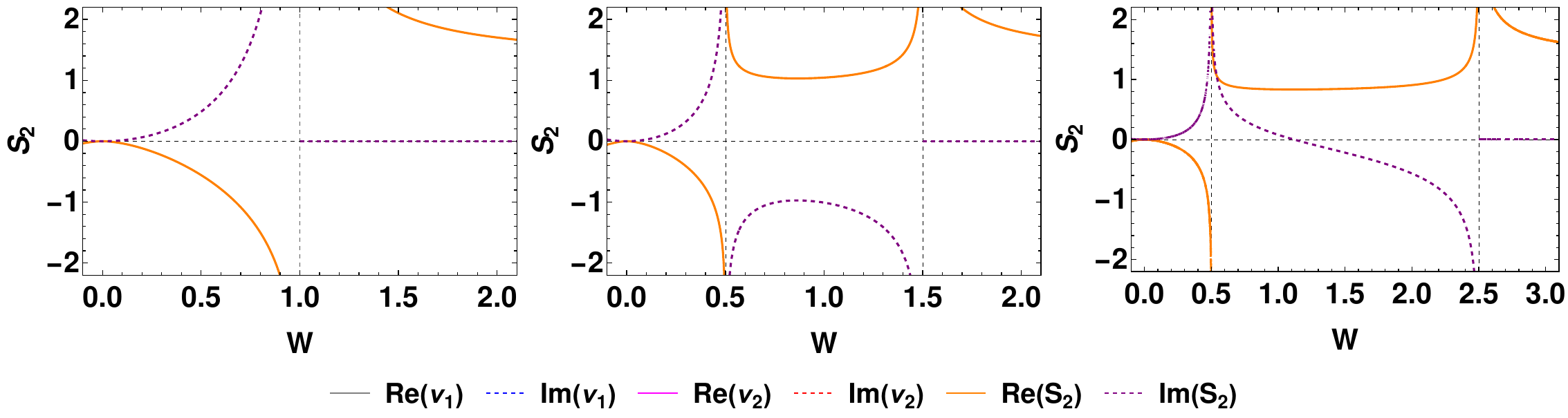"}
\caption{\label{fig:eigcorrsmallPT}Eigenvalues of the correlation matrix (top and middle panels)
and entanglement entropy (bottom panels)
for the $\mathcal{PT}$-symmetric two-cell model ($N=4$) with $u=1$, and fixed values of $V$ as indicated in the labels (a)-(c) as a function of $W$. $S_2$ is discontinuous at the phase transitions.}
\end{figure*}

As before, we start our analysis with the simple 2-cell model ($N=4$), which has the following correlation matrix,
\begin{widetext}
\begin{equation*}
C=\frac{1}{4}\left(
\begin{array}{cc}
 \frac{(V-W)^2}{(V-W)^2-1+\ii \sqrt{(V-W)^2-1}}+\frac{(V+W)^2}{(V+W)^2-1+\ii \sqrt{(V+W)^2-1}} & \frac{W-V}{\sqrt{(V-W)^2-1}}-\frac{V+W}{\sqrt{(V+W)^2-1}} \\
 \frac{W-V}{\sqrt{(V-W)^2-1}}-\frac{V+W}{\sqrt{(V+W)^2-1}} & \frac{(V-W)^2}{(V-W)^2-i \sqrt{(V-W)^2-1}-1}+\frac{(V+W)^2}{(V+W)^2-i \sqrt{(V+W)^2-1}-1} \\
\end{array}
\right)
\end{equation*}
\end{widetext}
associated with the filled states $-\sqrt{(V\mp W)^2-1}$ and empty states $\sqrt{(V\mp W)^2-1}$.
We plot in Fig.~\ref{fig:eigcorrsmallPT} the eigenvalues of $C$
and the associated entanglement entropy for some
fixed values of $V$ as a function of $W$,
covering all the four regions of the phase diagram.

For $V=0$ and $0<W<1$ that is, within the complex phase, see panel (a) in Fig.~\ref{fig:eigcorrsmallPT}, we observe that
the eigenvalues of the correlation matrix are real
with $\nu_j>1$ or
$\nu_j<0$ until the phase transition point $W=1$
is reached. As a consequence,
the entanglement entropy is complex in this phase. Crossing the transition point ($W=1$) towards the $\mathcal{PT}$-unbroken topological phase,
we note that the eigenvalues
have the form $1/2\pm \ii \alpha$, and the entanglement entropy is given by (\ref{S2formula}). In the
limit $W\rightarrow \infty$, $\alpha$ becomes negligible and the entanglement
entropy tends to $S_2\rightarrow 2\log(2)$, indeed characteristic of
a topological phase with winding $I=1$.

Increasing $V$ to $V=0.5$, see panel (b) in Fig.~\ref{fig:eigcorrsmallPT},
the behavior of the eigenvalues $\nu_j$ is the same
as for $V=0$ until the exceptional phase is reached at $W=0.5$. Within the exceptional phase,
the eigenvalues $\nu_j$ have
the form $\beta_{1,2}\pm \ii \alpha$, that is, the eigenvalues have the same imaginary part with opposite signs but different real parts
related by
$\beta_2=1-\beta_1$. As a consequence,
the entanglement entropy is complex in the exceptional phase. This phase ends at $W=1.5$ from which point the
$\mathcal{PT}$-unbroken topological phase is again reached being characterized by a real entanglement entropy which tends to $S_2\rightarrow2\log(2)$ deep inside the phase.

To conclude the analysis of the 2-cell model, we consider $V=1.5$
such that the $\mathcal{PT}$-unbroken trivial phase is considered. This phase for $V=1.5$ occurs when $0<W<0.5$, and we note that there the eigenvalues are real with 
$\nu_j>1$ or $\nu_j<0$, producing again a complex entanglement entropy except if $W=0$. However, deep inside the phase ($V\rightarrow\infty$), the eigenvalues
tend to $0$ and $1$, leading to a vanishing
entanglement entropy, as expected from a trivial phase. For $0.5<W<2.5$ we encounter once more the 
exceptional phase and for $W>2.5$ the 
$\mathcal{PT}$-unbroken topological phase. 

\begin{figure*}[!htbp]
\centering
\includegraphics[width=1.75\columnwidth]{"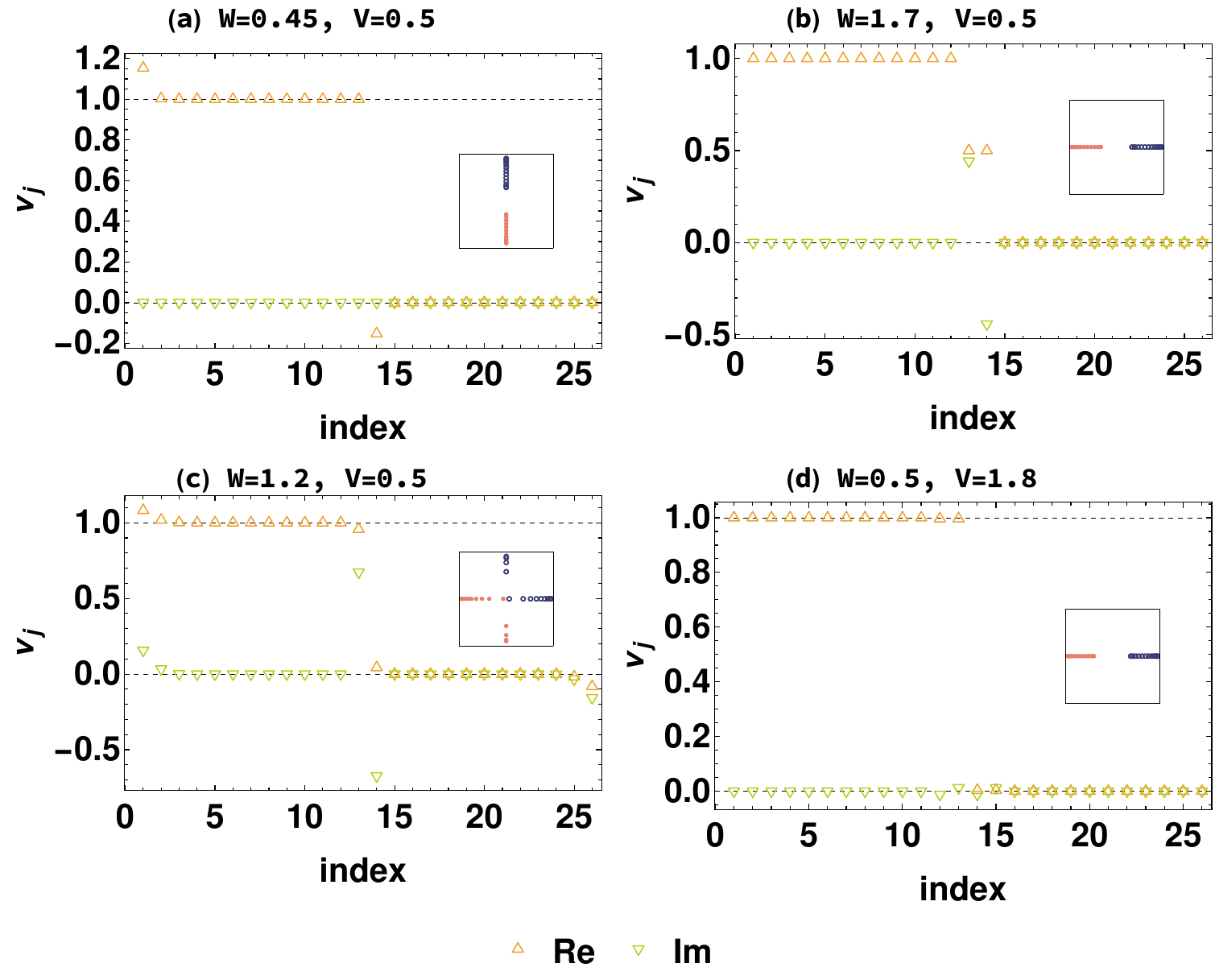"}
\caption{\label{fig:eigcorrPT}Eigenvalues of the correlation matrix for the
$\mathcal{PT}$ symmetric case with PBC and
$u=1$ for some fixed values of $W$ and $V$ indicated in the labels (a)-(d). 
The lattice size is $N=52$. In the insets, the complex spectrum is shown; occupied states are indicated by orange filled circles and unoccupied states by blue open circles.}
\end{figure*}

\begin{figure*}[!htbp]
\centering
\includegraphics[width=1.99\columnwidth]{"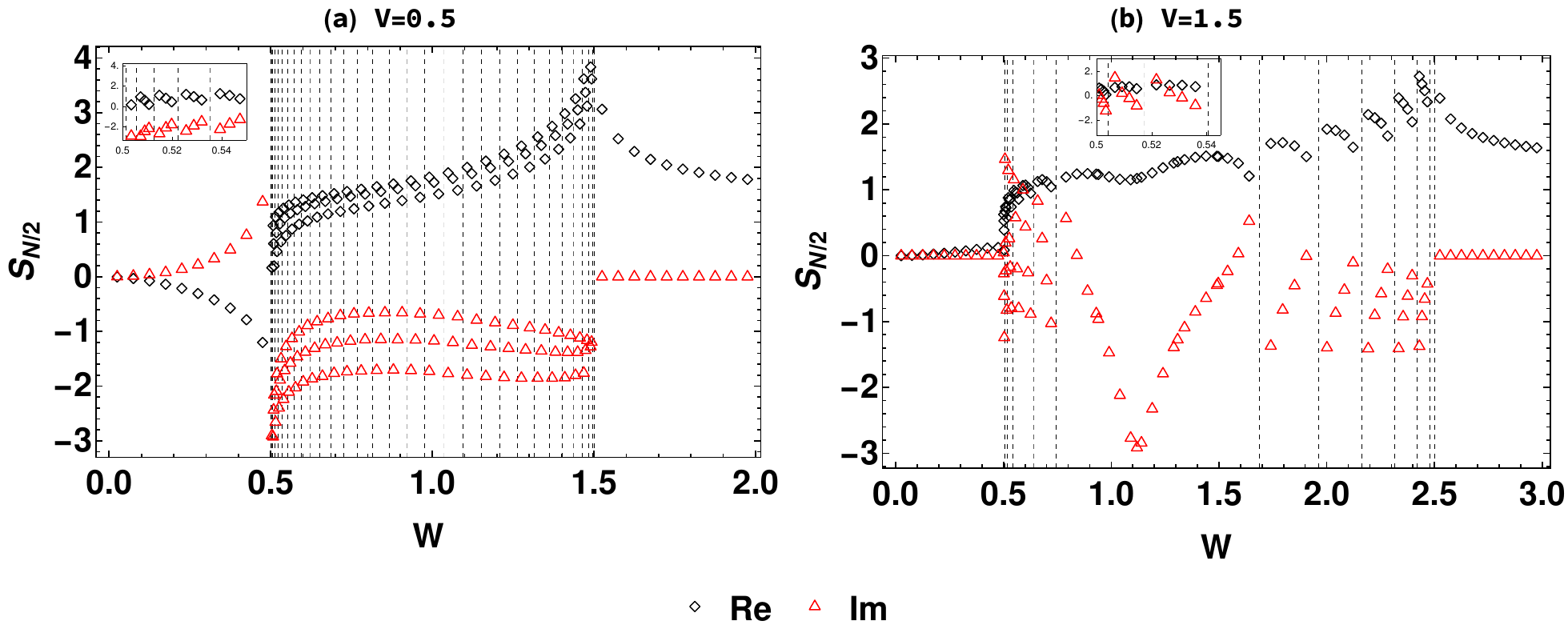"}
\caption{\label{fig:entPBCcutszerofieldimPT}Entanglement entropy for the
$\mathcal{PT}$ symmetric case with PBC and
$u=1$ 
as
a function of $W_L=W_R=W$ for two fixed values of $V_L=V_R=V$ as indicated in the labels. The lattice size is $N=120$. The dashed grid lines indicate the position of real exceptional points. In the insets, we zoom into regions where the exceptional points are very close to each other. In the thermodynamic limit, the exceptional points become close and $S_{N/2}$ is nowhere continuous.}
\end{figure*}

Finally, we consider the $\mathcal{PT}$ case for large lattice sizes.
Results for the correlation spectrum are collected in Fig.~\ref{fig:eigcorrPT} (where $N=52$) and for the entanglement entropy in Fig.~\ref{fig:entPBCcutszerofieldimPT} (where $N=120$).
The same four regions considered for the 2-cell model are analyzed.

Similarly to the 2-cell model, by fixing $0<V<1$,
we can reach the complex phase, the exceptional phase and the $\mathcal{PT}$-unbroken topological phase by varying $W$. On the other hand, by fixing $V>1$ and varying $W$ we can reach the $\mathcal{PT}$-unbroken trivial phase,
the exceptional phase and the $\mathcal{PT}$-unbroken topological phase.

Let us consider once more $V=0.5$. Within the complex phase ($0<W<0.5$ for $V=0.5$), all the eigenvalues of the correlation matrix jump above unity or below zero,
see panel (a) in Fig.~\ref{fig:eigcorrPT}
where the parameters $W=0.45$ and $V=0.5$ are used. This results
in a complex value for $S_{N/2}$,
see panel (a) in  Fig.~\ref{fig:entPBCcutszerofieldimPT}.
We note that two eigenvalues stand out,
being significantly greater than $1$ or less than $0$. All the eigenvalues of the correlation matrix in this phase seem to vary slowly with the size of the chain,
actually tending to stabilize. For example, the difference between the largest eigenvalues for the lattices with $N=52$ and with $N=120$ is of the order of $10^{-9}$. Therefore, the entanglement entropy in this phase seems to be finite. However, it is difficult to verify this property since the model parameters in this phase lie in a finite interval. Also,
the phase is surrounded by critical lines
that drive the entanglement entropy to high values. 

Moving to the $\mathcal{PT}$-unbroken topological phase, we note that the correlation spectrum is composed by eigenvalues close
to $1$ and
$0$, with some of them being slightly bigger than $1$ or smaller then $0$, as well as a conjugate pair of the form
$1/2\pm\ii \alpha$, see panel (b) in
Fig.~\ref{fig:eigcorrPT} where the parameters are set to $W=1.7,V=0.5$.
 As the lattice size increases,
we note that the eigenvalues tend to $1$ or $0$, while $\alpha$ decreases slowly and seems to stabilize with the length of the chain. The associated entanglement entropy is real and finite. Going deep into the phase by increasing $W$, as before, $\alpha$ decreases and the entanglement tends to  $S_{N/2}\rightarrow 2\log{2}$,
see both panels (a) and (b) in Fig.~\ref{fig:entPBCcutszerofieldimPT}.

In the $\mathcal{PT}$-unbroken trivial phase,
the correlation spectrum is also composed mostly of $0$'s and $1$'s, some
of them slightly bigger than $1$ or smaller than $0$. Also, some eigenvalues
carry an imaginary part in the form of conjugate pairs,
see panel (d) in Fig.~\ref{fig:eigcorrPT} for the model parameters $W=0.5$ and $V=1.8$. Interestingly, similarly to previous cases, both real and imaginary part of the eigenvalues vary slowly with the length of the chain. Nevertheless, deep inside this phase (large $V$), the eigenvalues tend to $0$ or $1$ and produces zero entanglement, as expected from a trivial phase; see panel (b) in Fig.~\ref{fig:entPBCcutszerofieldimPT}.

Let us now discuss the exceptional phase. In this phase we observe an intricate behavior of the correlation
spectrum, characterized by various complex eigenvalues, see panel (c) in
Fig.~\ref{fig:eigcorrPT} where the parameters are set to $V=0.5,\, W=1.2$. This behavior is reflected
in an intricate evolution of the entanglement entropy,
see panels (a) and (b) in Fig.~\ref{fig:entPBCcutszerofieldimPT}.
Interestingly, the entanglement entropy detects the presence of the real exceptional points (indicated by dashed lines in Fig.~\ref{fig:entPBCcutszerofieldimPT}).
For clarity, we only pick a few values of $W$ 
between the exceptional points. Note that in the thermodynamic limit the entanglement entropy is ill defined in the exceptional phase, since it is filled with exceptional points. For finite systems, we expect a scaling of $S_{N/2}$. This is numerically challenging and goes beyond the scope of this paper.

In brief, the entanglement entropy
has a rich behavior in the $\mathcal{PT}$ symmetric model and is capable of detecting both exceptional points and phase transitions. In the $\mathcal{PT}$-unbroken topological phase there are two protected eigenvalues at $1/2$, contributing $2\log 2$ to the entanglement entropy.

\section{Conclusions}
\label{Concl}
In this paper we have analytically studied a paradigmatic one-dimensional, non-Hermitian lattice model. Our goal was to illustrate many of the unique features of non-Hermitian systems---in particular, the proper bulk-boundary correspondence and the exceptional points---in a concrete model and to do so largely based on analytical instead of purely numerical results. A particular focus of our work was to clarify the recently described relation between topological winding numbers for periodic systems and topologically protected singular values as well as hidden zero modes in open systems \cite{MonkmanSirkerNH}. We have shown explicitly that the latter only become exact zero-energy eigenmodes in the limit of a semi-infinite chain. For a finite chain, they correspond to metastable states instead which is a crucial distinction for experiments.

We derived the eigensystems both for closed (periodic and anti-periodic) as well as open boundary conditions. Our analytical results show how the Bloch waves in the closed case change into localized eigenstates for open boundaries thus explicitly demonstrating the non-Hermitian skin effect. Another common feature of non-Hermitian systems are exceptional points where both eigenvalues and -vectors coalesce. In the considered model, we have shown that such points typically occur when the complex gap closes and that the location of exceptional points, in general, differs between the closed and the open case. Surprisingly, we found in the $\mathcal{PT}$-symmetric case that exceptional points can even become dense in the thermodynamic limit, forming an entire novel ``exceptional phase''. This phase has also other unique properties which are impossible in a Hermitian phase. We found, in particular, that transport is hyper-ballistic with a dynamical critical exponent $z=1/2$.

As already mentioned above, one of the central results of this work is an explicit demonstration of a proper bulk-boundary correspondence for a non-Hermitian system. This point has lead to some confusion in the literature. Here we have shown that the winding numbers for the periodic system determine the number of protected singular values which go to zero in the thermodynamic limit. This number also corresponds to the number of exact zero-energy edge modes for a semi-infinite chain. However, the eigenspectrum of the finite chain does, in general, not converge to that of the semi-infinite one. Thus, no bulk-boundary correspondence between winding numbers and the eigenspectrum of a finite system exists, in contrast to Hermitian systems. The exact zero-energy edge states in the semi-infinite non-Hermitian case are not exact eigenstates for a finite system. They become hidden which means that they correspond to metastable states which are, however, very long-lived and thus highly relevant for our understanding of non-Hermitian systems; see also Ref.~\cite{MonkmanSirkerNH} where this issue was discussed more generally. It is also important to note that even if zero-energy eigenstates for a finite system do exist, they are typically not topologically protected. For the studied model we demonstrated this explicitly by adding a small perturbation to the Hamiltonian and showing that the eigenenergies shift while the topologically protected singular values are stable. 

Another aspect of topological protection in non-Hermitian systems which we explored here is the entanglement spectrum. For cases with a line gap we could, in particular, define a half-filled system where occupied and unoccupied bands are separated by a gap. Since the studied system is Gaussian, the entanglement spectrum is fully determined by the spectrum of the single-particle correlation matrix. We found that phase transitions and exceptional points are indicated by discontinuities in the entanglement entropy. Furthermore, we demonstrated that in topologically non-trivial phases there are protected eigenvalues in the entanglement spectrum which lead to lower, non-trivial entanglement bounds. 

A non-trivial topology in non-Hermitian systems thus leads to two observable and experimentally relevant phenomena: (1) Extremely long-lived metastable zero energy states, and (2) non-trivial entanglement in systems with a gap between occupied and unoccupied energy bands. 

We end this section by mentioning some possible directions for future studies. First, it would be worthwhile to study the loci of exceptional points for other families of non-Hermitian free-fermionic models, including free fermions and parafermions
in disguise \cite{F19,AP20}. More importantly, it would be very interesting to extend these studies to the loci of exceptional points for non-Hermitian interacting models, see \textit{e.g.} Ref.~\cite{Wang_2023} for inspiration. Second, it would be important to go back to other models which have been studied in the literature using incorrect ideas of topology and to work out the proper bulk-boundary relations. For one specific example this has already been done in Ref.~\cite{MonkmanSirkerNH}. 
Finally, the scaling behavior of the entanglement entropy at the various critical lines observed in this work certainly deserves further investigations. The exceptional phase in the $\cal{PT}$-symmetric case is particularly intriguing, and studying this novel phase further might advance our understanding of critical behavior in non-Hermitian systems.

\section*{Acknowledgments}
R.A.P.~acknowledges support from the Rio Grande do Sul State Research Support Foundation (FAPERGS) and the National Council for Scientific and Technological Development (CNPq). J.S.~acknowledges support by NSERC via the Discovery grants program.
\appendix

\section{Discriminant of the characteristic polynomial}\label{app:EPpolynomial}

We recall that
a polynomial has repeated roots if and only if its discriminant 
\beq\label{disc}
\text{disc}(P_N(z))=\text{disc}
\left(\det_N \left(\mathcal{T}_N-z\right)\right)
\eeq
vanishes. The
discriminant is given by the determinant of the Sylvester matrix \cite{Gelfand1994}, and it is in general
an intricate expression of the parameters of the model. The vanishing of the discriminant \eqref{disc} is a necessary condition
for the existence of an exceptional point.
Of course, the discriminant vanishes
when the quasienergies collide. Additionally,
for
the model considered in this paper, the eigenvectors
always coalesce when the quasienergies collide,
recall the expressions (\ref{rvec},\ref{lvec},\ref{vecOBC}). Thus, the collision of quasienergies is also sufficient for the existence
of an exceptional point.

In the main text, the conditions for gap closing/exceptional points are determined
from the simple characterization of
the eigenvalues by (\ref{quasiCBC}) and (\ref{quasieOBC}).  Here these results are verified by explicitly computing the discriminant (\ref{disc}) using {\tt Mathematica} for different boundary conditions.

\subsection{Closed boundary conditions}

Let us consider $\gamma=1$ and $u=0$. For a lattice with $N=10$ sites the discriminant (\ref{disc})
is given by,
\bea\label{disc8PBC}
\disc(P_{10}^{\text{(PBC)}}(z))&=&10^{10}
\left(W_L^5+V_R^5\right) \left(V_L^5+W_R^5\right)
\non\\&\times&\left(V_L^5 W_L^5-V_R^5 W_R^5\right)^8.
\eea
The roots of the factors $\left(W_L^5+V_R^5\right) \left(V_L^5+W_R^5\right)
$ are clearly given by (\ref{CBCratios}). 
The extra factor comes from another type of
degeneracy in the spectrum which can
occur
within each band $\epsilon^\pm$ for different
quasimomentum $k$. Namely, one can impose
\beq\label{extradeg}
H_1(k)H_2(k)=H_1(k')H_2(k')
\eeq
which gives another constraint on the model
parameters. Note that these degeneracies
do not lead to exceptional points,
as the eigenvectors do not coalesce, recall
(\ref{rvec},\ref{lvec}). The extra factor in the discriminant is indeed given by (\ref{extradeg}) which vanishes for
\beq\label{extracon}
\frac{V_LW_L}{V_RW_R}=e^{\ii(k+k')}=
e^{\frac{2}{5} \ii \pi  (\ell+\ell')}
\eeq
with $\ell+\ell'=1,\dots,5$.

Now let us consider again $N=10$, $\gamma=1$, but with $u\neq 0$. In this case the discriminant is given
by,
\bea\label{disc8PBCunot0}
\disc(P_{10}^{\text{(PBC)}}(z))&=&10^{10}
\left(V_L^5 W_L^5-V_R^5 W_R^5\right)^8
\non\\ &\times& \left(\left(W_L+V_R\right) \left(V_L+W_R\right)-u^2\right) U(u)\non\\
\eea
where
\bea
U(u)&=&u^8+u^6 \left(V_LW_L+V_RW_R-4V_LV_R-4W_LW_R\right)
\non\\&+&u^4
   \big(6V_L^2V_R^2-3V_L^2V_RW_L+V_L^2W_L^2-3V_LV_R^2W_R
   \non\\&&+9V_LV_RW_LW_R-3V_LW_L^2W_R-
   3V_LW_L^2W_R
   \non\\&&
   +V_R^2W_R^2-3V_RW_LW_R^2
   +6W_L^2W_R^2\big)
   \non\\&+&
   u^2 \big(
   -4V_L^3V_R^3+3V_L^3V_R^2W_L-2V_L^3V_RW_L^2+V_L^3W_L^3\non\\
   &&+3V_L^2V_R^3W_R-6V_L^2V_R^2W_LW_R
   +4V_L^2V_RW_L^2W_R\non\\&&-2V_L^2W_L^3W_R
   -2V_LV_R^3W_R^2+4V_LV_R^2W_LW_R^2
   \non\\&&-6V_LV_RW_L^2W_R^2
   +3V_LW_L^3W_R^2+V_R^3W_R^3
   \non\\&&-2V_R^2W_LW_R^3
   +3V_RW_L^2W_R^3-4W_L^3W_R^3
   \big)
   \non\\&+&
   \left(-W_L V_R^3+W_L^2 V_R^2-W_L^3 V_R+W_L^4+V_R^4\right)\non\\ && \times \left(-V_L^3 W_R+V_L^2 W_R^2-V_L W_R^3+V_L^4+W_R^4\right).\non\\
\eea
The factor $\left(V_L^5 W_L^5-V_R^5 W_R^5\right)$ is also present
for the non-zero field case, and is solved by
(\ref{extracon}), however it does not lead
to exceptional points. To analyze the
factor
$\left(\left(W_L+V_R\right) \left(V_L+W_R\right)-u^2\right) U(u)$
we recall (\ref{gapclosingPBC}) for every
possible quantized $k$ (five for $N=10$). For example, for $k=2\pi$,
\beq
H_1(2\pi)H_2(2\pi)-u^2=\left(W_L+V_R\right) \left(V_L+W_R\right)-u^2.
\eeq
Actually, we can check by brute force the following
factorization,
\bea
&&\left(\left(W_L+V_R\right) \left(V_L+W_R\right)-u^2\right) U(u)=\non
\\
&&
\prod_{m=1}^5
H_1(k_m)H_2(k_m)-u^2
\eea
and confirm that the exceptional points give
the roots of the discriminant.

\subsection{Open boundary conditions}

Now let us consider $N=11$, $\gamma=0$ (OBC) and $u=0$. In this case the discriminant (\ref{disc})
is given by,
\beq\label{disc11OBC}
\disc(P_{11}^{\text{(OBC)}}(z))=
48922361856V_L^{20}V_R^{20}W_L^{20}W_R^{20}
f_1^3f_2^3f_3^3
\eeq
where
\bea\label{3f}
f_1&=&V_LV_R+W_LW_R\non\\
f_2&=&V_L^2V_R^2-V_LV_RW_LW_R+W_L^2W_R^2\non\\
f_3&=&V_L^2V_R^2+V_LV_RW_LW_R+W_L^2W_R^2.
\eea
Of course, the discriminant (\ref{disc11OBC}) vanishes when any of the couplings vanishes, consistent with our findings in the main text. The remaining factors can
be verified by rewriting
\beq\label{3f1}
f_1f_2f_3=V_L^5V_R^5\sum_{j=0}^5
\left(\frac{1}{\sqrt{\delta}}\right)^{2j}
\eeq
whose ten roots are given by (\ref{OBCratios2}).

Next, we turn on the field $u$. In this case, we have,
\bea
\label{disc11OBCu}
\disc(P_{11}^{\text{(OBC)}}(z))&=&
-48922361856 V_L^{20}V_R^{20}W_L^{20}W_R^{20}
\non\\
&\times&f_1^2f_2^2f_3^2g_1g_2g_3
\eea
where $f_j$ is given by (\ref{3f}) and
$g_j$ are given by
\bea
g_1&=&u^2-(V_LV_R+W_LW_R),\non\\
g_2&=&u^4-2u^2f_1+f_2,
\non\\
g_3&=&u^4-2u^2f_1+f_3.
\eea
Part of the roots remain the same as in the zero field
case, but field dependent factors $g_j$ appear.
These are associated with (\ref{gapclosingLodd1}).
Indeed, by brute force computation
we can verify
\beq
\prod_{m=1}^5u^2-\tilde{H}_1(\theta_m)\tilde{H}_2(\theta_m)=g_1g_2g_3
\eeq
where $\theta_m$ is given by (\ref{tquant}). Let us remark that the product $g_1g_2g_3$ only depends
on $\bar u$ and $\sqrt{\delta}$, up to an overall
factor.

Finally, let us consider $N=12$ with non-zero field
$u$. The discriminant is given by,
\beq\label{disc12OBCu}
\disc(P_{12}^{\text{(OBC)}}(z))=
4096V_L^{20}V_R^{20}W_L^{30}W_R^{30}U_0U_1(u)
\eeq
where 
\bea
U_0&=&153664V_L^5V_R^5+
98000V_L^4V_R^4W_LW_R
\non\\&+&
60152V_L^3V_R^3W_L^2W_R^2
+34709V_L^2V_R^2W_L^3W_R^3
\non\\&+&
17856V_LV_RW_L^4W_R^4+6912W_L^5W_R^5
\eea
and
\bea
U_1(u)&=&u^{12}-u^{10}(6V_LV_R+5W_LW_R)
\non\\&+&
5u^8(3V_L^2V_R^2+4V_LV_RW_LW_R+2W_L^2W_R^2)
\non\\&-&2u^6
(10V_L^3V_R^3+15V_L^2V_R^2W_LW_R+
12V_LV_RW_L^2W_R^2
\non\\&&+5W_L^3W_R^3)
\non\\
&+&u^4(15V_L^4V_R^4+20V_L^3V_R^3W_LW_R+18V_L^2V_R^2W_L^2W_R^2\non\\
&&+12V_LV_RW_L^3W_R^3+5W_L^4W_R^4)
\non\\&-&
u^2
(6V_L^5V_R^5+5V_L^4V_R^4W_LW_R+4V_L^3V_R^3W_L^2W_R^2
\non\\&&+3V_L^2V_R^2W_L^3W_R^3+2V_LV_RW_L^4W_R^4
+W_L^5W_R^5)\non\\
&+&V_L^6V_R^6.
\eea
Again, it is clear that the vanishing of
any of the coupling parameters leads
to a null discriminant (\ref{disc12OBCu}).
The factor $U_0$ can be written as
\bea
\label{U0}
&&U_0=153664V_L^5V_R^5
\Big(1+\frac{125}{196}\frac{1}{\sqrt{\delta}^{2}}+
\frac{7519}{19208}\frac{1}{\sqrt{\delta}^{4}}
\non\\&&+
\frac{34709}{153664}\frac{1}{\sqrt{\delta}^{6}}
+
\frac{279}{2401}\frac{1}{\sqrt{\delta}^{8}}
+
\frac{108}{2041}
\frac{1}{\sqrt{\delta}^{10}}\Big).
\eea
This factor is independent of the field $u$, and we can check numerically that the ten roots
are given by (\ref{HBcond},\ref{sqrtEPA}), see summary in Table \ref{tab1}. The field dependent factor $U_1(u)$ can also be written in terms
of $\sqrt{\delta}$ and $\bar u = u/(\sqrt{W_L}\sqrt{W_R})$, namely,
\bea
&&U_1(u)=V_L^6V_R^6
\Bigg(1-\frac{\bar u(3\bar u+1)}{\sqrt{\delta}^2}+
\frac{\bar u(\bar u+1)(3\bar u^2-\bar u-1)}{\sqrt{\delta}^4}\non\\&&-
\frac{\bar u(\bar u-1)^2(\bar u+1)^3}{\sqrt{\delta}^6}
\Bigg)
\Bigg(1-\frac{\bar u(3\bar u-1)}{\sqrt{\delta}^2}
\non\\&&+
\frac{\bar u(\bar u-1)(3\bar u^2+\bar u-1)}{\sqrt{\delta}^4}-
\frac{\bar u(\bar u+1)^2(\bar u-1)^3}{\sqrt{\delta}^6}
\Bigg).
\eea
We verify numerically that (\ref{trans2},\ref{sqrtEPB}) do provide
the roots of $U_1(u)=0$.
In Table \ref{tab2} we consider $\bar u=1/2$
and the associated solutions of (\ref{trans2}),
as well as the associated exceptional $\sqrt{\delta}$'s.

\begin{table}
\caption{The ten roots of Eq.~\eqref{U0}.\label{tab1}}
\begin{tabular}
{|c |c |} 
 \hline
 $\theta_{\ep}'$ & $\sqrt{\delta}$ \\ \hline
 $0.580505-0.209455\ii$ & $-0.628308+0.408561\ii$\\\hline
  $0.580505+0.209455\ii$ & $-0.628308-0.408561\ii$\\\hline
  $1.07762-0.24362\ii$ & $-0.344178+0.638086\ii$\\\hline
  $1.07762+0.24362\ii$ & $-0.344178-0.638086\ii$\\\hline
  $1.5708-0.252951\ii$ & $0.718376$
  \\\hline
  $1.5708+0.252951\ii$ & $-0.718376$
  \\\hline
  $2.06398-0.24362\ii$ & $0.344178+0.638086\ii$
    \\\hline
  $2.06398+0.24362\ii$ & $0.344178-0.638086\ii$
  \\\hline
  $2.56109-0.209455\ii$ & $0.628308+0.408561\ii$
   \\\hline
  $2.56109+0.209455\ii$ & $0.628308-0.408561\ii$\\\hline
\end{tabular}
\end{table}

\begin{table}
\caption{We consider $\bar u =0.5$ First six lines, $+$ solutions, last six lines $-$ solutions.\label{tab2}}
\begin{tabular}
{|c |c |} 
 \hline
 $\theta_{\ep}'$ & $\sqrt{\delta}$ \\ \hline
 $0.642557-0.0955613\ii$ & $-0.673088+0.294372\ii$\\\hline
$0.642557+0.0955613\ii$ & $-0.673088-0.294372\ii$\\\hline
$1.5708-0.191123\ii$ & $0.694828\ii$\\\hline
$1.5708+0.191123\ii$ & $-0.694828\ii$\\\hline
$2.49904-0.0955613\ii$ & $0.673088+0.294372\ii$\\\hline
$2.49904+0.0955613\ii$ & $0.673088-0.294372\ii$\\\hline\hline
$0.343297$ & $-1.31139$\\\hline
$1.10745-0.172812\ii$ &
$-0.36085+0.604218\ii$\\\hline
$1.10745+0.172812\ii$ &
$-0.36085-0.604218\ii$\\\hline
$2.03414-0.172812\ii$ &
$0.36085+0.604218\ii$\\\hline
$2.03414+0.172812\ii$ &
$0.36085-0.604218\ii$\\\hline
$2.7983$ & $1.31139$\\\hline
\end{tabular}
\end{table}

\section{Some details for the open boundary case}\label{app:OBC}

To make this paper self-contained,
we review some details of the diagonalization  
of $\mathcal{T}_N$ (\ref{tri}) for
open boundary conditions, that is, $\gamma=0$, following \cite{Gover94,shin_1997}.

The approach in \cite{Gover94} is based on recurrence relations satisfied
by the characteristic polynomial,
\beq\label{polch}
P_N(z)=\det_N \left(\mathcal{T}_N-z\right).
\eeq
The roots 
$P_N(z)=0$
are the quasienergies of the Hamiltonian (\ref{ham}). The characteristic polynomial (\ref{polch})
for $\gamma=0$ is special as it satisfies the recurrence relations
\bea\label{pol}
P_{2j}(z)&=&(-\ii u-z)P_{2j-1}(z)-V_LV_R P_{2j-2}(z),\non\\
P_{2j-1}(z)&=&(\ii u-z)P_{2j-2}(z)-W_LW_R P_{2j-3}(z)
\eea
depending on the parity of $N$.
Note that one has an additional recurrence relation for the
characteristic polynomial, namely
\bea\label{recP2}
P_j(z)&=&\left((z+\ii u)(z-\ii u)-(V_LV_R+W_LW_R)\right)P_{j-2}(z)
\non\\&-&V_LV_RW_LW_RP_{j-4}(z),
\eea
where appropriate initial conditions must be chosen. 

On the other hand, following \cite{shin_1997}, one notes that equations (\ref{difeq1},\ref{difeq2})
can be iterated to the same fourth order difference equation,
\bea\label{difeq3}
&&r_{j+2}+\frac{V_RW_R}{V_LW_L}r_{j-2}
\non\\&&=
\left(-\frac{(\ii u+\epsilon)(\ii u-\epsilon)+(V_LV_R+W_LW_R)}{V_LW_L}\right) r_j.
\eea
The difference equation (\ref{difeq3}) 
can be compared to (\ref{recP2}) leading to the solution (\ref{vecOBC}).

In addition, following \cite{shin_1997}, introduce
\beq
T_n(\alpha,\beta)= \frac{\alpha^{n+1}-\beta^{n+1}}{\alpha -\beta}
\eeq
and the sequence
\beq\label{sequence}
a_{n+2}-(\alpha+\beta)a_{n+1}+\alpha\beta a_{n}=0.
\eeq
Then, it follows that
\beq
a_n=T_{n-2}(\alpha,\beta)a_2-\alpha\beta T_{n-3}(\alpha,\beta)a_1
\eeq
is the solution of the difference equation (\ref{sequence}). Now choosing
\bea\label{cond}
&&\alpha+\beta = \frac{(\ii u-\epsilon)(-\ii u-\epsilon)-(V_LV_R+W_LW_R)}{V_LW_L},\\
&&\alpha\beta= \frac{V_RW_R}{V_LW_L},
\eea
it follows that
\bea
&&r_{2j}=T_{j-2}(\alpha,\beta)r_4-\alpha\beta T_{j-3}(\alpha,\beta)r_2,\label{vTeven}\\
&&r_{2j-1}=T_{j-2}(\alpha,\beta)r_3-\alpha\beta T_{j-3}(\alpha,\beta)r_1,\label{vTodd}
\eea
is a solution of (\ref{difeq1},\ref{difeq2},\ref{difeq3}).
The boundary conditions determine,
\bea\label{initialcond}
r_1&=&V_L, \quad r_2=\epsilon-\ii u,
\quad
r_3=W_R+V_L(\alpha+\beta),\non\\
r_4&=&(\epsilon-\ii u )\left(2\frac{W_R}{V_L}+\alpha+\beta\right).
\eea
where $r_1=V_L$ is an arbitrary choice.
The conditions
(\ref{cond}) give $\{\epsilon,\beta\}$ in terms of $\alpha$,
\bea\label{quasie}
\epsilon&=&\pm
\sqrt{V_LV_R+W_LW_R+V_LW_L(\alpha+\beta)-u^2}
   ,
   \non\\
   \beta&=&\frac{V_RW_R}{V_LW_L}\frac{1}{\alpha}.
\eea
Using the parametrization
\beq
\alpha=\frac{\sqrt{V_R}\sqrt{W_R}}{
\sqrt{V_L}\sqrt{W_L}}\exp(\ii \theta),
\eeq
and then
\beq
T_n\left(\alpha,\frac{V_RW_R}{V_LW_L}\frac{1}{\alpha}\right)=\frac{V_R^\frac{n}{2}W_R^\frac{n}{2}}{V_L^\frac{n}{2}W_L^\frac{n}{2}}\frac{\sin((n+1)\theta)}{\sin(\theta)}
\eeq
in (\ref{vTeven},\ref{vTodd}) we find
(\ref{vLeven},\ref{vLodd})
with the initial conditions (\ref{initialcond}).

The boundary condition at the end of the chain  gives a
condition on the parameter $\theta$, which depends on whether $N$ is odd or even.
For odd $N$,
the condition $x_{L+1}=0$ comes from (\ref{vLeven}) with $2n=L+1$ and gives the quantized solution (\ref{tquant}).
For even $N$,
the condition $x_{L+1}=0$ comes from (\ref{vLodd}) with $2n-1=L+1$ and leads to the
transcendental equation (\ref{trans}). The solution (\ref{vLevenL},\ref{vLoddL})
is obtained analogously.




\begin{thebibliography}{67}%
\makeatletter
\providecommand \@ifxundefined [1]{%
 \@ifx{#1\undefined}
}%
\providecommand \@ifnum [1]{%
 \ifnum #1\expandafter \@firstoftwo
 \else \expandafter \@secondoftwo
 \fi
}%
\providecommand \@ifx [1]{%
 \ifx #1\expandafter \@firstoftwo
 \else \expandafter \@secondoftwo
 \fi
}%
\providecommand \natexlab [1]{#1}%
\providecommand \enquote  [1]{``#1''}%
\providecommand \bibnamefont  [1]{#1}%
\providecommand \bibfnamefont [1]{#1}%
\providecommand \citenamefont [1]{#1}%
\providecommand \href@noop [0]{\@secondoftwo}%
\providecommand \href [0]{\begingroup \@sanitize@url \@href}%
\providecommand \@href[1]{\@@startlink{#1}\@@href}%
\providecommand \@@href[1]{\endgroup#1\@@endlink}%
\providecommand \@sanitize@url [0]{\catcode `\\12\catcode `\$12\catcode `\&12\catcode `\#12\catcode `\^12\catcode `\_12\catcode `\%12\relax}%
\providecommand \@@startlink[1]{}%
\providecommand \@@endlink[0]{}%
\providecommand \url  [0]{\begingroup\@sanitize@url \@url }%
\providecommand \@url [1]{\endgroup\@href {#1}{\urlprefix }}%
\providecommand \urlprefix  [0]{URL }%
\providecommand \Eprint [0]{\href }%
\providecommand \doibase [0]{https://doi.org/}%
\providecommand \selectlanguage [0]{\@gobble}%
\providecommand \bibinfo  [0]{\@secondoftwo}%
\providecommand \bibfield  [0]{\@secondoftwo}%
\providecommand \translation [1]{[#1]}%
\providecommand \BibitemOpen [0]{}%
\providecommand \bibitemStop [0]{}%
\providecommand \bibitemNoStop [0]{.\EOS\space}%
\providecommand \EOS [0]{\spacefactor3000\relax}%
\providecommand \BibitemShut  [1]{\csname bibitem#1\endcsname}%
\let\auto@bib@innerbib\@empty
\bibitem [{\citenamefont {Lindblad}(1976)}]{L1976}%
  \BibitemOpen
  \bibfield  {author} {\bibinfo {author} {\bibfnamefont {G.}~\bibnamefont {Lindblad}},\ }\bibfield  {title} {\bibinfo {title} {On the generators of quantum dynamical semigroups},\ }\href {https://doi.org/https://doi.org/10.1007/BF01608499} {\bibfield  {journal} {\bibinfo  {journal} {Communications in Mathematical Physics}\ }\textbf {\bibinfo {volume} {48}},\ \bibinfo {pages} {119} (\bibinfo {year} {1976})}\BibitemShut {NoStop}%
\bibitem [{\citenamefont {Roccati}\ \emph {et~al.}(2022)\citenamefont {Roccati}, \citenamefont {Palma}, \citenamefont {Bagarello},\ and\ \citenamefont {Ciccarello}}]{RPBC2022}%
  \BibitemOpen
  \bibfield  {author} {\bibinfo {author} {\bibfnamefont {F.}~\bibnamefont {Roccati}}, \bibinfo {author} {\bibfnamefont {G.~M.}\ \bibnamefont {Palma}}, \bibinfo {author} {\bibfnamefont {F.}~\bibnamefont {Bagarello}},\ and\ \bibinfo {author} {\bibfnamefont {F.}~\bibnamefont {Ciccarello}},\ }\bibfield  {title} {\bibinfo {title} {{Non-Hermitian Physics and Master Equations}},\ }\href {https://doi.org/10.1142/S1230161222500044} {\bibfield  {journal} {\bibinfo  {journal} {Open Systems \& Information Dynamics}\ }\textbf {\bibinfo {volume} {29}},\ \bibinfo {pages} {2250004} (\bibinfo {year} {2022})},\ \Eprint {https://arxiv.org/abs/2201.05367} {arXiv:2201.05367 [quant-ph]} \BibitemShut {NoStop}%
\bibitem [{\citenamefont {Miri}\ and\ \citenamefont {Alù}(2019)}]{MiriAlu}%
  \BibitemOpen
  \bibfield  {author} {\bibinfo {author} {\bibfnamefont {M.-A.}\ \bibnamefont {Miri}}\ and\ \bibinfo {author} {\bibfnamefont {A.}~\bibnamefont {Alù}},\ }\bibfield  {title} {\bibinfo {title} {Exceptional points in optics and photonics},\ }\href {https://doi.org/10.1126/science.aar7709} {\bibfield  {journal} {\bibinfo  {journal} {Science}\ }\textbf {\bibinfo {volume} {363}},\ \bibinfo {pages} {eaar7709} (\bibinfo {year} {2019})}\BibitemShut {NoStop}%
\bibitem [{\citenamefont {Su}\ \emph {et~al.}(2021)\citenamefont {Su}, \citenamefont {Estrecho}, \citenamefont {Biegańska}, \citenamefont {Huang}, \citenamefont {Wurdack}, \citenamefont {Pieczarka}, \citenamefont {Truscott}, \citenamefont {Liew}, \citenamefont {Ostrovskaya},\ and\ \citenamefont {Xiong}}]{SuEstrecho}%
  \BibitemOpen
  \bibfield  {author} {\bibinfo {author} {\bibfnamefont {R.}~\bibnamefont {Su}}, \bibinfo {author} {\bibfnamefont {E.}~\bibnamefont {Estrecho}}, \bibinfo {author} {\bibfnamefont {D.}~\bibnamefont {Biegańska}}, \bibinfo {author} {\bibfnamefont {Y.}~\bibnamefont {Huang}}, \bibinfo {author} {\bibfnamefont {M.}~\bibnamefont {Wurdack}}, \bibinfo {author} {\bibfnamefont {M.}~\bibnamefont {Pieczarka}}, \bibinfo {author} {\bibfnamefont {A.~G.}\ \bibnamefont {Truscott}}, \bibinfo {author} {\bibfnamefont {T.~C.~H.}\ \bibnamefont {Liew}}, \bibinfo {author} {\bibfnamefont {E.~A.}\ \bibnamefont {Ostrovskaya}},\ and\ \bibinfo {author} {\bibfnamefont {Q.}~\bibnamefont {Xiong}},\ }\bibfield  {title} {\bibinfo {title} {Direct measurement of a non-hermitian topological invariant in a hybrid light-matter system},\ }\href {https://doi.org/10.1126/sciadv.abj8905} {\bibfield  {journal} {\bibinfo  {journal} {Science Advances}\ }\textbf {\bibinfo {volume} {7}},\ \bibinfo {pages} {eabj8905} (\bibinfo {year} {2021})},\ \Eprint
  {https://arxiv.org/abs/https://www.science.org/doi/pdf/10.1126/sciadv.abj8905} {https://www.science.org/doi/pdf/10.1126/sciadv.abj8905} \BibitemShut {NoStop}%
\bibitem [{\citenamefont {Yang}\ \emph {et~al.}(2020)\citenamefont {Yang}, \citenamefont {Wang}, \citenamefont {Rao}, \citenamefont {Gui}, \citenamefont {Yao}, \citenamefont {Lu},\ and\ \citenamefont {Hu}}]{YangWang}%
  \BibitemOpen
  \bibfield  {author} {\bibinfo {author} {\bibfnamefont {Y.}~\bibnamefont {Yang}}, \bibinfo {author} {\bibfnamefont {Y.-P.}\ \bibnamefont {Wang}}, \bibinfo {author} {\bibfnamefont {J.~W.}\ \bibnamefont {Rao}}, \bibinfo {author} {\bibfnamefont {Y.~S.}\ \bibnamefont {Gui}}, \bibinfo {author} {\bibfnamefont {B.~M.}\ \bibnamefont {Yao}}, \bibinfo {author} {\bibfnamefont {W.}~\bibnamefont {Lu}},\ and\ \bibinfo {author} {\bibfnamefont {C.-M.}\ \bibnamefont {Hu}},\ }\bibfield  {title} {\bibinfo {title} {Unconventional singularity in anti-parity-time symmetric cavity magnonics},\ }\href {https://doi.org/10.1103/PhysRevLett.125.147202} {\bibfield  {journal} {\bibinfo  {journal} {Phys. Rev. Lett.}\ }\textbf {\bibinfo {volume} {125}},\ \bibinfo {pages} {147202} (\bibinfo {year} {2020})}\BibitemShut {NoStop}%
\bibitem [{\citenamefont {Naghiloo}\ \emph {et~al.}(2019)\citenamefont {Naghiloo}, \citenamefont {M.}, \citenamefont {Joglekar},\ and\ \citenamefont {Murch}}]{NaghilooAbbasi}%
  \BibitemOpen
  \bibfield  {author} {\bibinfo {author} {\bibfnamefont {M.}~\bibnamefont {Naghiloo}}, \bibinfo {author} {\bibfnamefont {A.}~\bibnamefont {M.}}, \bibinfo {author} {\bibfnamefont {Y.}~\bibnamefont {Joglekar}},\ and\ \bibinfo {author} {\bibfnamefont {K.}~\bibnamefont {Murch}},\ }\bibfield  {title} {\bibinfo {title} {Quantum state tomography across the exceptional point in a single dissipative qubit},\ }\href {https://doi.org/https://doi-org.uml.idm.oclc.org/10.1038/s41567-019-0652-z} {\bibfield  {journal} {\bibinfo  {journal} {Nat. Phys.}\ }\textbf {\bibinfo {volume} {15}},\ \bibinfo {pages} {1232} (\bibinfo {year} {2019})}\BibitemShut {NoStop}%
\bibitem [{\citenamefont {Bender}(2005)}]{Bender_2005}%
  \BibitemOpen
  \bibfield  {author} {\bibinfo {author} {\bibfnamefont {C.~M.}\ \bibnamefont {Bender}},\ }\bibfield  {title} {\bibinfo {title} {Introduction to pt-symmetric quantum theory},\ }\href {https://doi.org/10.1080/00107500072632} {\bibfield  {journal} {\bibinfo  {journal} {Contemporary Physics}\ }\textbf {\bibinfo {volume} {46}},\ \bibinfo {pages} {277–292} (\bibinfo {year} {2005})}\BibitemShut {NoStop}%
\bibitem [{\citenamefont {Ashida}\ \emph {et~al.}(2021)\citenamefont {Ashida}, \citenamefont {Gong},\ and\ \citenamefont {Ueda}}]{AGU2021}%
  \BibitemOpen
  \bibfield  {author} {\bibinfo {author} {\bibfnamefont {Y.}~\bibnamefont {Ashida}}, \bibinfo {author} {\bibfnamefont {Z.}~\bibnamefont {Gong}},\ and\ \bibinfo {author} {\bibfnamefont {M.}~\bibnamefont {Ueda}},\ }\bibfield  {title} {\bibinfo {title} {{Non-Hermitian physics}},\ }\href {https://doi.org/10.1080/00018732.2021.1876991} {\bibfield  {journal} {\bibinfo  {journal} {Adv. Phys.}\ }\textbf {\bibinfo {volume} {69}},\ \bibinfo {pages} {249} (\bibinfo {year} {2021})},\ \Eprint {https://arxiv.org/abs/2006.01837} {arXiv:2006.01837 [cond-mat.mes-hall]} \BibitemShut {NoStop}%
\bibitem [{\citenamefont {Bernard}\ and\ \citenamefont {LeClair}(2002)}]{BL2002}%
  \BibitemOpen
  \bibfield  {author} {\bibinfo {author} {\bibfnamefont {D.}~\bibnamefont {Bernard}}\ and\ \bibinfo {author} {\bibfnamefont {A.}~\bibnamefont {LeClair}},\ }\bibinfo {title} {A classification of non-hermitian random matrices},\ in\ \href {https://doi.org/10.1007/978-94-010-0514-2_19} {\emph {\bibinfo {booktitle} {Statistical Field Theories}}}\ (\bibinfo  {publisher} {Springer Netherlands},\ \bibinfo {year} {2002})\ p.\ \bibinfo {pages} {207–214}\BibitemShut {NoStop}%
\bibitem [{\citenamefont {{Esaki}}\ \emph {et~al.}(2011)\citenamefont {{Esaki}}, \citenamefont {{Sato}}, \citenamefont {{Hasebe}},\ and\ \citenamefont {{Kohmoto}}}]{ESHK2011}%
  \BibitemOpen
  \bibfield  {author} {\bibinfo {author} {\bibfnamefont {K.}~\bibnamefont {{Esaki}}}, \bibinfo {author} {\bibfnamefont {M.}~\bibnamefont {{Sato}}}, \bibinfo {author} {\bibfnamefont {K.}~\bibnamefont {{Hasebe}}},\ and\ \bibinfo {author} {\bibfnamefont {M.}~\bibnamefont {{Kohmoto}}},\ }\bibfield  {title} {\bibinfo {title} {{Edge states and topological phases in non-Hermitian systems}},\ }\href {https://doi.org/10.1103/PhysRevB.84.205128} {\bibfield  {journal} {\bibinfo  {journal} {\prb}\ }\textbf {\bibinfo {volume} {84}},\ \bibinfo {eid} {205128} (\bibinfo {year} {2011})},\ \Eprint {https://arxiv.org/abs/1107.2079} {arXiv:1107.2079 [cond-mat.mes-hall]} \BibitemShut {NoStop}%
\bibitem [{\citenamefont {{Lee}}(2016)}]{L2016}%
  \BibitemOpen
  \bibfield  {author} {\bibinfo {author} {\bibfnamefont {T.~E.}\ \bibnamefont {{Lee}}},\ }\bibfield  {title} {\bibinfo {title} {{Anomalous Edge State in a Non-Hermitian Lattice}},\ }\href {https://doi.org/10.1103/PhysRevLett.116.133903} {\bibfield  {journal} {\bibinfo  {journal} {\prl}\ }\textbf {\bibinfo {volume} {116}},\ \bibinfo {eid} {133903} (\bibinfo {year} {2016})},\ \Eprint {https://arxiv.org/abs/1603.05312} {arXiv:1603.05312 [quant-ph]} \BibitemShut {NoStop}%
\bibitem [{\citenamefont {{Lieu}}(2018)}]{L2018}%
  \BibitemOpen
  \bibfield  {author} {\bibinfo {author} {\bibfnamefont {S.}~\bibnamefont {{Lieu}}},\ }\bibfield  {title} {\bibinfo {title} {{Topological phases in the non-Hermitian Su-Schrieffer-Heeger model}},\ }\href {https://doi.org/10.1103/PhysRevB.97.045106} {\bibfield  {journal} {\bibinfo  {journal} {\prb}\ }\textbf {\bibinfo {volume} {97}},\ \bibinfo {eid} {045106} (\bibinfo {year} {2018})},\ \Eprint {https://arxiv.org/abs/1709.03788} {arXiv:1709.03788 [cond-mat.mes-hall]} \BibitemShut {NoStop}%
\bibitem [{\citenamefont {Martinez~Alvarez}\ \emph {et~al.}(2018)\citenamefont {Martinez~Alvarez}, \citenamefont {Barrios~Vargas},\ and\ \citenamefont {Foa~Torres}}]{MABVFT18}%
  \BibitemOpen
  \bibfield  {author} {\bibinfo {author} {\bibfnamefont {V.~M.}\ \bibnamefont {Martinez~Alvarez}}, \bibinfo {author} {\bibfnamefont {J.~E.}\ \bibnamefont {Barrios~Vargas}},\ and\ \bibinfo {author} {\bibfnamefont {L.~E.~F.}\ \bibnamefont {Foa~Torres}},\ }\bibfield  {title} {\bibinfo {title} {Non-hermitian robust edge states in one dimension: Anomalous localization and eigenspace condensation at exceptional points},\ }\bibfield  {journal} {\bibinfo  {journal} {Physical Review B}\ }\textbf {\bibinfo {volume} {97}},\ \href {https://doi.org/10.1103/physrevb.97.121401} {10.1103/physrevb.97.121401} (\bibinfo {year} {2018})\BibitemShut {NoStop}%
\bibitem [{\citenamefont {{Yao}}\ and\ \citenamefont {{Wang}}(2018)}]{YZ2018}%
  \BibitemOpen
  \bibfield  {author} {\bibinfo {author} {\bibfnamefont {S.}~\bibnamefont {{Yao}}}\ and\ \bibinfo {author} {\bibfnamefont {Z.}~\bibnamefont {{Wang}}},\ }\bibfield  {title} {\bibinfo {title} {{Edge States and Topological Invariants of Non-Hermitian Systems}},\ }\href {https://doi.org/10.1103/PhysRevLett.121.086803} {\bibfield  {journal} {\bibinfo  {journal} {\prl}\ }\textbf {\bibinfo {volume} {121}},\ \bibinfo {eid} {086803} (\bibinfo {year} {2018})},\ \Eprint {https://arxiv.org/abs/1803.01876} {arXiv:1803.01876 [cond-mat.mes-hall]} \BibitemShut {NoStop}%
\bibitem [{\citenamefont {{Yin}}\ \emph {et~al.}(2018)\citenamefont {{Yin}}, \citenamefont {{Jiang}}, \citenamefont {{Li}}, \citenamefont {{L{\"u}}},\ and\ \citenamefont {{Chen}}}]{YJLLC2018}%
  \BibitemOpen
  \bibfield  {author} {\bibinfo {author} {\bibfnamefont {C.}~\bibnamefont {{Yin}}}, \bibinfo {author} {\bibfnamefont {H.}~\bibnamefont {{Jiang}}}, \bibinfo {author} {\bibfnamefont {L.}~\bibnamefont {{Li}}}, \bibinfo {author} {\bibfnamefont {R.}~\bibnamefont {{L{\"u}}}},\ and\ \bibinfo {author} {\bibfnamefont {S.}~\bibnamefont {{Chen}}},\ }\bibfield  {title} {\bibinfo {title} {{Geometrical meaning of winding number and its characterization of topological phases in one-dimensional chiral non-Hermitian systems}},\ }\href {https://doi.org/10.1103/PhysRevA.97.052115} {\bibfield  {journal} {\bibinfo  {journal} {\pra}\ }\textbf {\bibinfo {volume} {97}},\ \bibinfo {eid} {052115} (\bibinfo {year} {2018})},\ \Eprint {https://arxiv.org/abs/1802.04169} {arXiv:1802.04169 [cond-mat.mes-hall]} \BibitemShut {NoStop}%
\bibitem [{\citenamefont {{Kawabata}}\ \emph {et~al.}(2019)\citenamefont {{Kawabata}}, \citenamefont {{Shiozaki}}, \citenamefont {{Ueda}},\ and\ \citenamefont {{Sato}}}]{KSUS2019}%
  \BibitemOpen
  \bibfield  {author} {\bibinfo {author} {\bibfnamefont {K.}~\bibnamefont {{Kawabata}}}, \bibinfo {author} {\bibfnamefont {K.}~\bibnamefont {{Shiozaki}}}, \bibinfo {author} {\bibfnamefont {M.}~\bibnamefont {{Ueda}}},\ and\ \bibinfo {author} {\bibfnamefont {M.}~\bibnamefont {{Sato}}},\ }\bibfield  {title} {\bibinfo {title} {{Symmetry and Topology in Non-Hermitian Physics}},\ }\href {https://doi.org/10.1103/PhysRevX.9.041015} {\bibfield  {journal} {\bibinfo  {journal} {Physical Review X}\ }\textbf {\bibinfo {volume} {9}},\ \bibinfo {eid} {041015} (\bibinfo {year} {2019})},\ \Eprint {https://arxiv.org/abs/1812.09133} {arXiv:1812.09133 [cond-mat.mes-hall]} \BibitemShut {NoStop}%
\bibitem [{\citenamefont {Bergholtz}\ \emph {et~al.}(2021)\citenamefont {Bergholtz}, \citenamefont {Budich},\ and\ \citenamefont {Kunst}}]{BBK2021}%
  \BibitemOpen
  \bibfield  {author} {\bibinfo {author} {\bibfnamefont {E.~J.}\ \bibnamefont {Bergholtz}}, \bibinfo {author} {\bibfnamefont {J.~C.}\ \bibnamefont {Budich}},\ and\ \bibinfo {author} {\bibfnamefont {F.~K.}\ \bibnamefont {Kunst}},\ }\bibfield  {title} {\bibinfo {title} {{Exceptional topology of non-Hermitian systems}},\ }\href {https://doi.org/10.1103/revmodphys.93.015005} {\bibfield  {journal} {\bibinfo  {journal} {Rev. Mod. Phys.}\ }\textbf {\bibinfo {volume} {93}},\ \bibinfo {pages} {015005} (\bibinfo {year} {2021})},\ \Eprint {https://arxiv.org/abs/1912.10048} {arXiv:1912.10048 [cond-mat.mes-hall]} \BibitemShut {NoStop}%
\bibitem [{\citenamefont {Ding}\ \emph {et~al.}(2022)\citenamefont {Ding}, \citenamefont {Fang},\ and\ \citenamefont {Ma}}]{DFM2022}%
  \BibitemOpen
  \bibfield  {author} {\bibinfo {author} {\bibfnamefont {K.}~\bibnamefont {Ding}}, \bibinfo {author} {\bibfnamefont {C.}~\bibnamefont {Fang}},\ and\ \bibinfo {author} {\bibfnamefont {G.}~\bibnamefont {Ma}},\ }\bibfield  {title} {\bibinfo {title} {{Non-Hermitian topology and exceptional-point geometries}},\ }\href {https://doi.org/10.1038/s42254-022-00516-5} {\bibfield  {journal} {\bibinfo  {journal} {Nature Rev. Phys.}\ }\textbf {\bibinfo {volume} {4}},\ \bibinfo {pages} {745} (\bibinfo {year} {2022})},\ \Eprint {https://arxiv.org/abs/2204.11601} {arXiv:2204.11601 [quant-ph]} \BibitemShut {NoStop}%
\bibitem [{\citenamefont {Altland}\ and\ \citenamefont {Zirnbauer}(1997)}]{AZ1997}%
  \BibitemOpen
  \bibfield  {author} {\bibinfo {author} {\bibfnamefont {A.}~\bibnamefont {Altland}}\ and\ \bibinfo {author} {\bibfnamefont {M.~R.}\ \bibnamefont {Zirnbauer}},\ }\bibfield  {title} {\bibinfo {title} {Nonstandard symmetry classes in mesoscopic normal-superconducting hybrid structures},\ }\href {https://doi.org/10.1103/physrevb.55.1142} {\bibfield  {journal} {\bibinfo  {journal} {Physical Review B}\ }\textbf {\bibinfo {volume} {55}},\ \bibinfo {pages} {1142–1161} (\bibinfo {year} {1997})}\BibitemShut {NoStop}%
\bibitem [{\citenamefont {Gong}\ \emph {et~al.}(2018)\citenamefont {Gong}, \citenamefont {Ashida}, \citenamefont {Kawabata}, \citenamefont {Takasan}, \citenamefont {Higashikawa},\ and\ \citenamefont {Ueda}}]{GAKTHU2018}%
  \BibitemOpen
  \bibfield  {author} {\bibinfo {author} {\bibfnamefont {Z.}~\bibnamefont {Gong}}, \bibinfo {author} {\bibfnamefont {Y.}~\bibnamefont {Ashida}}, \bibinfo {author} {\bibfnamefont {K.}~\bibnamefont {Kawabata}}, \bibinfo {author} {\bibfnamefont {K.}~\bibnamefont {Takasan}}, \bibinfo {author} {\bibfnamefont {S.}~\bibnamefont {Higashikawa}},\ and\ \bibinfo {author} {\bibfnamefont {M.}~\bibnamefont {Ueda}},\ }\bibfield  {title} {\bibinfo {title} {Topological phases of non-hermitian systems},\ }\href {https://doi.org/10.1103/PhysRevX.8.031079} {\bibfield  {journal} {\bibinfo  {journal} {Phys. Rev. X}\ }\textbf {\bibinfo {volume} {8}},\ \bibinfo {pages} {031079} (\bibinfo {year} {2018})},\ \Eprint {https://arxiv.org/abs/1802.07964} {arXiv:1802.07964 [cond-mat.mes-hall]} \BibitemShut {NoStop}%
\bibitem [{\citenamefont {Kunst}\ \emph {et~al.}(2018)\citenamefont {Kunst}, \citenamefont {Edvardsson}, \citenamefont {Budich},\ and\ \citenamefont {Bergholtz}}]{Kunst}%
  \BibitemOpen
  \bibfield  {author} {\bibinfo {author} {\bibfnamefont {F.~K.}\ \bibnamefont {Kunst}}, \bibinfo {author} {\bibfnamefont {E.}~\bibnamefont {Edvardsson}}, \bibinfo {author} {\bibfnamefont {J.~C.}\ \bibnamefont {Budich}},\ and\ \bibinfo {author} {\bibfnamefont {E.~J.}\ \bibnamefont {Bergholtz}},\ }\bibfield  {title} {\bibinfo {title} {Biorthogonal bulk-boundary correspondence in non-hermitian systems},\ }\href@noop {} {\bibfield  {journal} {\bibinfo  {journal} {Physical review letters}\ }\textbf {\bibinfo {volume} {121}},\ \bibinfo {pages} {026808} (\bibinfo {year} {2018})}\BibitemShut {NoStop}%
\bibitem [{\citenamefont {Monkman}\ and\ \citenamefont {Sirker}(2025)}]{MonkmanSirkerNH}%
  \BibitemOpen
  \bibfield  {author} {\bibinfo {author} {\bibfnamefont {K.}~\bibnamefont {Monkman}}\ and\ \bibinfo {author} {\bibfnamefont {J.}~\bibnamefont {Sirker}},\ }\bibfield  {title} {\bibinfo {title} {Hidden zero modes and topology of multiband non-hermitian systems},\ }\href {https://doi.org/10.1103/PhysRevLett.134.056601} {\bibfield  {journal} {\bibinfo  {journal} {Phys. Rev. Lett.}\ }\textbf {\bibinfo {volume} {134}},\ \bibinfo {pages} {056601} (\bibinfo {year} {2025})}\BibitemShut {NoStop}%
\bibitem [{\citenamefont {Heiss}(2012)}]{H2012}%
  \BibitemOpen
  \bibfield  {author} {\bibinfo {author} {\bibfnamefont {W.~D.}\ \bibnamefont {Heiss}},\ }\bibfield  {title} {\bibinfo {title} {{The physics of exceptional points}},\ }\href {https://doi.org/10.1088/1751-8113/45/44/444016} {\bibfield  {journal} {\bibinfo  {journal} {J. Phys. A}\ }\textbf {\bibinfo {volume} {45}},\ \bibinfo {pages} {444016} (\bibinfo {year} {2012})},\ \Eprint {https://arxiv.org/abs/1210.7536} {arXiv:1210.7536 [quant-ph]} \BibitemShut {NoStop}%
\bibitem [{\citenamefont {Su}\ \emph {et~al.}(1979)\citenamefont {Su}, \citenamefont {Schrieffer},\ and\ \citenamefont {Heeger}}]{SSH1979}%
  \BibitemOpen
  \bibfield  {author} {\bibinfo {author} {\bibfnamefont {W.~P.}\ \bibnamefont {Su}}, \bibinfo {author} {\bibfnamefont {J.~R.}\ \bibnamefont {Schrieffer}},\ and\ \bibinfo {author} {\bibfnamefont {A.~J.}\ \bibnamefont {Heeger}},\ }\bibfield  {title} {\bibinfo {title} {Solitons in polyacetylene},\ }\href {https://doi.org/10.1103/PhysRevLett.42.1698} {\bibfield  {journal} {\bibinfo  {journal} {Phys. Rev. Lett.}\ }\textbf {\bibinfo {volume} {42}},\ \bibinfo {pages} {1698} (\bibinfo {year} {1979})}\BibitemShut {NoStop}%
\bibitem [{\citenamefont {Ling}\ \emph {et~al.}(2015)\citenamefont {Ling}, \citenamefont {Xiao}, \citenamefont {Chan}, \citenamefont {Yu},\ and\ \citenamefont {Fung}}]{Ling}%
  \BibitemOpen
  \bibfield  {author} {\bibinfo {author} {\bibfnamefont {C.}~\bibnamefont {Ling}}, \bibinfo {author} {\bibfnamefont {M.}~\bibnamefont {Xiao}}, \bibinfo {author} {\bibfnamefont {C.~T.}\ \bibnamefont {Chan}}, \bibinfo {author} {\bibfnamefont {S.~F.}\ \bibnamefont {Yu}},\ and\ \bibinfo {author} {\bibfnamefont {K.~H.}\ \bibnamefont {Fung}},\ }\bibfield  {title} {\bibinfo {title} {Topological edge plasmon modes between diatomic chains of plasmonic nanoparticles},\ }\href@noop {} {\bibfield  {journal} {\bibinfo  {journal} {Optics express}\ }\textbf {\bibinfo {volume} {23}},\ \bibinfo {pages} {2021} (\bibinfo {year} {2015})}\BibitemShut {NoStop}%
\bibitem [{\citenamefont {{Longhi}}\ \emph {et~al.}(2015)\citenamefont {{Longhi}}, \citenamefont {{Gatti}},\ and\ \citenamefont {{Valle}}}]{Longhi}%
  \BibitemOpen
  \bibfield  {author} {\bibinfo {author} {\bibfnamefont {S.}~\bibnamefont {{Longhi}}}, \bibinfo {author} {\bibfnamefont {D.}~\bibnamefont {{Gatti}}},\ and\ \bibinfo {author} {\bibfnamefont {G.~D.}\ \bibnamefont {{Valle}}},\ }\bibfield  {title} {\bibinfo {title} {{Robust light transport in non-Hermitian photonic lattices}},\ }\href {https://doi.org/10.1038/srep13376} {\bibfield  {journal} {\bibinfo  {journal} {Scientific Reports}\ }\textbf {\bibinfo {volume} {5}},\ \bibinfo {eid} {13376} (\bibinfo {year} {2015})},\ \Eprint {https://arxiv.org/abs/1503.08787} {arXiv:1503.08787 [physics.optics]} \BibitemShut {NoStop}%
\bibitem [{\citenamefont {Hatano}\ and\ \citenamefont {Nelson}(1996)}]{HN1996}%
  \BibitemOpen
  \bibfield  {author} {\bibinfo {author} {\bibfnamefont {N.}~\bibnamefont {Hatano}}\ and\ \bibinfo {author} {\bibfnamefont {D.~R.}\ \bibnamefont {Nelson}},\ }\bibfield  {title} {\bibinfo {title} {Localization transitions in non-hermitian quantum mechanics},\ }\href {https://doi.org/10.1103/physrevlett.77.570} {\bibfield  {journal} {\bibinfo  {journal} {Physical Review Letters}\ }\textbf {\bibinfo {volume} {77}},\ \bibinfo {pages} {570} (\bibinfo {year} {1996})}\BibitemShut {NoStop}%
\bibitem [{\citenamefont {{Herviou}}\ \emph {et~al.}(2019)\citenamefont {{Herviou}}, \citenamefont {{Regnault}},\ and\ \citenamefont {{Bardarson}}}]{HRB2019}%
  \BibitemOpen
  \bibfield  {author} {\bibinfo {author} {\bibfnamefont {L.}~\bibnamefont {{Herviou}}}, \bibinfo {author} {\bibfnamefont {N.}~\bibnamefont {{Regnault}}},\ and\ \bibinfo {author} {\bibfnamefont {J.~H.}\ \bibnamefont {{Bardarson}}},\ }\bibfield  {title} {\bibinfo {title} {{Entanglement spectrum and symmetries in non-Hermitian fermionic non-interacting models}},\ }\href {https://doi.org/10.21468/SciPostPhys.7.5.069} {\bibfield  {journal} {\bibinfo  {journal} {SciPost Physics}\ }\textbf {\bibinfo {volume} {7}},\ \bibinfo {eid} {069} (\bibinfo {year} {2019})},\ \Eprint {https://arxiv.org/abs/1908.09852} {arXiv:1908.09852 [cond-mat.mes-hall]} \BibitemShut {NoStop}%
\bibitem [{\citenamefont {Chang}\ \emph {et~al.}(2020)\citenamefont {Chang}, \citenamefont {You}, \citenamefont {Wen},\ and\ \citenamefont {Ryu}}]{CYWR2020}%
  \BibitemOpen
  \bibfield  {author} {\bibinfo {author} {\bibfnamefont {P.-Y.}\ \bibnamefont {Chang}}, \bibinfo {author} {\bibfnamefont {J.-S.}\ \bibnamefont {You}}, \bibinfo {author} {\bibfnamefont {X.}~\bibnamefont {Wen}},\ and\ \bibinfo {author} {\bibfnamefont {S.}~\bibnamefont {Ryu}},\ }\bibfield  {title} {\bibinfo {title} {{Entanglement spectrum and entropy in topological non-Hermitian systems and nonunitary conformal field theory}},\ }\href {https://doi.org/10.1103/PhysRevResearch.2.033069} {\bibfield  {journal} {\bibinfo  {journal} {Phys. Rev. Res.}\ }\textbf {\bibinfo {volume} {2}},\ \bibinfo {pages} {033069} (\bibinfo {year} {2020})},\ \Eprint {https://arxiv.org/abs/1909.01346} {arXiv:1909.01346 [cond-mat.str-el]} \BibitemShut {NoStop}%
\bibitem [{\citenamefont {Guo}\ \emph {et~al.}(2021)\citenamefont {Guo}, \citenamefont {Yu}, \citenamefont {Huang}, \citenamefont {Yang}, \citenamefont {Chi}, \citenamefont {Liao},\ and\ \citenamefont {Xiang}}]{GYHYCLX2021}%
  \BibitemOpen
  \bibfield  {author} {\bibinfo {author} {\bibfnamefont {Y.-B.}\ \bibnamefont {Guo}}, \bibinfo {author} {\bibfnamefont {Y.-C.}\ \bibnamefont {Yu}}, \bibinfo {author} {\bibfnamefont {R.-Z.}\ \bibnamefont {Huang}}, \bibinfo {author} {\bibfnamefont {L.-P.}\ \bibnamefont {Yang}}, \bibinfo {author} {\bibfnamefont {R.-Z.}\ \bibnamefont {Chi}}, \bibinfo {author} {\bibfnamefont {H.-J.}\ \bibnamefont {Liao}},\ and\ \bibinfo {author} {\bibfnamefont {T.}~\bibnamefont {Xiang}},\ }\bibfield  {title} {\bibinfo {title} {{Entanglement entropy of non-Hermitian free fermions}},\ }\href {https://doi.org/10.1088/1361-648X/ac216e} {\bibfield  {journal} {\bibinfo  {journal} {J. Phys. Condens. Matter}\ }\textbf {\bibinfo {volume} {33}},\ \bibinfo {pages} {475502} (\bibinfo {year} {2021})},\ \Eprint {https://arxiv.org/abs/2105.09793} {arXiv:2105.09793 [cond-mat.mes-hall]} \BibitemShut {NoStop}%
\bibitem [{\citenamefont {Lee}(2022)}]{L2022}%
  \BibitemOpen
  \bibfield  {author} {\bibinfo {author} {\bibfnamefont {C.~H.}\ \bibnamefont {Lee}},\ }\bibfield  {title} {\bibinfo {title} {{Exceptional Bound States and Negative Entanglement Entropy}},\ }\href {https://doi.org/10.1103/PhysRevLett.128.010402} {\bibfield  {journal} {\bibinfo  {journal} {Phys. Rev. Lett.}\ }\textbf {\bibinfo {volume} {128}},\ \bibinfo {pages} {010402} (\bibinfo {year} {2022})},\ \Eprint {https://arxiv.org/abs/2011.09505} {arXiv:2011.09505 [cond-mat.quant-gas]} \BibitemShut {NoStop}%
\bibitem [{\citenamefont {Chen}\ \emph {et~al.}(2022)\citenamefont {Chen}, \citenamefont {Peng}, \citenamefont {Lu},\ and\ \citenamefont {Lu}}]{CPLL2022}%
  \BibitemOpen
  \bibfield  {author} {\bibinfo {author} {\bibfnamefont {W.}~\bibnamefont {Chen}}, \bibinfo {author} {\bibfnamefont {L.}~\bibnamefont {Peng}}, \bibinfo {author} {\bibfnamefont {H.}~\bibnamefont {Lu}},\ and\ \bibinfo {author} {\bibfnamefont {X.}~\bibnamefont {Lu}},\ }\bibfield  {title} {\bibinfo {title} {Characterizing bulk-boundary correspondence of one-dimensional non-hermitian interacting systems by edge entanglement entropy},\ }\bibfield  {journal} {\bibinfo  {journal} {Physical Review B}\ }\textbf {\bibinfo {volume} {105}},\ \href {https://doi.org/10.1103/physrevb.105.075126} {10.1103/physrevb.105.075126} (\bibinfo {year} {2022})\BibitemShut {NoStop}%
\bibitem [{\citenamefont {Hu}\ \emph {et~al.}(2023)\citenamefont {Hu}, \citenamefont {Fu},\ and\ \citenamefont {Zhang}}]{HFZ2023}%
  \BibitemOpen
  \bibfield  {author} {\bibinfo {author} {\bibfnamefont {S.-X.}\ \bibnamefont {Hu}}, \bibinfo {author} {\bibfnamefont {Y.}~\bibnamefont {Fu}},\ and\ \bibinfo {author} {\bibfnamefont {Y.}~\bibnamefont {Zhang}},\ }\bibfield  {title} {\bibinfo {title} {{Nontrivial worldline winding in non-Hermitian quantum systems}},\ }\href {https://doi.org/10.1103/PhysRevB.108.245114} {\bibfield  {journal} {\bibinfo  {journal} {Phys. Rev. B}\ }\textbf {\bibinfo {volume} {108}},\ \bibinfo {pages} {245114} (\bibinfo {year} {2023})},\ \Eprint {https://arxiv.org/abs/2307.01260} {arXiv:2307.01260 [quant-ph]} \BibitemShut {NoStop}%
\bibitem [{\citenamefont {Fossati}\ \emph {et~al.}(2023)\citenamefont {Fossati}, \citenamefont {Ares},\ and\ \citenamefont {Calabrese}}]{FAC2023}%
  \BibitemOpen
  \bibfield  {author} {\bibinfo {author} {\bibfnamefont {M.}~\bibnamefont {Fossati}}, \bibinfo {author} {\bibfnamefont {F.}~\bibnamefont {Ares}},\ and\ \bibinfo {author} {\bibfnamefont {P.}~\bibnamefont {Calabrese}},\ }\bibfield  {title} {\bibinfo {title} {{Symmetry-resolved entanglement in critical non-Hermitian systems}},\ }\href {https://doi.org/10.1103/PhysRevB.107.205153} {\bibfield  {journal} {\bibinfo  {journal} {Phys. Rev. B}\ }\textbf {\bibinfo {volume} {107}},\ \bibinfo {pages} {205153} (\bibinfo {year} {2023})},\ \Eprint {https://arxiv.org/abs/2303.05232} {arXiv:2303.05232 [cond-mat.stat-mech]} \BibitemShut {NoStop}%
\bibitem [{\citenamefont {{Le Gal}}\ \emph {et~al.}(2023)\citenamefont {{Le Gal}}, \citenamefont {{Turkeshi}},\ and\ \citenamefont {{Schir{\`o}}}}]{GTS2023}%
  \BibitemOpen
  \bibfield  {author} {\bibinfo {author} {\bibfnamefont {Y.}~\bibnamefont {{Le Gal}}}, \bibinfo {author} {\bibfnamefont {X.}~\bibnamefont {{Turkeshi}}},\ and\ \bibinfo {author} {\bibfnamefont {M.}~\bibnamefont {{Schir{\`o}}}},\ }\bibfield  {title} {\bibinfo {title} {{Volume-to-area law entanglement transition in a non-Hermitian free fermionic chain}},\ }\href {https://doi.org/10.21468/SciPostPhys.14.5.138} {\bibfield  {journal} {\bibinfo  {journal} {SciPost Physics}\ }\textbf {\bibinfo {volume} {14}},\ \bibinfo {eid} {138} (\bibinfo {year} {2023})},\ \Eprint {https://arxiv.org/abs/2210.11937} {arXiv:2210.11937 [cond-mat.stat-mech]} \BibitemShut {NoStop}%
\bibitem [{\citenamefont {Brunelli}\ \emph {et~al.}(2023)\citenamefont {Brunelli}, \citenamefont {Wanjura},\ and\ \citenamefont {Nunnenkamp}}]{BWN2023}%
  \BibitemOpen
  \bibfield  {author} {\bibinfo {author} {\bibfnamefont {M.}~\bibnamefont {Brunelli}}, \bibinfo {author} {\bibfnamefont {C.~C.}\ \bibnamefont {Wanjura}},\ and\ \bibinfo {author} {\bibfnamefont {A.}~\bibnamefont {Nunnenkamp}},\ }\bibfield  {title} {\bibinfo {title} {{Restoration of the non-Hermitian bulk-boundary correspondence via topological amplification}},\ }\href {https://doi.org/10.21468/SciPostPhys.15.4.173} {\bibfield  {journal} {\bibinfo  {journal} {SciPost Phys.}\ }\textbf {\bibinfo {volume} {15}},\ \bibinfo {pages} {173} (\bibinfo {year} {2023})},\ \Eprint {https://arxiv.org/abs/2207.12427} {arXiv:2207.12427 [quant-ph]} \BibitemShut {NoStop}%
\bibitem [{\citenamefont {Herviou}\ \emph {et~al.}(2019)\citenamefont {Herviou}, \citenamefont {Bardarson},\ and\ \citenamefont {Regnault}}]{HBR2019}%
  \BibitemOpen
  \bibfield  {author} {\bibinfo {author} {\bibfnamefont {L.}~\bibnamefont {Herviou}}, \bibinfo {author} {\bibfnamefont {J.~H.}\ \bibnamefont {Bardarson}},\ and\ \bibinfo {author} {\bibfnamefont {N.}~\bibnamefont {Regnault}},\ }\bibfield  {title} {\bibinfo {title} {Defining a bulk-edge correspondence for non-hermitian hamiltonians via singular-value decomposition},\ }\bibfield  {journal} {\bibinfo  {journal} {Physical Review A}\ }\textbf {\bibinfo {volume} {99}},\ \href {https://doi.org/10.1103/physreva.99.052118} {10.1103/physreva.99.052118} (\bibinfo {year} {2019}),\ \Eprint {https://arxiv.org/abs/1901.00010} {arXiv:1901.00010 [cond-mat.mes-hall]} \BibitemShut {NoStop}%
\bibitem [{\citenamefont {Arouca}\ \emph {et~al.}(2020)\citenamefont {Arouca}, \citenamefont {Lee},\ and\ \citenamefont {Morais~Smith}}]{ALMS2020}%
  \BibitemOpen
  \bibfield  {author} {\bibinfo {author} {\bibfnamefont {R.}~\bibnamefont {Arouca}}, \bibinfo {author} {\bibfnamefont {C.~H.}\ \bibnamefont {Lee}},\ and\ \bibinfo {author} {\bibfnamefont {C.}~\bibnamefont {Morais~Smith}},\ }\bibfield  {title} {\bibinfo {title} {Unconventional scaling at non-hermitian critical points},\ }\bibfield  {journal} {\bibinfo  {journal} {Physical Review B}\ }\textbf {\bibinfo {volume} {102}},\ \href {https://doi.org/10.1103/physrevb.102.245145} {10.1103/physrevb.102.245145} (\bibinfo {year} {2020}),\ \Eprint {https://arxiv.org/abs/2009.03541} {arXiv:2009.03541 [cond-mat.stat-mech]} \BibitemShut {NoStop}%
\bibitem [{\citenamefont {Halder}\ \emph {et~al.}(2023)\citenamefont {Halder}, \citenamefont {Ganguly},\ and\ \citenamefont {Basu}}]{HGB2023}%
  \BibitemOpen
  \bibfield  {author} {\bibinfo {author} {\bibfnamefont {D.}~\bibnamefont {Halder}}, \bibinfo {author} {\bibfnamefont {S.}~\bibnamefont {Ganguly}},\ and\ \bibinfo {author} {\bibfnamefont {S.}~\bibnamefont {Basu}},\ }\bibfield  {title} {\bibinfo {title} {{Properties of the non-Hermitian SSH model: role of symmetry}},\ }\href {https://doi.org/10.1088/1361-648X/acadc5} {\bibfield  {journal} {\bibinfo  {journal} {J. Phys. Condens. Matter}\ }\textbf {\bibinfo {volume} {35}},\ \bibinfo {pages} {105901} (\bibinfo {year} {2023})},\ \Eprint {https://arxiv.org/abs/2209.13838} {arXiv:2209.13838 [quant-ph]} \BibitemShut {NoStop}%
\bibitem [{\citenamefont {{Aquino}}\ \emph {et~al.}(2023)\citenamefont {{Aquino}}, \citenamefont {{Lopes}},\ and\ \citenamefont {{Barci}}}]{ALB2023}%
  \BibitemOpen
  \bibfield  {author} {\bibinfo {author} {\bibfnamefont {R.}~\bibnamefont {{Aquino}}}, \bibinfo {author} {\bibfnamefont {N.}~\bibnamefont {{Lopes}}},\ and\ \bibinfo {author} {\bibfnamefont {D.~G.}\ \bibnamefont {{Barci}}},\ }\bibfield  {title} {\bibinfo {title} {{Critical and noncritical non-Hermitian topological phase transitions in one-dimensional chains}},\ }\href {https://doi.org/10.1103/PhysRevB.107.035424} {\bibfield  {journal} {\bibinfo  {journal} {\prb}\ }\textbf {\bibinfo {volume} {107}},\ \bibinfo {eid} {035424} (\bibinfo {year} {2023})},\ \Eprint {https://arxiv.org/abs/2208.14400} {arXiv:2208.14400 [cond-mat.str-el]} \BibitemShut {NoStop}%
\bibitem [{\citenamefont {{Kitagawa}}\ \emph {et~al.}(2010)\citenamefont {{Kitagawa}}, \citenamefont {{Berg}}, \citenamefont {{Rudner}},\ and\ \citenamefont {{Demler}}}]{KBRD2010}%
  \BibitemOpen
  \bibfield  {author} {\bibinfo {author} {\bibfnamefont {T.}~\bibnamefont {{Kitagawa}}}, \bibinfo {author} {\bibfnamefont {E.}~\bibnamefont {{Berg}}}, \bibinfo {author} {\bibfnamefont {M.}~\bibnamefont {{Rudner}}},\ and\ \bibinfo {author} {\bibfnamefont {E.}~\bibnamefont {{Demler}}},\ }\bibfield  {title} {\bibinfo {title} {{Topological characterization of periodically driven quantum systems}},\ }\href {https://doi.org/10.1103/PhysRevB.82.235114} {\bibfield  {journal} {\bibinfo  {journal} {\prb}\ }\textbf {\bibinfo {volume} {82}},\ \bibinfo {eid} {235114} (\bibinfo {year} {2010})},\ \Eprint {https://arxiv.org/abs/1010.6126} {arXiv:1010.6126 [cond-mat.mes-hall]} \BibitemShut {NoStop}%
\bibitem [{\citenamefont {Leykam}\ \emph {et~al.}(2017)\citenamefont {Leykam}, \citenamefont {Bliokh}, \citenamefont {Huang}, \citenamefont {Chong},\ and\ \citenamefont {Nori}}]{LBHCN2017}%
  \BibitemOpen
  \bibfield  {author} {\bibinfo {author} {\bibfnamefont {D.}~\bibnamefont {Leykam}}, \bibinfo {author} {\bibfnamefont {K.~Y.}\ \bibnamefont {Bliokh}}, \bibinfo {author} {\bibfnamefont {C.}~\bibnamefont {Huang}}, \bibinfo {author} {\bibfnamefont {Y.~D.}\ \bibnamefont {Chong}},\ and\ \bibinfo {author} {\bibfnamefont {F.}~\bibnamefont {Nori}},\ }\bibfield  {title} {\bibinfo {title} {{Edge Modes, Degeneracies, and Topological Numbers in Non-Hermitian Systems}},\ }\href {https://doi.org/10.1103/PhysRevLett.118.040401} {\bibfield  {journal} {\bibinfo  {journal} {Phys. Rev. Lett.}\ }\textbf {\bibinfo {volume} {118}},\ \bibinfo {pages} {040401} (\bibinfo {year} {2017})},\ \Eprint {https://arxiv.org/abs/1610.04029} {arXiv:1610.04029 [cond-mat.mes-hall]} \BibitemShut {NoStop}%
\bibitem [{\citenamefont {Shen}\ \emph {et~al.}(2018)\citenamefont {Shen}, \citenamefont {Zhen},\ and\ \citenamefont {Fu}}]{SZF2018}%
  \BibitemOpen
  \bibfield  {author} {\bibinfo {author} {\bibfnamefont {H.}~\bibnamefont {Shen}}, \bibinfo {author} {\bibfnamefont {B.}~\bibnamefont {Zhen}},\ and\ \bibinfo {author} {\bibfnamefont {L.}~\bibnamefont {Fu}},\ }\bibfield  {title} {\bibinfo {title} {Topological band theory for non-hermitian hamiltonians},\ }\href {https://doi.org/10.1103/PhysRevLett.120.146402} {\bibfield  {journal} {\bibinfo  {journal} {Phys. Rev. Lett.}\ }\textbf {\bibinfo {volume} {120}},\ \bibinfo {pages} {146402} (\bibinfo {year} {2018})}\BibitemShut {NoStop}%
\bibitem [{\citenamefont {{Garrison}}\ and\ \citenamefont {{Wright}}(1988)}]{GW1988}%
  \BibitemOpen
  \bibfield  {author} {\bibinfo {author} {\bibfnamefont {J.~C.}\ \bibnamefont {{Garrison}}}\ and\ \bibinfo {author} {\bibfnamefont {E.~M.}\ \bibnamefont {{Wright}}},\ }\bibfield  {title} {\bibinfo {title} {{Complex geometrical phases for dissipative systems}},\ }\href {https://doi.org/10.1016/0375-9601(88)90905-X} {\bibfield  {journal} {\bibinfo  {journal} {Physics Letters A}\ }\textbf {\bibinfo {volume} {128}},\ \bibinfo {pages} {177} (\bibinfo {year} {1988})}\BibitemShut {NoStop}%
\bibitem [{\citenamefont {Dattoli}\ \emph {et~al.}(1990)\citenamefont {Dattoli}, \citenamefont {Mignani},\ and\ \citenamefont {Torre}}]{DMT1990}%
  \BibitemOpen
  \bibfield  {author} {\bibinfo {author} {\bibfnamefont {G.}~\bibnamefont {Dattoli}}, \bibinfo {author} {\bibfnamefont {R.}~\bibnamefont {Mignani}},\ and\ \bibinfo {author} {\bibfnamefont {A.}~\bibnamefont {Torre}},\ }\bibfield  {title} {\bibinfo {title} {Geometrical phase in the cyclic evolution of non-hermitian systems},\ }\href {https://doi.org/10.1088/0305-4470/23/24/020} {\bibfield  {journal} {\bibinfo  {journal} {Journal of Physics A: Mathematical and General}\ }\textbf {\bibinfo {volume} {23}},\ \bibinfo {pages} {5795} (\bibinfo {year} {1990})}\BibitemShut {NoStop}%
\bibitem [{\citenamefont {{Mostafazadeh}}(1999)}]{M1999}%
  \BibitemOpen
  \bibfield  {author} {\bibinfo {author} {\bibfnamefont {A.}~\bibnamefont {{Mostafazadeh}}},\ }\bibfield  {title} {\bibinfo {title} {{A new class of adiabatic cyclic states and geometric phases for non-Hermitian Hamiltonians}},\ }\href {https://doi.org/10.1016/S0375-9601(99)00790-2} {\bibfield  {journal} {\bibinfo  {journal} {Physics Letters A}\ }\textbf {\bibinfo {volume} {264}},\ \bibinfo {pages} {11} (\bibinfo {year} {1999})},\ \Eprint {https://arxiv.org/abs/quant-ph/9911003} {arXiv:quant-ph/9911003 [quant-ph]} \BibitemShut {NoStop}%
\bibitem [{\citenamefont {Liang}\ and\ \citenamefont {Huang}(2013)}]{LH2013}%
  \BibitemOpen
  \bibfield  {author} {\bibinfo {author} {\bibfnamefont {S.-D.}\ \bibnamefont {Liang}}\ and\ \bibinfo {author} {\bibfnamefont {G.-Y.}\ \bibnamefont {Huang}},\ }\bibfield  {title} {\bibinfo {title} {Topological invariance and global berry phase in non-hermitian systems},\ }\href {https://doi.org/10.1103/PhysRevA.87.012118} {\bibfield  {journal} {\bibinfo  {journal} {Phys. Rev. A}\ }\textbf {\bibinfo {volume} {87}},\ \bibinfo {pages} {012118} (\bibinfo {year} {2013})}\BibitemShut {NoStop}%
\bibitem [{\citenamefont {{Jiang}}\ \emph {et~al.}(2018)\citenamefont {{Jiang}}, \citenamefont {{Yang}},\ and\ \citenamefont {{Chen}}}]{JYC2018}%
  \BibitemOpen
  \bibfield  {author} {\bibinfo {author} {\bibfnamefont {H.}~\bibnamefont {{Jiang}}}, \bibinfo {author} {\bibfnamefont {C.}~\bibnamefont {{Yang}}},\ and\ \bibinfo {author} {\bibfnamefont {S.}~\bibnamefont {{Chen}}},\ }\bibfield  {title} {\bibinfo {title} {{Topological invariants and phase diagrams for one-dimensional two-band non-Hermitian systems without chiral symmetry}},\ }\href {https://doi.org/10.1103/PhysRevA.98.052116} {\bibfield  {journal} {\bibinfo  {journal} {\pra}\ }\textbf {\bibinfo {volume} {98}},\ \bibinfo {eid} {052116} (\bibinfo {year} {2018})},\ \Eprint {https://arxiv.org/abs/1809.00850} {arXiv:1809.00850 [cond-mat.mes-hall]} \BibitemShut {NoStop}%
\bibitem [{\citenamefont {Gover}(1994)}]{Gover94}%
  \BibitemOpen
  \bibfield  {author} {\bibinfo {author} {\bibfnamefont {M.}~\bibnamefont {Gover}},\ }\bibfield  {title} {\bibinfo {title} {The eigenproblem of a tridiagonal 2-{T}oeplitz matrix},\ }\href {https://doi.org/https://doi.org/10.1016/0024-3795(94)90481-2} {\bibfield  {journal} {\bibinfo  {journal} {Linear Algebra and its Applications}\ }\textbf {\bibinfo {volume} {197-198}},\ \bibinfo {pages} {63} (\bibinfo {year} {1994})}\BibitemShut {NoStop}%
\bibitem [{\citenamefont {Shin}(1997)}]{shin_1997}%
  \BibitemOpen
  \bibfield  {author} {\bibinfo {author} {\bibfnamefont {B.~C.}\ \bibnamefont {Shin}},\ }\bibfield  {title} {\bibinfo {title} {A formula for eigenpairs of certain symmetric tridiagonal matrices},\ }\href {https://doi.org/10.1017/S0004972700033918} {\bibfield  {journal} {\bibinfo  {journal} {Bulletin of the Australian Mathematical Society}\ }\textbf {\bibinfo {volume} {55}},\ \bibinfo {pages} {249–254} (\bibinfo {year} {1997})}\BibitemShut {NoStop}%
\bibitem [{\citenamefont {Ikramov}(1999)}]{Ikramov_1999}%
  \BibitemOpen
  \bibfield  {author} {\bibinfo {author} {\bibfnamefont {K.~D.}\ \bibnamefont {Ikramov}},\ }\bibfield  {title} {\bibinfo {title} {Shin’s formulas for eigenpairs of symmetric tridiagonal 2-toeplitz matrices},\ }\href {https://doi.org/10.1017/S0004972700032664} {\bibfield  {journal} {\bibinfo  {journal} {Bulletin of the Australian Mathematical Society}\ }\textbf {\bibinfo {volume} {59}},\ \bibinfo {pages} {119–120} (\bibinfo {year} {1999})}\BibitemShut {NoStop}%
\bibitem [{\citenamefont {Sirker}\ \emph {et~al.}(2014)\citenamefont {Sirker}, \citenamefont {Maiti}, \citenamefont {Konstantinidis},\ and\ \citenamefont {Sedlmayr}}]{SMKS2014}%
  \BibitemOpen
  \bibfield  {author} {\bibinfo {author} {\bibfnamefont {J.}~\bibnamefont {Sirker}}, \bibinfo {author} {\bibfnamefont {M.}~\bibnamefont {Maiti}}, \bibinfo {author} {\bibfnamefont {N.~P.}\ \bibnamefont {Konstantinidis}},\ and\ \bibinfo {author} {\bibfnamefont {N.}~\bibnamefont {Sedlmayr}},\ }\bibfield  {title} {\bibinfo {title} {Boundary fidelity and entanglement in the symmetry protected topological phase of the {SSH} model},\ }\href {https://doi.org/10.1088/1742-5468/2014/10/p10032} {\bibfield  {journal} {\bibinfo  {journal} {Journal of Statistical Mechanics: Theory and Experiment}\ }\textbf {\bibinfo {volume} {2014}},\ \bibinfo {pages} {P10032} (\bibinfo {year} {2014})},\ \Eprint {https://arxiv.org/abs/1406.7832} {arXiv:1406.7832 [cond-mat.str-el]} \BibitemShut {NoStop}%
\bibitem [{\citenamefont {{Cheng}}\ \emph {et~al.}(2024)\citenamefont {{Cheng}}, \citenamefont {{Batchelor}},\ and\ \citenamefont {{Cocks}}}]{CBC2023arXiv}%
  \BibitemOpen
  \bibfield  {author} {\bibinfo {author} {\bibfnamefont {E.}~\bibnamefont {{Cheng}}}, \bibinfo {author} {\bibfnamefont {M.~T.}\ \bibnamefont {{Batchelor}}},\ and\ \bibinfo {author} {\bibfnamefont {D.}~\bibnamefont {{Cocks}}},\ }\bibfield  {title} {\bibinfo {title} {{Topological analysis of the complex SSH model using the quantum geometric tensor}},\ }\href {https://doi.org/10.1088/1751-8121/ad5d2e} {\bibfield  {journal} {\bibinfo  {journal} {Journal of Physics A Mathematical General}\ }\textbf {\bibinfo {volume} {57}},\ \bibinfo {eid} {305001} (\bibinfo {year} {2024})},\ \Eprint {https://arxiv.org/abs/2308.04626} {arXiv:2308.04626 [cond-mat.str-el]} \BibitemShut {NoStop}%
\bibitem [{\citenamefont {Pfeuty}(1970)}]{P70}%
  \BibitemOpen
  \bibfield  {author} {\bibinfo {author} {\bibfnamefont {P.}~\bibnamefont {Pfeuty}},\ }\bibfield  {title} {\bibinfo {title} {The one-dimensional {I}sing model with a transverse field},\ }\href {https://doi.org/10.1016/0003-4916(70)90270-8} {\bibfield  {journal} {\bibinfo  {journal} {Annals of Physics}\ }\textbf {\bibinfo {volume} {57}},\ \bibinfo {pages} {79} (\bibinfo {year} {1970})}\BibitemShut {NoStop}%
\bibitem [{\citenamefont {Henry}\ and\ \citenamefont {Batchelor}(2023)}]{HB2023}%
  \BibitemOpen
  \bibfield  {author} {\bibinfo {author} {\bibfnamefont {R.~A.}\ \bibnamefont {Henry}}\ and\ \bibinfo {author} {\bibfnamefont {M.~T.}\ \bibnamefont {Batchelor}},\ }\bibfield  {title} {\bibinfo {title} {{Exceptional points in the Baxter-Fendley free parafermion model}},\ }\href {https://doi.org/10.21468/SciPostPhys.15.1.016} {\bibfield  {journal} {\bibinfo  {journal} {SciPost Phys.}\ }\textbf {\bibinfo {volume} {15}},\ \bibinfo {pages} {016} (\bibinfo {year} {2023})},\ \Eprint {https://arxiv.org/abs/2301.11031} {arXiv:2301.11031 [cond-mat.stat-mech]} \BibitemShut {NoStop}%
\bibitem [{\citenamefont {Ryu}\ \emph {et~al.}(2010)\citenamefont {Ryu}, \citenamefont {Schnyder}, \citenamefont {Furusaki},\ and\ \citenamefont {Ludwig}}]{RyuSchnyder}%
  \BibitemOpen
  \bibfield  {author} {\bibinfo {author} {\bibfnamefont {S.}~\bibnamefont {Ryu}}, \bibinfo {author} {\bibfnamefont {A.~P.}\ \bibnamefont {Schnyder}}, \bibinfo {author} {\bibfnamefont {A.}~\bibnamefont {Furusaki}},\ and\ \bibinfo {author} {\bibfnamefont {A.~W.~W.}\ \bibnamefont {Ludwig}},\ }\bibfield  {title} {\bibinfo {title} {Topological insulators and superconductors: tenfold way and dimensional hierarchy},\ }\href@noop {} {\bibfield  {journal} {\bibinfo  {journal} {New Journal of Physics}\ }\textbf {\bibinfo {volume} {12}},\ \bibinfo {pages} {065010} (\bibinfo {year} {2010})}\BibitemShut {NoStop}%
\bibitem [{\citenamefont {Kawabata}\ \emph {et~al.}(2023)\citenamefont {Kawabata}, \citenamefont {Xiao}, \citenamefont {Ohtsuki},\ and\ \citenamefont {Shindou}}]{KawabataXiao}%
  \BibitemOpen
  \bibfield  {author} {\bibinfo {author} {\bibfnamefont {K.}~\bibnamefont {Kawabata}}, \bibinfo {author} {\bibfnamefont {Z.}~\bibnamefont {Xiao}}, \bibinfo {author} {\bibfnamefont {T.}~\bibnamefont {Ohtsuki}},\ and\ \bibinfo {author} {\bibfnamefont {R.}~\bibnamefont {Shindou}},\ }\bibfield  {title} {\bibinfo {title} {Singular-value statistics of non-hermitian random matrices and open quantum systems},\ }\href {https://doi.org/10.1103/PRXQuantum.4.040312} {\bibfield  {journal} {\bibinfo  {journal} {PRX Quantum}\ }\textbf {\bibinfo {volume} {4}},\ \bibinfo {pages} {040312} (\bibinfo {year} {2023})}\BibitemShut {NoStop}%
\bibitem [{\citenamefont {Porras}\ and\ \citenamefont {Fern\'andez-Lorenzo}(2019)}]{Porras1}%
  \BibitemOpen
  \bibfield  {author} {\bibinfo {author} {\bibfnamefont {D.}~\bibnamefont {Porras}}\ and\ \bibinfo {author} {\bibfnamefont {S.}~\bibnamefont {Fern\'andez-Lorenzo}},\ }\bibfield  {title} {\bibinfo {title} {Topological amplification in photonic lattices},\ }\href {https://doi.org/10.1103/PhysRevLett.122.143901} {\bibfield  {journal} {\bibinfo  {journal} {Phys. Rev. Lett.}\ }\textbf {\bibinfo {volume} {122}},\ \bibinfo {pages} {143901} (\bibinfo {year} {2019})}\BibitemShut {NoStop}%
\bibitem [{\citenamefont {B{\"o}ttcher}\ and\ \citenamefont {Silbermann}(1999)}]{BoettcherSilbermann}%
  \BibitemOpen
  \bibfield  {author} {\bibinfo {author} {\bibfnamefont {A.}~\bibnamefont {B{\"o}ttcher}}\ and\ \bibinfo {author} {\bibfnamefont {B.}~\bibnamefont {Silbermann}},\ }\href@noop {} {\emph {\bibinfo {title} {Introduction to large truncated Toeplitz matrices}}}\ (\bibinfo  {publisher} {Springer (New York)},\ \bibinfo {year} {1999})\BibitemShut {NoStop}%
\bibitem [{\citenamefont {Monkman}\ and\ \citenamefont {Sirker}(2023)}]{MonkmanSirker4}%
  \BibitemOpen
  \bibfield  {author} {\bibinfo {author} {\bibfnamefont {K.}~\bibnamefont {Monkman}}\ and\ \bibinfo {author} {\bibfnamefont {J.}~\bibnamefont {Sirker}},\ }\bibfield  {title} {\bibinfo {title} {Symmetry-resolved entanglement: general considerations, calculation from correlation functions, and bounds for symmetry-protected topological phases},\ }\href {https://doi.org/10.1088/1751-8121/ad086d} {\bibfield  {journal} {\bibinfo  {journal} {Journal of Physics A: Mathematical and Theoretical}\ }\textbf {\bibinfo {volume} {56}},\ \bibinfo {pages} {495001} (\bibinfo {year} {2023})}\BibitemShut {NoStop}%
\bibitem [{\citenamefont {{Peschel}}(2003)}]{P2003}%
  \BibitemOpen
  \bibfield  {author} {\bibinfo {author} {\bibfnamefont {I.}~\bibnamefont {{Peschel}}},\ }\bibfield  {title} {\bibinfo {title} {{LETTER TO THE EDITOR: Calculation of reduced density matrices from correlation functions}},\ }\href {https://doi.org/10.1088/0305-4470/36/14/101} {\bibfield  {journal} {\bibinfo  {journal} {Journal of Physics A Mathematical General}\ }\textbf {\bibinfo {volume} {36}},\ \bibinfo {pages} {L205} (\bibinfo {year} {2003})},\ \Eprint {https://arxiv.org/abs/cond-mat/0212631} {arXiv:cond-mat/0212631 [cond-mat]} \BibitemShut {NoStop}%
\bibitem [{\citenamefont {{Vidal}}\ \emph {et~al.}(2003)\citenamefont {{Vidal}}, \citenamefont {{Latorre}}, \citenamefont {{Rico}},\ and\ \citenamefont {{Kitaev}}}]{VLRK2003}%
  \BibitemOpen
  \bibfield  {author} {\bibinfo {author} {\bibfnamefont {G.}~\bibnamefont {{Vidal}}}, \bibinfo {author} {\bibfnamefont {J.~I.}\ \bibnamefont {{Latorre}}}, \bibinfo {author} {\bibfnamefont {E.}~\bibnamefont {{Rico}}},\ and\ \bibinfo {author} {\bibfnamefont {A.}~\bibnamefont {{Kitaev}}},\ }\bibfield  {title} {\bibinfo {title} {{Entanglement in Quantum Critical Phenomena}},\ }\href {https://doi.org/10.1103/PhysRevLett.90.227902} {\bibfield  {journal} {\bibinfo  {journal} {\prl}\ }\textbf {\bibinfo {volume} {90}},\ \bibinfo {eid} {227902} (\bibinfo {year} {2003})},\ \Eprint {https://arxiv.org/abs/quant-ph/0211074} {arXiv:quant-ph/0211074 [quant-ph]} \BibitemShut {NoStop}%
\bibitem [{\citenamefont {Shi}\ \emph {et~al.}(2024)\citenamefont {Shi}, \citenamefont {Dong}, \citenamefont {Bao},\ and\ \citenamefont {Guo}}]{ShiDong}%
  \BibitemOpen
  \bibfield  {author} {\bibinfo {author} {\bibfnamefont {S.}~\bibnamefont {Shi}}, \bibinfo {author} {\bibfnamefont {L.}~\bibnamefont {Dong}}, \bibinfo {author} {\bibfnamefont {J.}~\bibnamefont {Bao}},\ and\ \bibinfo {author} {\bibfnamefont {B.}~\bibnamefont {Guo}},\ }\bibfield  {title} {\bibinfo {title} {Entanglement entropy and topological properties in a long-range non-hermitian su–schrieffer–heeger model},\ }\href@noop {} {\bibfield  {journal} {\bibinfo  {journal} {Physica B: Condensed Matter}\ }\textbf {\bibinfo {volume} {674}},\ \bibinfo {pages} {415601} (\bibinfo {year} {2024})}\BibitemShut {NoStop}%
\bibitem [{\citenamefont {Fendley}(2019)}]{F19}%
  \BibitemOpen
  \bibfield  {author} {\bibinfo {author} {\bibfnamefont {P.}~\bibnamefont {Fendley}},\ }\bibfield  {title} {\bibinfo {title} {Free fermions in disguise},\ }\href {https://doi.org/10.1088/1751-8121/ab305d} {\bibfield  {journal} {\bibinfo  {journal} {Journal of Physics A: Mathematical and Theoretical}\ }\textbf {\bibinfo {volume} {52}},\ \bibinfo {pages} {335002} (\bibinfo {year} {2019})}\BibitemShut {NoStop}%
\bibitem [{\citenamefont {Alcaraz}\ and\ \citenamefont {Pimenta}(2020)}]{AP20}%
  \BibitemOpen
  \bibfield  {author} {\bibinfo {author} {\bibfnamefont {F.~C.}\ \bibnamefont {Alcaraz}}\ and\ \bibinfo {author} {\bibfnamefont {R.~A.}\ \bibnamefont {Pimenta}},\ }\bibfield  {title} {\bibinfo {title} {Free fermionic and parafermionic quantum spin chains with multispin interactions},\ }\bibfield  {journal} {\bibinfo  {journal} {Physical Review B}\ }\textbf {\bibinfo {volume} {102}},\ \href {https://doi.org/10.1103/physrevb.102.121101} {10.1103/physrevb.102.121101} (\bibinfo {year} {2020})\BibitemShut {NoStop}%
\bibitem [{\citenamefont {Wang}\ \emph {et~al.}(2023)\citenamefont {Wang}, \citenamefont {Li}, \citenamefont {Song},\ and\ \citenamefont {Wang}}]{Wang_2023}%
  \BibitemOpen
  \bibfield  {author} {\bibinfo {author} {\bibfnamefont {H.-R.}\ \bibnamefont {Wang}}, \bibinfo {author} {\bibfnamefont {B.}~\bibnamefont {Li}}, \bibinfo {author} {\bibfnamefont {F.}~\bibnamefont {Song}},\ and\ \bibinfo {author} {\bibfnamefont {Z.}~\bibnamefont {Wang}},\ }\bibfield  {title} {\bibinfo {title} {Scale-free non-hermitian skin effect in a boundary-dissipated spin chain},\ }\bibfield  {journal} {\bibinfo  {journal} {SciPost Physics}\ }\textbf {\bibinfo {volume} {15}},\ \href {https://doi.org/10.21468/scipostphys.15.5.191} {10.21468/scipostphys.15.5.191} (\bibinfo {year} {2023})\BibitemShut {NoStop}%
\bibitem [{\citenamefont {Gelfand}\ \emph {et~al.}(1994)\citenamefont {Gelfand}, \citenamefont {Kapranov},\ and\ \citenamefont {Zelevinsky}}]{Gelfand1994}%
  \BibitemOpen
  \bibfield  {author} {\bibinfo {author} {\bibfnamefont {I.~M.}\ \bibnamefont {Gelfand}}, \bibinfo {author} {\bibfnamefont {M.~M.}\ \bibnamefont {Kapranov}},\ and\ \bibinfo {author} {\bibfnamefont {A.~V.}\ \bibnamefont {Zelevinsky}},\ }\href {https://doi.org/10.1007/978-0-8176-4771-1} {\emph {\bibinfo {title} {Discriminants, Resultants, and Multidimensional Determinants}}}\ (\bibinfo  {publisher} {Birkhäuser Boston},\ \bibinfo {year} {1994})\BibitemShut {NoStop}%
\end{thebibliography}

%

\end{document}